\documentclass[final,1p,times,authoryear]{elsarticle}

\pdfoutput=1
\usepackage{geometry}                % See geometry.pdf to learn the layout options. There are lots.
\usepackage{amssymb}
\usepackage{longtable}
\usepackage{natbib}  % no idea why
\usepackage{epstopdf}

\usepackage{color}
\usepackage{amsmath}
\usepackage[normalem]{ulem}
\usepackage{wasysym}  % a symbol package that contains \permil and many useful astronomical symbols
\journal{Geochimica Cosmochimica Acta}

\begin{document}

\begin{frontmatter}



\title{Strange messenger: A new history of hydrogen on Earth, as told by Xenon}

\author{Kevin J.\ Zahnle\corref{cor1}\fnref{label1}}
\ead{Kevin.J.Zahnle@NASA.gov}

\cortext[cor1]{Corresponding author}
\address[label1]{Space Science Division, NASA Ames Research Center, MS 245-3, Moffett Field CA 94035 USA}

\author{Marko Gacesa\fnref{label2}}

\address[label2]{Space Sciences Lab, UC Berkeley CA 94720 USA}

\author{David C.\ Catling\fnref{label3}}
%\ead{anthony.r.dobrovolskis@nasa.gov}

\address[label3]{Department of Geology, University of Washington, Seattle WA , USA}

\begin{abstract}

Atmospheric xenon is strongly mass fractionated,
the result of a process that apparently continued through the Archean and perhaps beyond.
Previous models that explain Xe fractionation by hydrodynamic hydrogen escape cannot gracefully explain how Xe escaped
when Ar and Kr did not, nor allow Xe to escape in the Archean. 
Here we show that Xe is the only noble gas that can escape as an ion in a photo-ionized hydrogen wind, 
% This works because Xe ionizes more easily than hydrogen, and as an ion it couples strongly to other ions.  
possible in the absence of a geomagnetic field or along polar magnetic field lines that open into interplanetary space.
To quantify the hypothesis we construct new 1-D models of hydrodynamic 
diffusion-limited hydrogen escape from highly-irradiated CO$_2$-H$_2$-H atmospheres.
The models reveal three minimum requirements for Xe escape:  solar EUV irradiation needs to exceed $10\times$ that of the modern Sun; the total hydrogen mixing ratio in the atmosphere needs to exceed 1\% (equiv.\ to $0.5\%$ CH$_4$);
and transport amongst the ions in the lower ionosphere needs to lift the Xe ions to the base of the outflowing hydrogen corona.
The long duration of Xe escape implies that, if a constant process, Earth lost the hydrogen from at least one ocean of water,
 roughly evenly split between the Hadean and the Archean. 
However, to account for both Xe's fractionation and also its depletion 
with respect to Kr and primordial $^{244}$Pu, Xe escape 
must have been limited to small apertures or short episodes, 
which suggests that Xe escape was restricted to polar windows by a geomagnetic field,
or dominated by outbursts of high solar activity, or limited to transient episodes of abundant hydrogen, or a combination of these.
  Xenon escape stopped when 
the hydrogen (or methane) mixing ratio became too small, or
 EUV radiation from the aging Sun became too weak, or
charge exchange between Xe$^+$ and O$_2$ rendered Xe neutral.
%{\color{red} \sout{ All three are plausible.}}
\newpage

\end{abstract}

 \begin{keyword}  % seek for these

 Earth atmospheric evolution, noble gases %\sep Earth thermal evolution 


% 96.12.Jt  Atmospheres
 
 Accepted 17 Sept 2018


  \end{keyword}

\end{frontmatter}


\section{Introduction}
\label{Introduction}

Hydrogen escape offers a plausible explanation for the oxygenation of Earth's atmosphere \citep{Catling2001,Zahnle2013}.
The mechanism is straightforward: hydrogen escape irreversibly oxidizes the Earth, beginning with the atmosphere,
and then working its way down.
The importance of hydrogen in Earth's early atmosphere in promoting the 
formation of molecules suitable to the origin of life was recognized before the dawn of the space age \citep{Urey1952}. 
The subsequent importance of hydrogen escape in promoting chemical evolution in a direction
suitable to creating life was also recognized by \citet{Urey1952}. 
The potential importance of hydrogen escape in driving biological evolution toward oxygen-using
aerobic ecologies has not been as fully appreciated, but it is quite clear that the bias is there.

The hydrogen that escapes derives mostly from water.
It can be liberated from water by photolysis, by photosynthesis
followed by fermentation or diagenesis of organic matter releasing H$_2$ or CH$_4$, or by oxidation of the crust and mantle. 
A smaller source was the hydrogen in hydrocarbons delivered by asteroids and comets.
Earth's tiny Ne/N ratio, two elements that are equally abundant in the Sun, argues persuasively against a significant primary reservoir of gravitationally-captured solar nebular gases \citep{Aston1924,Brown1949,Zahnle2010}. 
Hydrogen is common to many atmospheric gases, 
but above Earth's water vapor cold trap only H$_2$ or CH$_4$ can be abundant.
At still higher altitudes, atmospheric photochemistry decomposes CH$_4$ into H$_2$ and other products;
thus, even if $\mathrm{CH}_4$ were more abundant near the surface,
$\mathrm{H}_2$ would be the more abundant species at the top of the atmosphere.

Neither hydrogen nor methane leave much of a signal in the geologic record.
The D/H ratio in old rocks has been interpreted as suggesting considerable hydrogen escape early in Earth's history \citep{Pope2012},
but the argument does not appear to have been widely accepted \citep[e.g.,][]{Korenaga2017}, because 
the D/H ratio is relatively susceptible to alteration under diagenesis.
In the late Archean ca 2.6-2.8 Ga there is circumstantial evidence that biogenic methane was present in the atmosphere,
implicated in the generation of isotopically light carbon \citep{Hayes1994,Hinrichs2002,Zerkle2012},
and in the generation of ``mass-independent'' isotopic fractionations of sulfur \citep[``MIF-S'',][]{Zahnle2006}.

Here we discuss new evidence from a strange messenger that suggests that Earth's atmosphere, for much of the first half of its history, contained a great deal of hydrogen or methane, 
and that the amount of hydrogen escape may have been much greater than has been appreciated.

\subsection{Xenon's story}

Xenon is the heaviest gas found in natural planetary atmospheres.  
It would therefore seem the least likely to escape to space.
Yet there is more circumstantial evidence that xenon has escaped from Earth than for any element other than helium.
% Even for hydrogen, which we know escapes from Earth today, the evidence for historical escape is no stronger. 

\begin{figure}[!htb]
   \centering
    \includegraphics[width=0.95\textwidth]{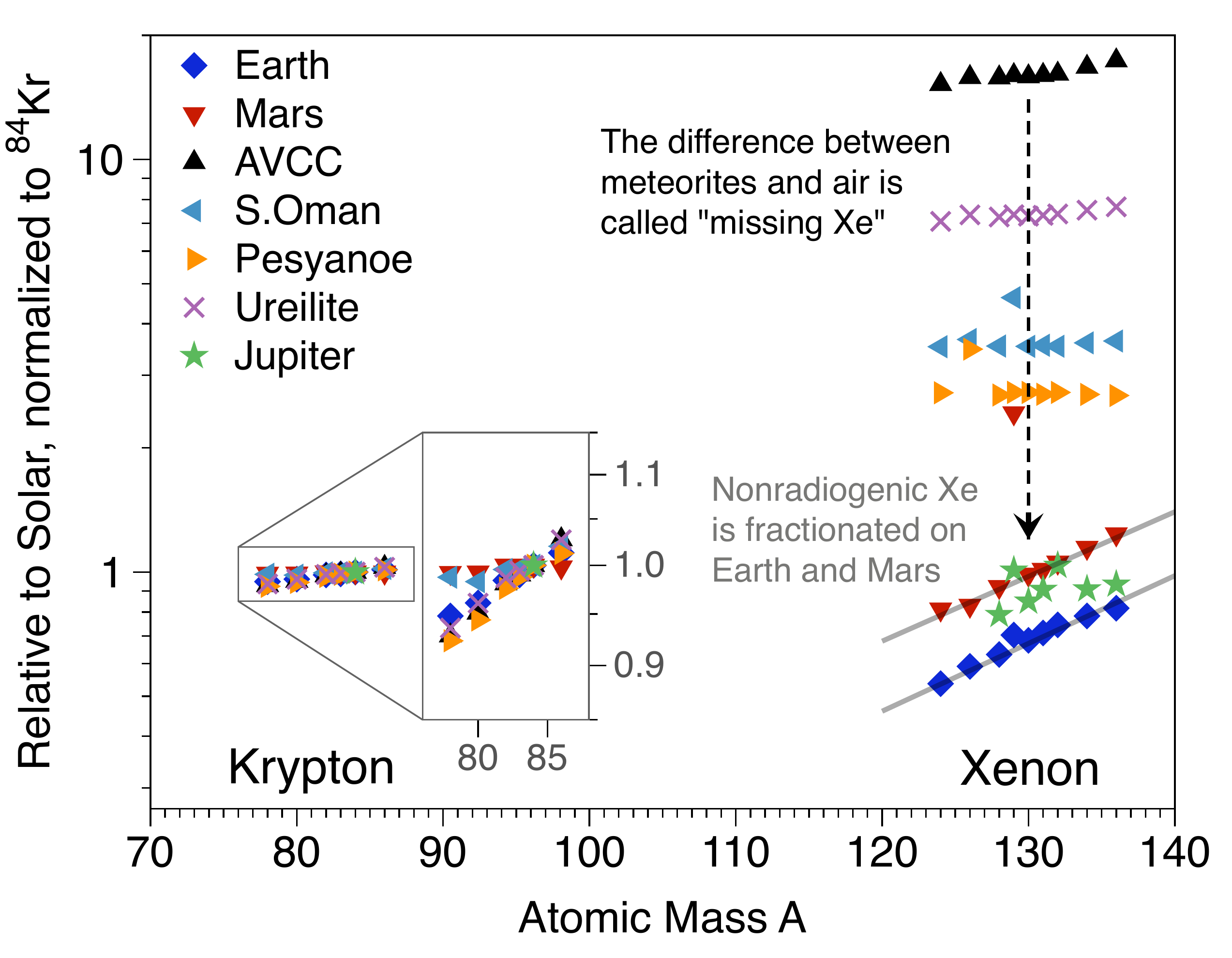} 
   \caption{
   Xenon and krypton isotopes of Earth (air), Mars (atmosphere), average carbonaceous chondritic meteorites (AVCC),
 a gas-rich enstatite chondrite (S.\ Oman), a gas-rich enstatite achondrite (Pesyanoe), a ureilite (Havero), and Jupiter,
 all normalized against solar abundances and plotted on a scale where $^{84}$Kr is one.
   Meteorite data are from \citet{Pepin1991}, Mars from SNC meteorites \citep{Swindle1997} and Jupiter from the {\em Galileo Probe} \citep{Atreya2003}.
   Solar isotopes are from \citet{Meshik2014}.  
   The straightforward interpretation of Xe's depletion and mass fractionation 
   is that Xe has preferentially escaped from the atmospheres of Earth and Mars.
   The inset shows Kr isotopes, the striking features of which are the kinship between Mars and solar on one hand, and the kinship between Earth and meteorites on the other. 
   The modest fractionations seen in Kr all fall within the range of known meteoritical sources and
   hence cannot be attributed unambiguously to escape.
   } 
\label{EarthMars2}
\end{figure}

The evidence is of three kinds.
First, the nine stable isotopes of atmospheric xenon are strongly mass fractionated compared to any known solar system source.
The magnitude of the fractionation -- about 4\% per amu, or 60\% from $^{124}$Xe to $^{136}$Xe -- is very great.  
The fractionation is easily seen on Figure \ref{EarthMars2}, which compares Xe to Kr  
in a variety of solar system materials.

Second, atmospheric Xe is depleted with respect to meteorites by a factor of 4-20 when compared to krypton.
 The elemental depletion is also obvious on Figure \ref{EarthMars2}. %, is called the ``missing xenon'' problem \citep{Ozima1983,Ozima2002}.
     Xenon's low abundance relative to krypton, when compared to carbonaceous chondrites, has been called the ``missing xenon'' problem \citep{Ozima1983}.

Third, xenon's radiogenic isotopes are much less abundant than they should be when
measured against the presumed cosmic abundances of two extinct parents.
The disappearance of radiogenic Xe does not demand fractionating xenon escape;
 it can be accommodated by a wide range of speculative escape mechanisms, such as giant impacts,
 but it does set some limits. %,
 
No more than 7\% of the $^{129}$Xe in Earth's atmosphere 
 derives from decay of $^{129}$I (15.7 Myr half life) \citep{Pepin2006}. 
This is only 1\% of what Earth could have, given the estimated abundance of iodine in the
Earth and the inferred primordial relative abundance of $^{129}$I in meteorites \citep{Tolstikhin2014}.
Moreover, some or much of the excess $^{129}$Xe could be cometary \citep{Marty2017}. 
 Clearly xenon escape was pervasive early in the history of Earth or the materials it was made from.
 However, because $^{129}$Xe escape may not be pertinent to Xe fractionation,
 we will not make use of the $^{129}$I-$^{129}$Xe system in this study.
 
A more useful constraint comes from spontaneous fission of $^{244}$Pu (half life 80 Myr),
 which spawns a distinctive spectrum of heavy xenon isotopes.  The initial abundance of $^{244}$Pu in Earth
 can be estimated from Earth's U abundance and the Pu/U ratio in meteorites.
 The amount of fissiogenic Xe in the atmosphere is determined from the difference between a smooth
 mass fractionation process acting on Earth's primordial Xe (U-Xe) and what is actually in the atmosphere \citep{Pepin2000,Pepin2006}.
 It turns out that only about 20\% of the expected amount of fissiogenic Xe is in the atmosphere,
 and there is much less in the mantle \citep{Tolstikhin2014}.
 Retention of 20\% of fissiogenic Xe sets a bound on Xe escape taking
place after plutonium's daughters were degassed.   
 
 Spontaneous fission of $^{238}$U (half life 4.47 Gyr) has generated about 5\% as much fissiogenic Xe as Pu.
 It is seen in mantle samples, but taking into account that most Xe from $^{238}$U must still be in the mantle,
 it cannot be responsible for more than $\sim\!10$\% of the fissiogenic Xe in the atmosphere,
 which is small compared to the other uncertainties,

Until recently, despite the hint from $^{244}$Pu that not all Xe loss was early,
it had generally been supposed that xenon was fractionated and lost through
 an energetic process unique to the early solar system,
probably hydrodynamic hydrogen escape
powered by copious EUV (``extreme ultraviolet radiation'') emitted by the active young Sun
\citep{Sekiya1980, Hunten1987, Sasaki1988, Pepin1991, Pepin2006, Tolstikhin1994, Dauphas2003, Dauphas2014}.
Vigorous hydrodynamic hydrogen escape can be an effective way to mass fractionate heavy gases \citep{Sekiya1980,Zahnle1986,Hunten1987,Sasaki1988,Pepin1991,Dauphas2014}. 
Heavy atoms escape because collisions with the outbound hydrogen push them outwards faster than gravity can pull them back.
Because lighter atoms escape preferentially, the rump atmosphere becomes enriched in heavier atoms and isotopes.
As Xe is the heaviest gas, all the other gases must also escape. 
But with the possible exception of neon \citep{Sasaki1988},
the other gases display no isotopic evidence of having done so.
%, although the Ar/Kr elemental ratio may be accounted for in this way \citep[][present some examples]{Dauphas2014}. 
Workarounds have been to propose that Kr and the other noble gases escaped quantitatively during the hydrodynamic escape episode, 
while leaving some fractionated Xe behind.
The lighter noble gases would later be resupplied by a process
that did not supply much xenon; several different processes have been suggested 
\citep{Sasaki1988, Pepin1991,Tolstikhin1994,Dauphas2003,Dauphas2014}.
The discovery that atmospheric Kr is isotopically lighter than the Kr found in the mantle \citep{Holland2009}
appears to contradict models that resupply the atmosphere by degassing the mantle but fits well with models
that resupply the atmosphere with (hypothetical) Xe-depleted comets \citep{Dauphas2014}.

\begin{figure}[!htb] 
  \centering
  \includegraphics[width=1.0\textwidth]{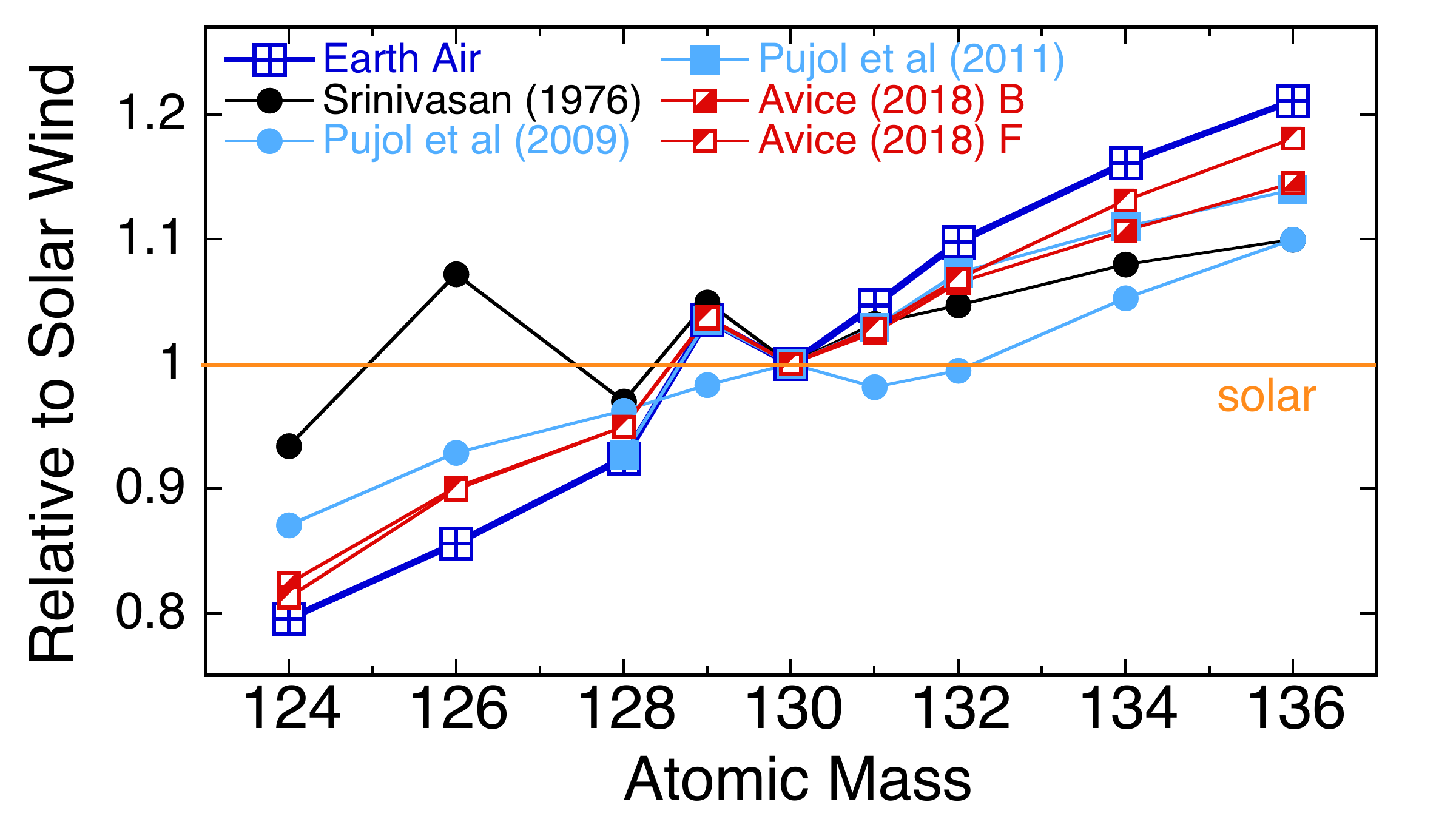} 
\caption{\small Xenon trapped in Archean rocks \citep{Srinivasan1976,Pujol2009, Pujol2011,Avice2018}
is less fractionated than Xe in modern air.
Isotopic abundances are each normalized to solar \citep{Meshik2014},
 and Xe$^{130}$ is normalized to one.
``Avice (2018) B'' refers to a Barberton quartz dated to $3.2\pm 0.1$ Ga.
``Avice (2018) F'' refers to a Fortescue quartz with an assigned age of 2.7 Ga.
Evidently Xe was escaping from Earth during the Archean.}
\label{Figure2}
\end{figure}

\begin{figure}[!htb] 
  \centering
  \includegraphics[width=1.0\textwidth]{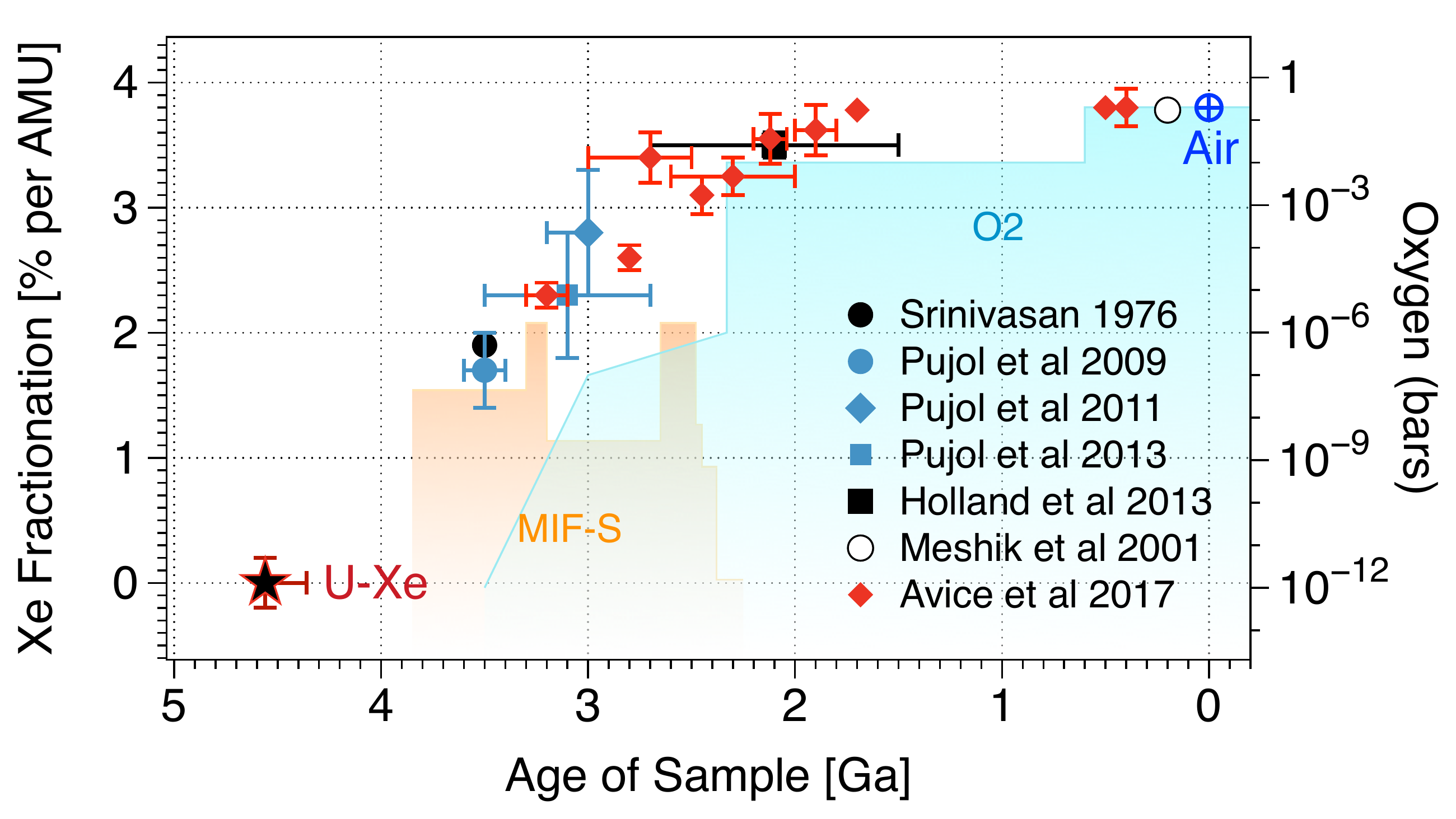} 
\caption{\small The history of xenon mass fractionation on Earth compared to 
schematic histories of mass-independent fractionation of sulfur (S-MIF, orange) and to the inferred history of oxygen (O$_2$, blue) in air.
Earth's original isotopic structure is labeled ``U-Xe,'' 
a kind of primordial Xe that is like solar Xe but depleted in Xe's two heaviest isotopes \citep{Pepin2006}.
Adapted from \citet{Avice2018}.}
\label{Avice}
\end{figure}

A feature common to all previous hydrodynamic escape models is that hydrogen escape fluxes large enough to power Xe escape would have been limited to Earth's first $\sim\! 100$ Myrs when the young Sun was still an enormous EUV source \citep{Pepin2013}.
New evidence now indicates that xenon's mass fractionation evolved over the first half of Earth's history, 
only converging with modern air between 1.8 Ga and 2.5 Ga \citep{Pujol2011,Pujol2013,Avice2018,Warr2018}.
The new evidence comes from the isotopic compositions of trapped atmospheric xenon recovered from several ancient (2-3.5 Ga) rocks
\citep{Srinivasan1976, Pujol2009,Pujol2011,Pujol2013,Holland2013, Avice2018,Bekaert2018}.
The recovered samples of Archean xenon resemble modern air, but they are less strongly mass fractionated. 
We show some examples in Figure \ref{Figure2}.
%In the two 3.5 Ga barites, the Xe fractionation is about half of what it is today \citep{Srinivasan1976, Pujol2009}.
%The differences from air are not readily accounted for as contamination by modern air,
%nor as mixtures of air and a once abundant unfractionated mantle xenon that is no longer extant. 
%Rather, the straightforward interpretation is that the trapped xenon truly does sample ancient air.
%A second argument in favor of this interpretation is in the xenon trapped in the fluid inclusions of hydrothermal quartz
%of $3.0\pm 0.2$ Ga K-Ar closure age \citep{Pujol2011}.  If argon remained trapped, it is reasonable to think that
%the xenon also dates to $3.0\pm 0.2$ Ga.  This particular xenon was about three-quarters of the way to becoming modern air. 
The full suite of data compiled by \citet{Avice2018} are plotted in summary form in Figure \ref{Avice}, where we have compared xenon's story to oxygen's and sulfur's.

\medskip

The evidence that Xe alone escaped among the noble gases requires a mechanism unique to Xe.
The evidence that Xe escape continued through the Archean requires a mechanism that can work
in Earth's Archean atmosphere, and therefore one that can work at more modest levels of solar EUV.
Hydrodynamic xenon escape as an ion is such a mechanism. 
%Ion escape 
It retains the advantages of traditional hydrodynamic escape 
without the two disadvantages of requiring too much solar EUV energy and predicting even greater amounts
Kr and Ar fractionation.
In particular, the fractionation mechanism is the
same --- collisions push ions outwards, and gravity pulls them  back --- as in neutral hydrodynamic escape.
The difference is that the collisional cross sections are much bigger between ions.

What makes it possible for Xe alone of the noble gases to escape as an ion in hydrogen is that Xe
is the only noble gas more easily ionized than hydrogen (Table \ref{Energies}).
% {\color{red} \sout{(12.13 eV vs.\ 13.60 eV for H and 14.00 eV for Kr). }}
\begin{table}[htb]
\caption{Some interesting ionization thresholds (NIST)}
\begin{center}
\begin{tabular}{lcclcclcc}
  & $\mathrm{I_P}$ [eV] & $\lambda$ [nm] &  & $\mathrm{I_P}$ [eV] & $\lambda$ [nm]&  & $\mathrm{I_P}$ [eV] & $\lambda$ [nm]\\
Ar & 15.76  &  \phantom{1}78.67          & ~ CO$_2$ & 13.78  &  \phantom{1}89.98    &~ CH$_4$ & 12.61  &  \phantom{1}98.33   \\
N$_2$  & 15.58  &  \phantom{1}79.59  & ~ O & 13.62  &  \phantom{1}91.04               &~ Xe & 12.13  & 102.2\phantom{3}  \\
H$_2$ & 15.43  &  \phantom{1}80.36  & ~ H & 13.60  &  \phantom{1}91.17                &~ O$_2$ & 12.07  & 102.7\phantom{3}  \\
N & 14.53  &  \phantom{1}85.34          & ~ HCN & 13.60 &  \phantom{1}91.17           &~ C$_2$H$_2$ & 11.40  & 108.8\phantom{3}\\
CO & 14.01  &  \phantom{1}88.48        &~ OH & 13.02 &  \phantom{1}95.24              &~ NO & \phantom{1}9.26 & 133.9\phantom{3} \\
Kr & 14.00  &  \phantom{1}88.57         &~ H$_2$O & 12.62  &  \phantom{1}98.26      &~ HCO & \phantom{1}8.12 & 152.7\phantom{3}  \\
\end{tabular} 
\end{center}
\label{Energies}
\end{table}%
Hydrogen will be partially photo-ionized. 
In a hydrogen-dominated hydrodynamic wind, gases that are more difficult to ionize than hydrogen will tend to be present as neutrals.
But Xe, being more easily ionized than H or H$_2$, will tend to be present as an ion.
First, Xe can be photo-ionized by UV radiation ($91.2<\lambda<102.3$ nm) to which H and H$_2$ are transparent. 
Second, as we shall show, Xe can be ionized by charge exchange with H$^+$.
Third, Xe$^{+}$ can be slow to recombine in hydrogen.
In particular, the important reaction ${\rm Kr}^{+} + {\rm H}_2 \rightarrow {\rm KrH}^{+} + {\rm H}$ is fast. 
The KrH$^{+}$ ion quickly dissociatively recombines: ${\rm KrH}^{+} + {\rm e}^{-} \rightarrow {\rm Kr} + {\rm H}$.
The corresponding reaction ${\rm Xe}^{+} + {\rm H}_2$ does not occur \citep{Anicich1993}.
Thus Xe$^{+}$ tends to persist in H$_2$ whereas Kr$^+$ is quickly neutralized.

Ions interact strongly with each other through the Coulomb force, especially at low temperatures.
If the escaping hydrogen is significantly ionized, and if the ions are also escaping,
the strong Coulomb interactions between ions permit Xe$^+$ to escape at hydrogen
escape fluxes well below what would be required for neutral Kr or even neutral Ne to escape.
Under these circumstances fractionating hydrodynamic escape can apply uniquely to Xe among the noble gases over a wide range of hydrogen escape fluxes, despite Xe's greater mass.  

\medskip

Parenthetically, it has been speculated that trapping Xe$^+$ in organic hazes on Archean Earth could lead to a way to fractionate
Xe in the atmosphere \citep{Hebrard2014,Avice2018}.
Indeed, it has recently been shown that isotopically ancient Xe was trapped, and has remained trapped,
 in ancient organic matter on Earth \citep{Bekaert2018}.
Ionized Xe can be chemically incorporated into organic material
\citep{Frick1979,Marrocchi2011,Marrocchi2013}.
The trapped Xe is mass fractionated with a $\sim\!1\%$ per amu preference for
the heavier isotopes \citep{Frick1979,Marrocchi2011,Marrocchi2013}. 
% Krypton exhibits the same behavior \citep{Frick1979}.
For trapping to work as a fractionating mechanism,
most of Earth's Xe must have resided in organic matter through much of the Hadean and Archean,
and the process must have run through several rock weathering cycles to build up the fractionation,
yet there still needs to be a Xe-specific escape process, as otherwise
when the trapped Xe is released by weathering the atmosphere would regain its original unfractionated isotopic composition.
We think it simpler to assign both depletion and fractionation to the hydrodynamic escape process, so we have not pursued this
more complicated scenario here.

\medskip

Martian atmospheric Xe as determined from SNC meteorites \citep{Swindle1997}
superficially resembles Xe in Earth's air, which makes it tempting to imagine that Earth and Mars received
their Xe from a common fractionated source.  
However, martian Xe has been less impacted by escape.
Martian Xe is 50\% less depleted than Earth's (well-seen in Figure \ref{EarthMars2})
and it has been fractionated by about 2.5\% per amu from what was initially solar Xe, whilst Earth's was
 fractionated by about 4\% per amu from U-Xe. %, a kind of primordial Xe that is depleted in Xe's heaviest isotopes \citep{Pepin2006}.
 Moreover, Xe fractionation on Mars took place very early \citep{Cassata2017}.
These differences imply parallel evolution rather than a common source. 

\medskip

To assess Xe$^+$ escape from Earth's ancient atmosphere, we first need to develop a model
of irradiation-fueled diffusion-limited hydrodynamic hydrogen escape that includes a full energy budget and that computes 
temperature and ionization as well as the escape flux.
This project is described in detail in Appendix A.
A subset of results pertinent to Xe escape are summarized in Section \ref{section two}.
Section \ref{section three} addresses Xe chemistry and Xe escape,
 although the details of the model are relegated to Appendix B, as the notation and development
  follow directly from the hydrogen escape model developed in Appendix A.
Section \ref{section four} addresses histories of atmospheric hydrogen and hydrogen escape that best
reproduce the observed history of Xe fractionation and depletion.
Section \ref{section five} recapitulates the chief results and chief caveats, and suggests directions for further research.
Appendix C provides a complete alphabetized table of symbols used in the text and Appendices.

\section{Hydrogen escape from a CO$_2$-rich atmosphere}
\label{section two}

 %The model is fully described in the Supplemental Material.
  The general problem of irradiation-driven thermal escape from planetary atmosphere can get very complicated.
Here we wish to develop a description of hydrogen escape in the presence of a static background atmosphere
suitable for investigating Xe escape.
The heavy gases in Earth's Archean atmosphere were likely N$_2$ and CO$_2$.
We simplify the problem by considering a CO$_2$-H$_2$ atmosphere.
We chose CO$_2$ rather than N$_2$ because we wished to consider a case
in which both radiative heating and radiative cooling are important to the energy budget \citep{Kulikov2007}.
%{\color{blue}\bf We are aware that N$_2$-H$_2$ would be simpler and N$_2$-CO$_2$-H$_2$ more realistic.}
% We expect that escape from an N$_2$-H$_2$ atmosphere would be easier than from a CO$_2$-H$_2$ atmosphere. 
 We will argue that, to first approximation,
 photochemistry in the CO$_2$-H$_2$ atmosphere allows CO$_2$ to persist while H$_2$ persists.
Our minimum system therefore comprises only H, H$_2$ and CO$_2$ as neutral species.
% The neutral number densities are denoted $n_1$, $n_2$, and $n_3$, respectively.  
We include 5 ions: H$^+$, H$_2^+$, H$_3^+$, CO$_2^+$, and HCO$^+$.
% --- with densities denoted $x_i$, with $i$ running from 1 to 5, respectively.
% which form in consequence to photo-ionization of O. 
We assume local photochemical equilibrium for the ions, which is a good approximation for the molecular ions, less good
 for H$^+$ at high altitudes.
This minimal system omits N$_2$, N, CO, O, O$_2$, NO, and their ions.
The chemistry is fully described and summarized in Table \ref{Chemistry Table} in Appendix A.

\subsection{Radiation}

The two essential free parameters in irradiation-driven thermal hydrogen escape are 
solar irradiation $S$ and the hydrogen mixing ratio $f_{\mathrm{H}_2}$.
Hydrogen efficiently absorbs radiation at wavelengths $\lambda < 91.2$ nm.
This serves as a practical definition of EUV.
Water vapor and CO$_2$ absorb efficiently at wavelengths shorter than $\lambda < 200$ nm.
This is a useful definition of far ultraviolet radiation (FUV).
The EUV is often lumped together with X-rays as XUV, a convenience that exploits the relative availability of stellar X-ray luminosities. 

It is  observed that older sunlike stars emit less X-ray 
and FUV radiation \citep{Zahnle1982,Ribas2005,Claire2012,Tu2015}.
The observations are sparse enough to be fit to power laws of the form $F_{\mathrm{xuv}} \propto t^{-a}$, where $t$ is the age of the star, and the power $a$ is of order unity.
A popular parameterization of the average solar $F_{\mathrm{xuv}}$ at Earth is
\begin{equation}
\label{XUV}
 F_{\mathrm{xuv}}  = 5 \left(4.5/t\right)^{1.24} \quad {\rm ergs}\,{\rm cm}^{-2}{\rm s}^{-1}
\end{equation}
where $t$ is the age of the Sun in Gyrs \citep{Ribas2005}.
At very early times, say $t<0.1$ Gyr, this saturates to $F_{\mathrm{xuv}}  \approx 400$ ergs cm$^{-2}$s$^{-1}$.
FUV radiation does not decay as quickly as XUV, but for present purposes we will ignore this detail,
and lump the entire XUV and FUV emission together as a single multiple $S$ of the modern Sun,
\begin{equation}
\label{S}
S(t) \equiv F_{\mathrm{xuv}}/F_{\mathrm{xuv}\odot} .
\end{equation}
\citet{Tu2015} compare several different models; they estimate
that $5<S<10$ at 3.5 Ga and $10<S<40$ at 4.0 Ga, corresponding to $1<a<1.7$.
This does not take into account the factor five variation in XUV between active and quiet Sun.
With variability included, we might expect $2<S<20$ at 3.5 Ga and $1<S<10$ at 2.5 Ga.

\subsection{Vertical structure, method of solution, and the outer boundary conditions}

Vertical structure and transport equations are simplified from the self-consistent 5 moment approximation to multi-component hydrodynamic
flow presented by \citet{Schunk1980}. 
We merge this description with the description of two component diffusion given by \citet{Hunten1973}
to express collision terms as binary diffusion coefficients
and to include parameterized Eddy diffusivity in the lower atmosphere.
% This is described in Appendix
 We make several other major simplifications: 
 (i) We assume spherical symmetry.
 (ii) We ignore diurnal cycles and latitudinal differences.
 (iii) We presume that H and H$_2$ flow outward at the same velocity $u$. 
 (iv) As we are considering a relatively dense gas, we use a single temperature $T$ for all species.
 % This greatly simplifies the mathematics with no major consequences.
% (iv) We ignore photodissociation of CO$_2$ into CO and O,
% which we justify by showing that CO$_2$ rapidly recombines when H$_2$ is present, and thus is likely to persist while H$_2$ persists.
 (v) We presume that CO$_2$ does not escape, and thus that hydrogen must diffuse through the CO$_2$.
 (vi) We neglect thermal conduction, which becomes a relatively small term in the energy budget 
 at the high levels of solar irradiation needed if Xe is to escape.
 (vii) We neglect the solar wind, collisional ionization by exogenous particles, 
 and energy flows between different regions of the magnetosphere.

The equations and approximations are developed and fully described in Appendix A.
The five basic equations to be solved are Eq \ref{E5} for the hydrogen velocity $u(r)$; Eqs \ref{B9} and \ref{B10} for the atomic and molecular hydrogen number densities $n_1(r)$ and $n_2(r)$, respectively;
 Eq \ref{CCF three} for the CO$_2$ number density $n_3(r)$; and Eq \ref{C4} for the temperature $T(r)$.
The system is solved with the shooting method, integrating upward from a lower boundary density $n(r_0)=1\times 10^{13}$ cm$^{-3}$,
which is below the homopause.
The total H$_2$ mixing ratio $f_{\mathrm{H}_2} (= 0.5f_{\mathrm{H}}  + f_{\mathrm{H}_2}  + 2f_{\mathrm{CH}_4} + \ldots)$
and the total hydrogen escape flux $\phi_{\mathrm{H}_2} (= 0.5\phi_{\mathrm{H}} + \phi_{\mathrm{H}_2} + 2\phi_{\mathrm{CH}_4} + \ldots)$ at the lower boundary 
are treated as independent free parameters.  %These are the inputs to the model.

We seek the unique transonic solution that has just enough energy at the critical point to escape,
in keeping with the philosophy that nothing that happens beyond the critical point of a transonic wind can influence the atmosphere at the lower boundary.
The energy criterion is given by Eq \ref{stopping condition}.
The transonic solution makes the simplifying assumption that conditions far from Earth
are ignorable. This assumption is probably very good for calculating the hydrogen escape flux,
which is determined by conditions much deeper in the atmosphere where the bulk of XUV radiation is absorbed.   
Subsonic solutions require additional parameters to describe the conditions of interplanetary space.
As a practical matter, differences between the transonic solution and a relevant subsonic solution are negligible
save at great distances \citep{Kasting1983}. 
We solve for the solar irradiation $S$ required to support $\phi_{\mathrm{H}_2}$ by iterating $S$ using bisection.

\subsection{Snapshots taken from a particular model for purposes of illustration}

  \begin{figure}[!htb] %  figure placement: here, top, bottom
    \centering
 \begin{minipage}[c]{0.49\textwidth}
   \centering
  \includegraphics[width=1.0\textwidth]{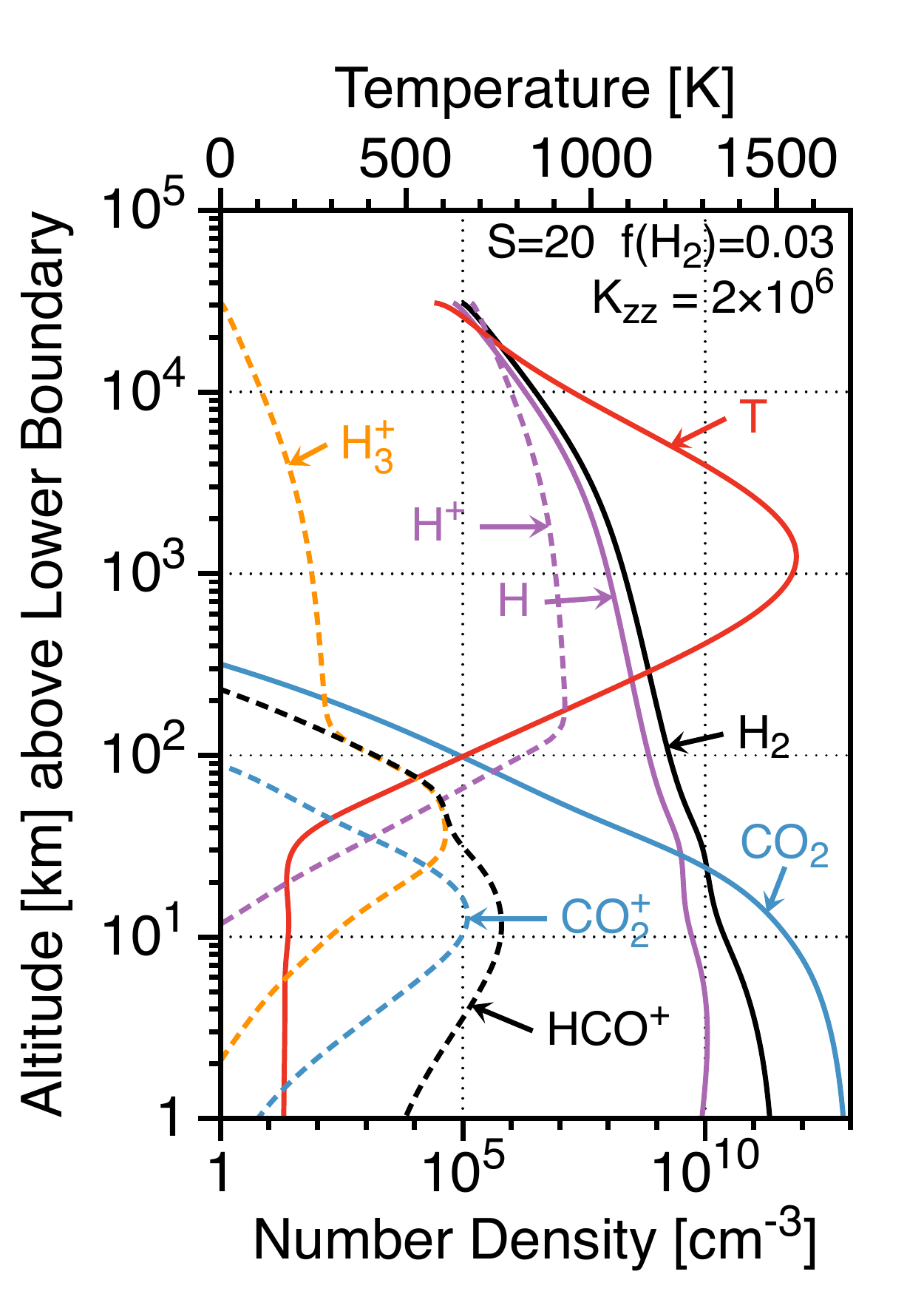} 
 \end{minipage}
 \begin{minipage}[c]{0.49\textwidth}
   \centering
  \includegraphics[width=1.0\textwidth]{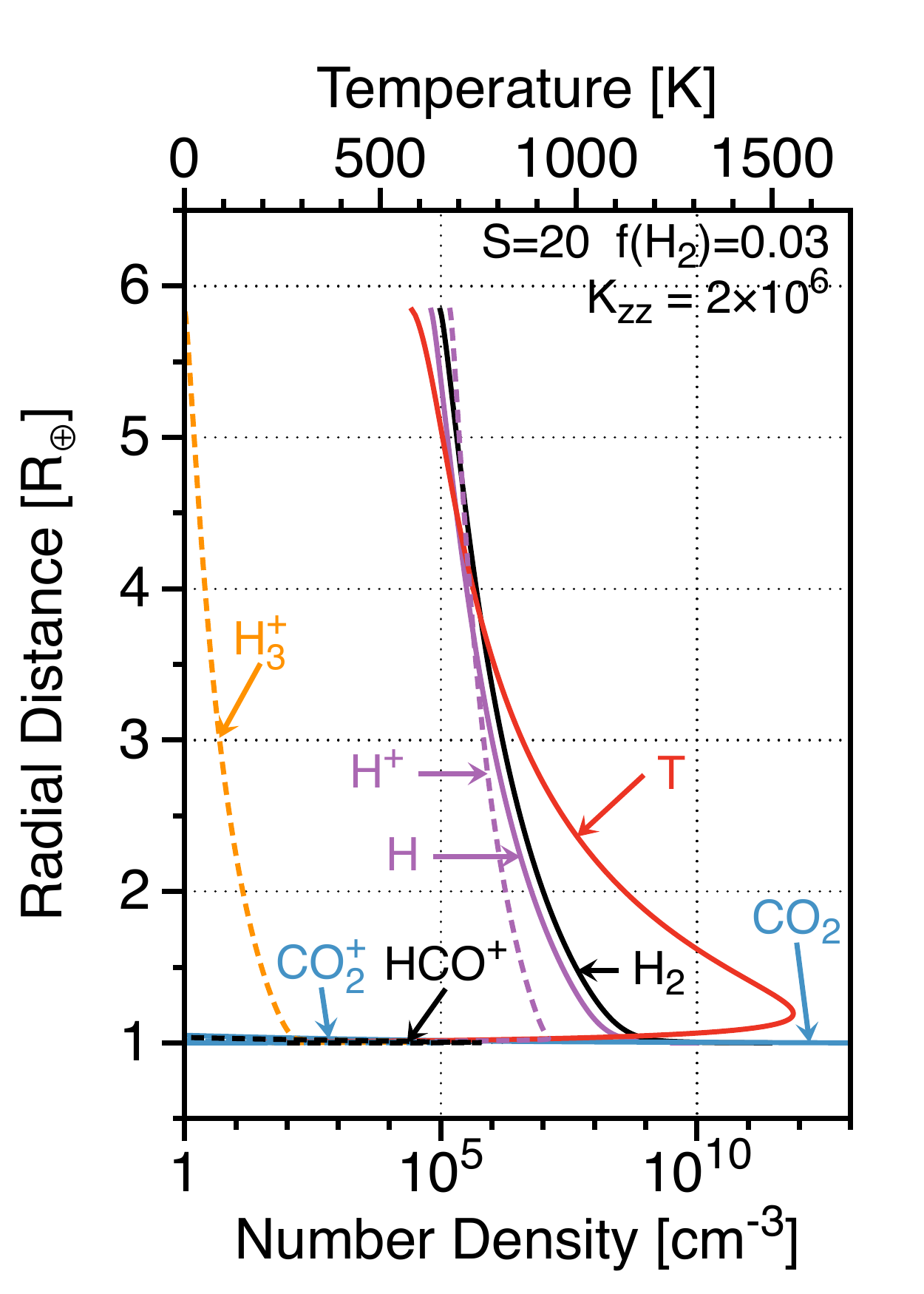} 
 \end{minipage}
\begin{minipage}[c]{1.0\textwidth}
   \centering
\caption{\small Temperature and densities of neutrals and ions as a function of altitude in an exemplary
model, which we call the ``nominal'' model ($S=20$, $f_{\mathrm{H}_2}=0.03$, and $K_{zz}=2\times 10^6$ cm$^2$s$^{-1}$,
for which $\phi_{\mathrm{H}_2}=7.2\times 10^{11}$ cm$^{-2}$s$^{-1}$.) 
This nominal model was chosen to illustrate conditions in which Xe$^+$ escape is expected; 
the high $S$ of the nominal model would be a rare occurrence later than 3.5 Ga.
Altitude is measured from an arbitrary lower boundary where the total density $n(r_0)=1\times 10^{13}$ cm$^{-3}$.
The logarithmic scale best illustrates the overall structure of the upper atmosphere, which features a warm 
ionized escaping hydrogen exosphere above a cold layer of molecular ions.
The linear scale better illustrates the extent of the hydrogen exosphere. }
 \label{Densities}
 \end{minipage}
\end{figure}

It is helpful to illustrate some properties of a particular model.
For this purpose we have chosen a model (hereafter referred to as the ``nominal'' model)
that lies well within the field of models in which Xe escape is predicted to take place.
The key parameters are a relatively high EUV flux ($S=20$) and a relatively high hydrogen
mixing ratio ($f_{\mathrm{H}_2}=0.03$).
The high $S$ of the nominal model would be typical before 4.0 Ga but rare after 3.5 Ga.
Other nominal parameters are a lower boundary density  
$n(r_0)=1\times 10^{13}$ cm$^{-3}$,
neutral eddy diffusivity $K_{zz} = 2\times 10^{6}$ cm$^2$s$^{-1}$,
and spherical symmetry.
Our models of hydrogen escape are not very sensitive to these other parameters.
The hydrogen escape flux in this particular model is $\phi_{\mathrm{H}_2}=7.2\times 10^{11}$ cm$^{-2}$s$^{-1}$,
equivalent to 82\% of the diffusion-limited flux (Eq \ref{diffusion-limited-2}).
  
 Figure \ref{Densities} shows
 the temperature and the densities of the ions and neutrals as a function of altitude in the nominal model;
 the two panels differ only on how the altitude axis is scaled.
 The linear scale effectively illustrates the extent
 of the hydrogen atmosphere, while hiding everything else.
 The logarithmic scaling of altitude 
 focuses attention on the structure of the atmosphere, in which a cold CO$_2$-rich layer supporting
 a client population of cold molecular ions is overlain by
 warm hydrogen and atomic ions.
 Other aspects of the nominal model and other models are discussed in more detail in Appendix A.

\subsection{Many solutions}

\begin{figure}[!htb] 
  \centering
  \includegraphics[width=1.0\textwidth]{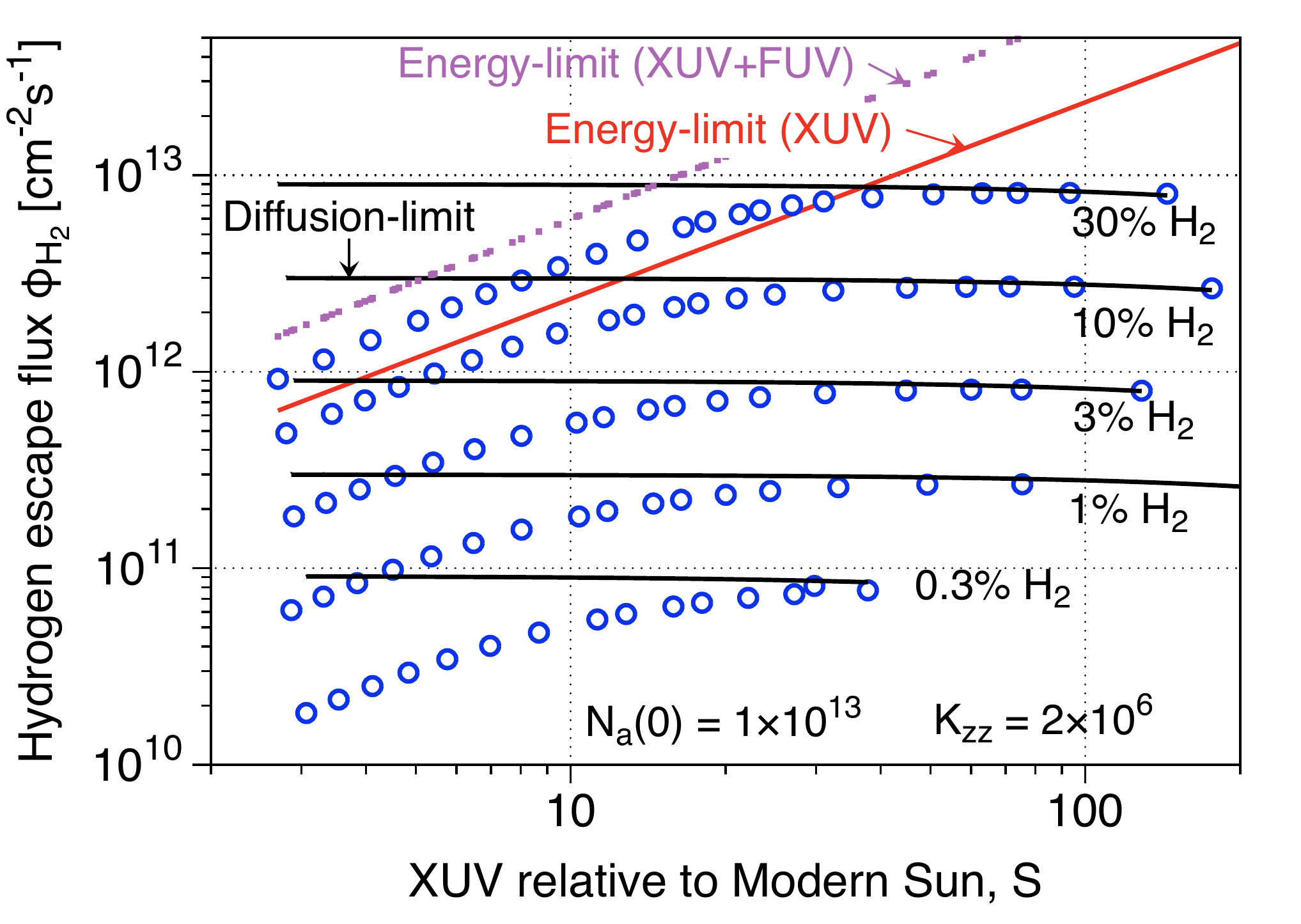} 
\caption{\small Hydrogen escape fluxes $\phi_{\mathrm{H}_2}$
from terrestrial CO$_2$-H$_2$ atmospheres as a function of XUV and FUV irradiation $S$,
relative to the modern Sun.
% The lower boundary conditions are the same as in Figures \ref{Densities}-\ref{Figure8}.
Computed $\phi_{\mathrm{H}_2}$ are shown for
five H$_2$ mixing ratios ranging from 0.1\% to 30\% by volume (blue circles).
These are compared to the diffusion-limited flux for the same five hydrogen mixing ratios (horizontal black lines)
and to estimates of the energy-limited escape flux, computed either from the EUV energy that hydrogen
can absorb (solid red diagonal line) or to the total XUV and FUV energy absorbed by H$_2$ and CO$_2$
(broken purple line) above the (arbitrary) lower boundary of the model.
The computed escape fluxes asymptote to the appropriate diffusion limits at high $S$.
Smaller values of $S$ and $f_{\mathrm{H}_2}$ are not shown because two basic assumptions of this study --- 
that thermal conduction can be ignored and that a hydrodynamic description is appropriate, respectively --- begin to break down. }
\label{Figure9}
\end{figure}

Figure \ref{Figure9} presents results from a basic parameter survey of CO$_2$-H$_2$ atmospheres.
The plot shows how the total hydrogen escape flux $\phi_{\mathrm{H}_2}$ changes in response to
changing solar irradiation $S$ and hydrogen mixing ratios $f_{\mathrm{H}_2}$.
Our results (blue circles) in Figure \ref{Figure9} are compared
to two limits often encountered in the literature.
The so-called energy-limited flux 
compares the XUV energy absorbed to the energy required to lift a given mass out of Earth's potential well
and into space \citep{Watson1981}.
Details are lumped together in an efficiency factor $\eta$ that is often taken to lie between 0.1 and 0.6
\citep{Lammer2013,Bolmont2017}. 
In one version of the energy limit, all XUV photons that can be directly absorbed
by hydrogen ($\lambda<91.2$ nm) contribute to escape.
This limit is labeled ``Energy-limit (XUV)'' on Figure \ref{Figure9}.
A second energy limit includes all solar radiation absorbed above a fixed lower boundary.
% which might be identified with an optical depth at some fiducial wavelength, with the homopause, or something else. 
This includes FUV radiation absorbed by CO$_2$.
This outer limit is labeled ``Energy-limit (XUV+FUV)'' on Figure \ref{Figure9}.
The relevant equations, Eq \ref{energy-limited-1} and Eq \ref{energy-limited-2} in Appendix A,
 are evaluated
 with $\eta=0.5$ to facilitate comparisons with the detailed model. % to the fluxes we compute with the full model.  
The diffusion-limited flux, the upper bound on how fast hydrogen
can diffuse through a hydrostatic atmosphere of CO$_2$, is derived in Appendix A in the limit of constant mixing ratios \citep{Zahnle1986,Hunten1987}.
Equation \ref{diffusion-limited-2} is plotted on Figure \ref{Figure9} for each $f_{\mathrm{H}_2}$.

It is apparent from Figure \ref{Figure9} that the diffusion limit 
is well-obeyed as an upper limit at all levels of irradiation we consider, and it closely
approximates the actual escape flux at higher levels of solar irradiation.
By contrast the energy-limited flux is ambiguously defined and not obviously well-obeyed,
although at low $S$ the slope is correct. 
% The ambiguity of definition is the problem. 
The more restricted ``energy-limited (XUV)'' flux can underestimate escape because it neglects FUV absorbed by CO$_2$,
whilst the higher ``energy-limited (XUV+FUV)'' flux overestimates escape.
Key points are that FUV absorbed by molecules other than H or H$_2$ can
be important \citep{Sekiya1981}, and that both the energy limit and the diffusion limit overestimate
hydrogen escape when the hydrogen above the homopause is optically thin to EUV,
as discussed by \citet{Tian2005b}. 

Our results in Figure \ref{Figure9} can be roughly fit by 
\begin{equation}
\label{fit}
\phi_{\mathrm{H}_2} \approx \frac{2\times 10^{12} f_{\mathrm{H}_2} S}{ \sqrt{1 + 0.006 S^2} }\quad \mathrm{cm}^{-2}\mathrm{s}^{-1},
\end{equation}
good for $f_{\mathrm{H}_2} < 0.2$. 
Equation \ref{fit} asymptotes to the diffusion-limited flux at large $S$ and asymptotes to an
appropriate energy-limited flux for small $S$.

 \begin{figure}[!htb]
   \centering
    \includegraphics[width=0.9\textwidth]{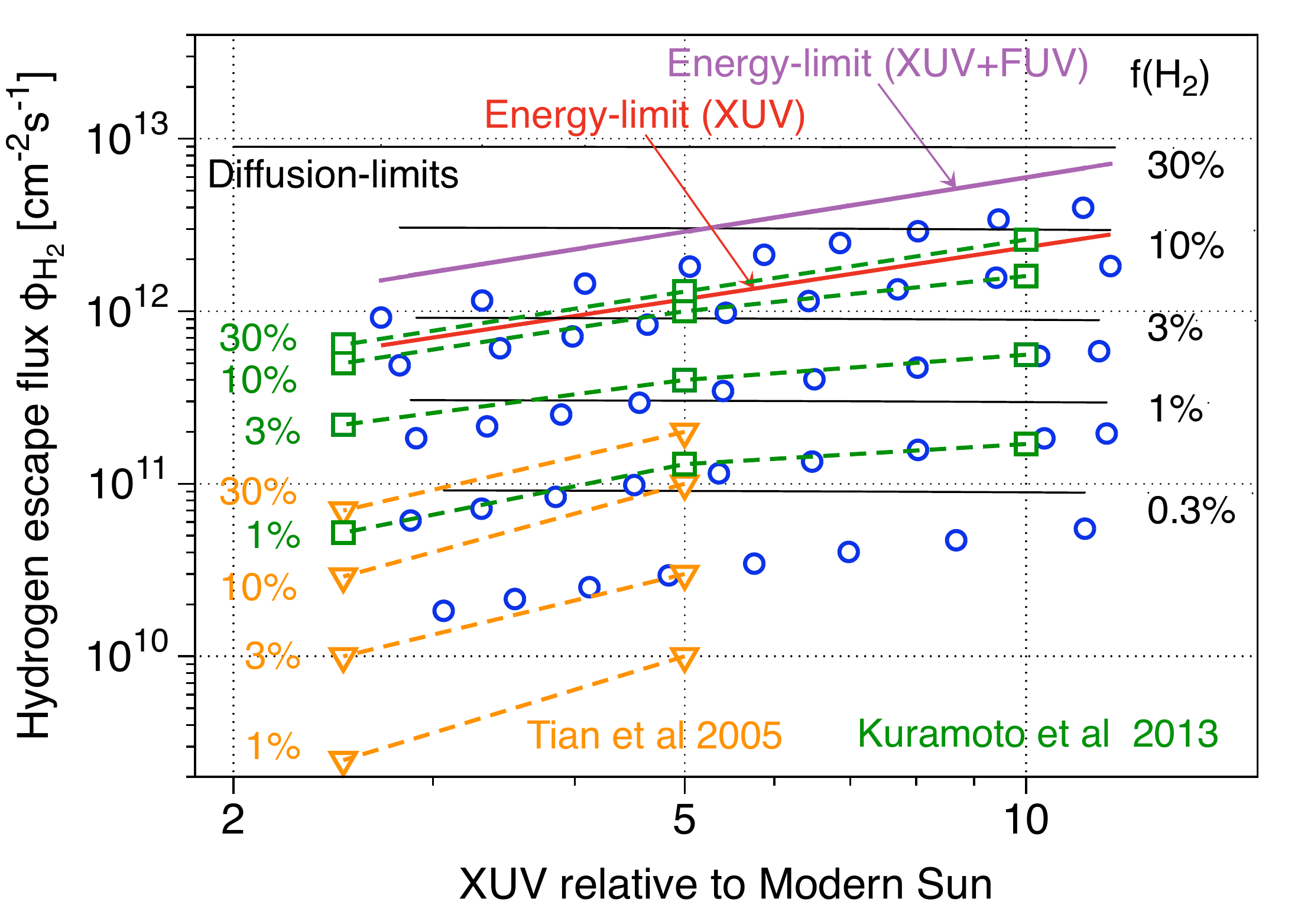} 
   \caption{Comparison between our model (blue circles, from Figure \ref{Figure9})
   and published results from two different hydrocodes.
   The orange triangles denote \citet{Tian2005b}, the green squares
   denote \citet{Kuramoto2013}; each is plotted for four hydrogen mixing ratios.
   The \citet{Kuramoto2013} results are quite similar to ours, with minor differences attributable to absorption of solar XUV 
   by CO$_2$, which is accounted for in our model but neglected by \citet{Kuramoto2013}.
%   Note that this plot is restricted to XUV fluxes smaller than that required for Xe escape. 
} 
\label{Figure9a}
\end{figure}

It is interesting to compare our results to those obtained by time-stepping hydrocode models.
In Figure \ref{Figure9a} we compare our results % for small values of $S$
to those obtained by \citet{Tian2005b} and \citet{Kuramoto2013}.
Both hydrocode models explore transonic escape of pure H$_2$ atmospheres from Earth in response to enhanced
levels of incident EUV radiation.  Both presume that hydrogen diffusively separates from a lower
atmosphere comprising unspecified radiatively active heavy molecules, implicitly CO$_2$, although neither model
actually includes CO$_2$.
The lower boundary is held to a fixed temperature and serves as an infinite heat sink.  
Both hydrocode models include thermal conduction. 

The comparison reveals no significant differences in the predicted hydrogen escape fluxes
between our shooting code and one of the hydrocodes over the limited EUV range explored by \citet{Kuramoto2013}. 
Both our model and \citet{Kuramoto2013} predict significantly more hydrogen escape than does \citet{Tian2005b}
for the range of $S$ considered.
Why the two hydrocode models differ is not known to us.
But if the agreement between our model and \citet{Kuramoto2013} has meaning,
the indication is that the physics common to the two models are determining the
outcome, and that processes treated differently (e.g., thermal conduction, optical depth, radiative cooling) by the two models are not particularly important to  hydrogen escape.

\subsection{The magnetic field and the solar wind} 

\begin{figure}[!htb] 
  \centering
  \includegraphics[width=1.0\textwidth]{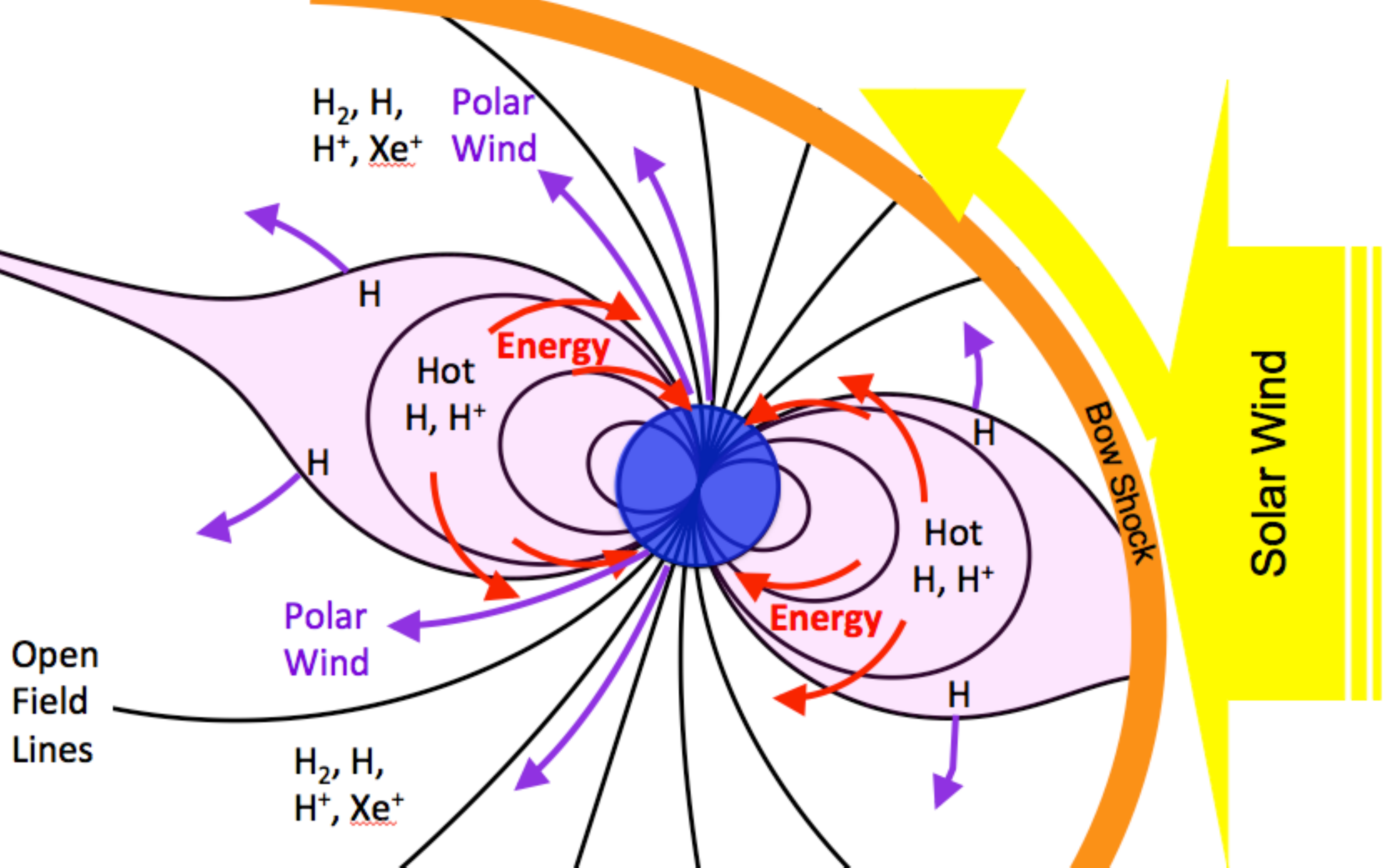} 
\caption{\small Geometry of a polar wind in a geomagnetic field. 
The magnetic field prevents ions from escaping from the tropics.  Neutral hydrogen diffuses through
the captured ions and eventually escapes, but escape is enough impeded that the corona
becomes rather dense and very hot.  By contrast, the geomagnetic field does not prevent ion escape
along the open polar field lines, and as a consequence of free escape the gas 
remains relatively cool while also retaining the capacity to drag Xe ions with it.
The flow of energy from the hot dense trapped corona to the poles has the potential of greatly
enhancing the effective $S$ pertinent to the polar wind.
The solar wind complicates the flow but,
regarded over the globe as a whole, is more likely to aid escape than to impede it.
 }
\label{cartoon2}
\end{figure}

In a more realistic setting of Earth, the specific properties of the outer boundary conditions
 would be determined by the details of the interaction with the solar wind and the weakening of
the geomagnetic field with distance (Figure \ref{cartoon2}).  
As most of the bulk properties of the hydrogen wind are determined near the homopause 
where most of the XUV and FUV radiation is absorbed (shown for the nominal model in Figure \ref{Figure7} in Appendix A),
the details of the outer boundary
 are not likely to be very important either to hydrogen escape or to xenon escape.

It is illuminating to compare the ram pressure of the incident solar wind to the ram
pressure of hydrodynamically escaping hydrogen and to the strength of the geomagnetic field.
The ram pressure of the solar wind is of the order of $\dot M_{sw}v_{sw} \approx 10^{-8}$ dynes cm$^{-2}$ at Earth.
This can be compared to the ram pressure of the hydrodynamic wind
% $\phi_{{\rm H}_2}u(r/R_{\oplus})^{\beta}$, which in the nominal model is $5\times 10^{-7}(r/R_{\oplus})^{\beta}$ dynes cm$^{-2}$.
% The parameter $\beta=2$ in spherical symmetry and $\beta=3$ in dipolar symmetry.
$\phi_{{\rm H}_2}u(r/R_{\oplus})^{2}$, which in the nominal model is $5\times 10^{-7}(r/R_{\oplus})^{2}$ dynes cm$^{-2}$.
If we presume that $\dot M_{sw}$ scales with $S$, the two rams would butt heads at $r\approx 1.6 R_{\oplus}$.

The more pertinent comparison is to the much greater strength of Earth's magnetic field for $r<4R_{\oplus}$.
A dipole field falls off as ${(0.25B_{\oplus}^2/\pi})({R_{\oplus}/ r})^6 \approx 1.5\times 10^{-7} ({6R_{\oplus}/ r})^6 $ dynes cm$^{-2}$
using $B_{\oplus}=0.31$ Gauss for Earth today.
These crude estimates suggest that the solar wind will play a major role in determining the outer boundary
conditions in the absence of a geomagnetic field, but would have less consequence for hydrodynamic
escape channeled along magnetic field lines in the presence of a field, even one as small as $0.1B_{\oplus}$. 
Whether early Earth had a significant magnetic field is debated \citep{Ozima2005,Tarduno2014,Biggin2015,Weiss2018}.

\section{Xenon escape as ion}
\label{section three}

\begin{figure}[!htb] %  figure placement: here, top, bottom
   \centering
   \label{Figure16}
 \begin{minipage}[c]{0.54\textwidth}
   \centering
  \includegraphics[width=1.0\textwidth]{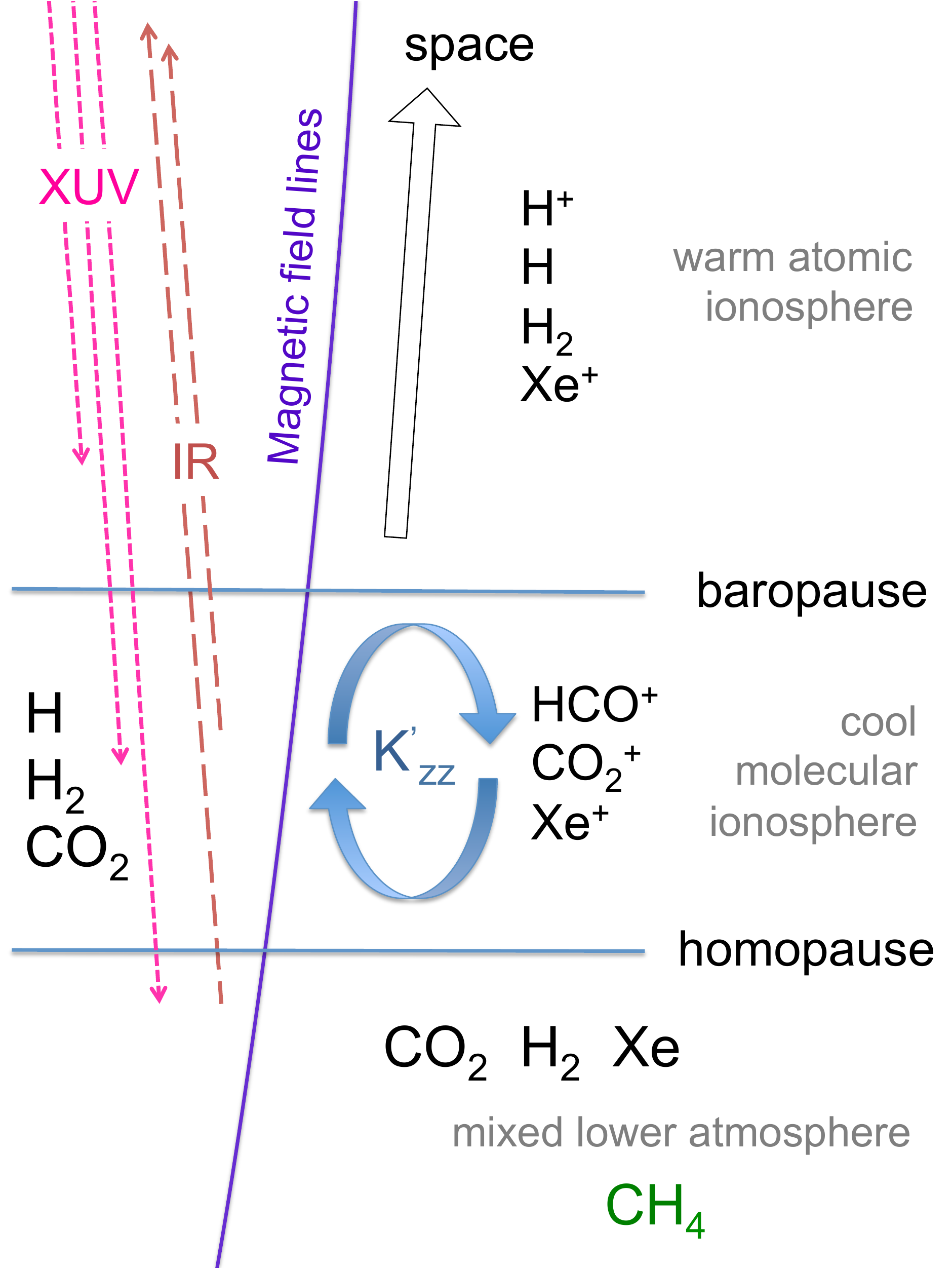} 
 \end{minipage}
\begin{minipage}[c]{0.45\textwidth}
   \centering
\caption{\footnotesize 
Structure of the CO$_2$-H$_2$ upper atmosphere pertinent to Xe escape.
There are three layers: a mixed lower atmosphere; 
a relatively thin cool molecular ionosphere with HCO$^+$ and CO$_2^+$ the major ions; and a
greatly distended outflowing upper ionosphere dominated by hydrogen and its ions. 
The divisions between the layers are marked homopause and ``baropause.''
The latter denotes the highest altitude where heavy elements are important.
There are three barriers to Xe escape as an ion.
First, ion escape is blocked by a geomagnetic field save near the poles where field lines open into space.
Second, hydrogen escape from the atomic ionosphere must be vigorous enough that Xe ions are swept away.
Quantifying this requirement determines what $S$ and $f_{\mathrm{H}_2}$ must be for Xe escape.
Third, Xe ions need to be transported through the molecular ionosphere to the baropause,
probably by vertical movements of molecular ions themselves. 
This too depends on $S$.
We model this in 1-D as an ion eddy diffusivity $K'_{zz}$, which we treat as a third parameter in the model.
}
  \label{expt5}
 \end{minipage}
\end{figure}

We use the solutions for H escape obtained in Section \ref{section two} and Appendix A to construct 
models of Xe escape.
We assume that xenon is a trace constituent that has no effect on the background atmosphere.
Our purpose is to determine the minimum requirements for Xe to escape, which is equivalent
to determining whether Xe$^+$ escape is possible.
We therefore focus on Xe$^+$.
In this section we first discuss how we compute Xe$^+$ escape for a given atmosphere and level of solar irradiation.
Some general considerations pertinent to Xe$^+$ escape are outlined in Figure \ref{expt5}.
% We then discuss Xe fractionation in the atmosphere.

\subsection{Xenon chemistry}

Xenon can be directly photo-ionized \citep{Huebner1992}, 
\begin{equation}
\tag{J4}\label{JXe}
\mathrm{Xe} + h\nu\,(\lambda < 102.3\,\mathrm{nm}) \rightarrow \mathrm{Xe}^+ + \mathrm{e}^-,
\end{equation}
and it can be ionized by chemical reactions with other ions.
In the idealized H-H$_2$-CO$_2$ atmosphere, the chief possibilities are reactions of neutral Xe with
the primary ions H$^+$, H$^+_2$, and CO$^+_2$.
The charge exchange reaction with CO$_2^+$ \citep{Anicich1986} 
\begin{equation}
\tag{R15}\label{R15}
\mathrm{CO}_2^+ + \mathrm{Xe} \rightarrow \mathrm{CO}_2 + \mathrm{Xe}^+ \qquad \qquad k_{15}=6\times 10^{-10} \,\mathrm{cm}^3\mathrm{s}^{-1}
\end{equation} 
is fast and important at low altitudes where CO$_2$ is photo-ionized. 
The reaction of Xe with H$_2^+$ 
\begin{equation}
\mathrm{H}_2^+ + \mathrm{Xe} \rightarrow \mathrm{HXe}^+ + \mathrm{H} 
\end{equation} 
occurs but it is not a source of Xe$^+$ because the XeH$^+$ ion dissociatively recombines; we ignore it.

Charge exchange with H$^+$ presents the most interesting case.
In general, charge exchange reactions between light atoms are slow
save when the reaction is nearly resonant \citep{Huntress1977}.
The reaction into the Xe$^+$ ground state is far from resonance,
\begin{equation}
\tag{R16a}\label{R16a}
\mathrm{H}^+ + \mathrm{Xe} \rightarrow \mathrm{Xe}^+(^2\mathrm{P}_{3\over 2}) + \mathrm{H} + 1.47\,\mathrm{~eV},
\end{equation} 
 and is therefore unlikely to be fast.
On the other hand, Xe$^+$ has
 a low-lying electronically-excited state for which charge exchange is exothermic yet not far from resonance \citep{Shakeshaft1972},
\begin{equation}
\tag{R16b}\label{R16b}
\mathrm{H}^+ + \mathrm{Xe} \rightarrow \mathrm{Xe}^+(^2\mathrm{P}_{1\over 2}) + \mathrm{H} + 0.16\,\mathrm{~eV.}
\end{equation} 
There appear to be no relevant measurements, while
a calculation by \citet{Sterling2011} does not take into account fine structure and thus uses large asymptotic energy separation between different charge arrangements.  
We therefore revisited the calculation of the rate of R16b, as described in Appendix B.
An empirical curve fit to $k_{16b}$ good for temperatures between 50 and $10^5$ K is
\begin{equation}
\label{k16b}
k_{16b} = 3.83\times 10^{-8} T^{0.386} \exp{\left(-55.8/T^{0.326}\right)} \qquad \mathrm{cm}^{3} \mathrm{s}^{-1}.
\end{equation}
The rate $k_{16b}$ is of the order of $10^{-11}$ cm$^3$s$^{-1}$ at 300 K and $10^{-10}$ cm$^3$s$^{-1}$ at 800 K,
rates that are fast enough that % neutral Xe is rapidly ionized where H$^+$ is abundant,
charge exchange becomes an important source of Xe$^+$ at high altitudes where H$^+$ is abundant.
The reverse of R16b is 0.16 eV endothermic and therefore smaller than $k_{16b}$ by
a factor of the order of $e^{-1860/T}$, which ensures that the excited $\mathrm{Xe}^+(^2\mathrm{P}_{1\over 2})$ ion
 relaxes radiatively or collisionally to the ground state before it can lose its charge.  

The Xe$^+$ ion once made does not react with H, H$_2$, or CO$_2$. 
A potentially important reaction at lower altitudes is the nearly resonant charge exchange with O$_2$
\citep{Anicich1993}
\begin{equation}
\tag{R17}\label{R17}
\mathrm{Xe}^+ + \mathrm{O}_2 \rightarrow \mathrm{Xe} + \mathrm{O}^+_2 + 0.06\,\mathrm{eV} \quad \qquad \qquad \qquad k_{17}=1.2\times 10^{-10} \,\mathrm{cm}^3\mathrm{s}^{-1}
\end{equation} 
\begin{equation}
\tag{R17r}\label{R17r}
\mathrm{Xe} + \mathrm{O}^+_2 \rightarrow \mathrm{Xe}^+ + \mathrm{O}_2 - 0.06\,\mathrm{eV}\quad \qquad k_{17r}=3.0\times 10^{-10}e^{-500/T} \,\mathrm{cm}^3\mathrm{s}^{-1}
\end{equation} 
R17r can be a source of Xe$^+$ in an atmosphere without much O$_2$,
because O$_2^+$ can be abundant in a CO$_2$ atmosphere without O$_2$ being abundant \citep[cf., Venus;][]{Fegley2003},
but R17 becomes an important sink of Xe$^+$ when O$_2$ is abundant. 

Xe$^+$ also reacts with small hydrocarbons other than CH$_4$ \citep[e.g., C$_2$H$_2$ and C$_2$H$_6$,][]{Anicich1993}
to form HXe$^+$, which then dissociatively recombines. 
This suggests that Xe escape could be suppressed when conditions favor
 formation of high altitude hydrocarbon hazes (as seen on Titan, Triton, and Pluto).
Organic hazes have been a topic of extensive speculation for Archean Earth
\citep{Domagal-Goldman2008,Zerkle2012,Hebrard2014,Arney2016,Izon2017}.
% Hydrocarbon hazes have been suggested as Xe traps on Titan \citep{Jacovi2008}.
It might be tempting to link a possible hiatus in Xe isotopic evolution noted by \citet{Avice2018} between 3.2 Ga and 2.7 Ga
to Archean organic hazes.
It has recently been shown that isotopically ancient Xe was trapped 
 in ancient organic matter on Earth \citep{Bekaert2018}; however, it is not yet known
if the trapping was by a high altitude haze.

\begin{figure}[!htb] 
  \centering
  \includegraphics[width=1.0\textwidth]{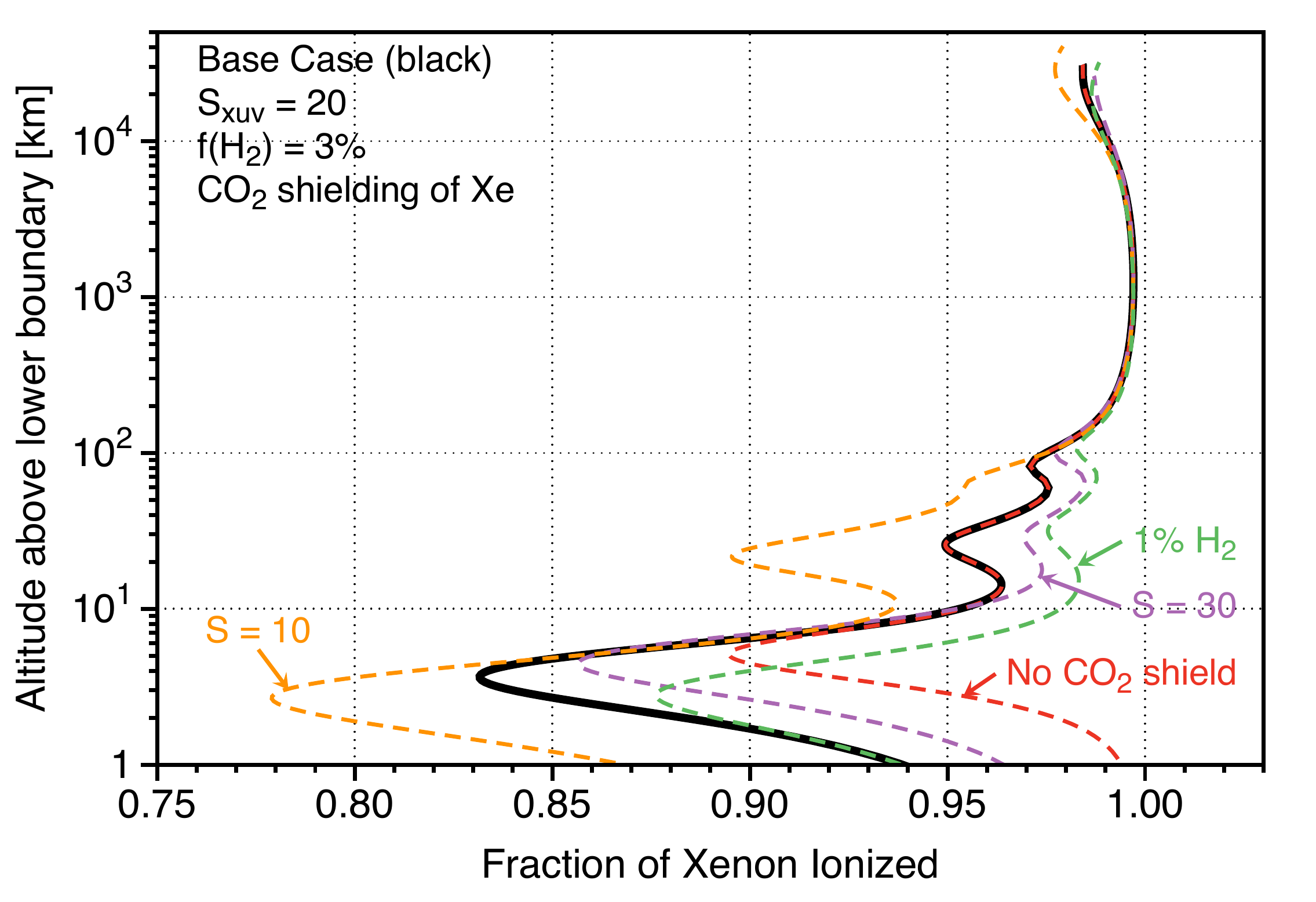} 
\caption{\small Equilibrium fractional ionization of xenon for several exemplary cases.
The broken curves are each labeled according to how they differ from the base case.
Altitude is measured from an arbitrary lower boundary where the total density $n(r_0)=1\times 10^{13}$ cm$^{-3}$.
The variant ``no CO$_2$ shield'' (red dashed line) presumes that Xe is photo-ionized through windows in CO$_2$'s opacity. }
\label{Ionization_fraction}
\end{figure}

Radiative recombination of Xe$^+$ is the unavoidable sink.
We assume that it is neither much faster nor much slower than radiative recombination of H$^+$, 
\begin{equation}
\tag{R18}\label{R18}
\mathrm{Xe}^+ + e^- \rightarrow \mathrm{Xe} + h\nu \qquad \qquad k_{18}=1.4\times 10^{-10}T^{-0.7} \,\mathrm{cm}^3\mathrm{s}^{-1}
\end{equation} 
In Figure \ref{Ionization_fraction} we show the equilibrium ionization computed from
\begin{equation}
\label{Xe ionization}
k_{18} x_j n_e = J_{\mathrm{Xe}}n_j + k_{15}n_3n_j + k_{16b}x_1n_j,
\end{equation}
where we have denoted the number densities of different Xe and Xe$^+$ isotopes by $n_j$ and $x_j$, respectively.
Figure \ref{Ionization_fraction} plots the fractional ionization $x_j/(n_j+x_j)$ in the nominal model and several variants
on the nominal model.  
Variants include $f_{\mathrm{H}_2}=0.01$, $S=10$, and $S=30$.
``CO$_2$ shielding'' refers to the overlap between Xe's absorption and CO$_2$'s absorption, both of which are spiky between
91.2 and 102.3 nm; the base case presumes that CO$_2$'s absorption and Xe's absorption fully overlap,
so that CO$_2$ shields Xe from photo-ionization.
The variant, labeled ``no CO$_2$ shield,'' presumes no spectral overlap between CO$_2$ and Xe,
in which case Xe is photo-ionized through windows in CO$_2$'s opacity.
In all cases Xe's equilibrium ionization typically exceeds $90\%$.

\subsection{Xenon transport}

The forces acting on Xe$^+$ ions are the collisions with hydrogen and H$^+$ that push
Xe$^+$ outwards, the collisions with CO$_2$ that block it, the Coulomb interactions with  
molecular ions that are not escaping, the electric field that tethers the electrons to the ions, and the force of gravity that pulls
the Xe ions back to Earth.  To these we add eddy mixing as described in Appendix B.
% The governing equation for Xe$^+$ escape is Eq \ref{Xe velocity} in Appendix B.

The molecular ions, here CO$_2^+$ and HCO$^+$, present a barrier to Xe$^+$ escape,
because Xe$^+$ is strongly coupled to them by the strong Coulomb interaction and the relatively high
reduced mass (compared to H$^+$) that makes collisions proportionately more effective at transferring momenta.
Molecular ions dominate Xe$^+$ transport in the nominal model from the homopause at 10 km above the lower boundary to the baropause at 80 km above the lower boundary, Figure \ref{Collision}. 
Where the molecular ions flow upward, Xe$^+$ is carried up with them and may gain the opportunity to escape;
 where the ions flow downward or sideways or sit quietly, Xe$^+$ cannot escape. 
 On Earth today at relevant thermospheric altitudes, 
vertical winds of the order of 10 to 20 m s$^{-1}$ are often observed, and often sustained for an hour or more.
% {\bf This paraphrasing is detestable, dishonest, and demanded}    
\citep{Ishii2005,Larsen2012}.
We will assume that these vertical velocities, which are measured in the Doppler shifts of narrow forbidden lines of atomic oxygen \citep{Ishii2005}, are also pertinent to the molecular ions. 
These winds imply rapid and considerable vertical transport.
An upward velocity of 20 m/s sustained for an hour suffices to lift Xe$^+$ by 70 km, which is enough to carry
it through the molecular ionosphere to where it can be handed off to protons and hydrogen escape.
% (Figure \ref{Recombination}).
The molecular ions themselves cannot get very far, because
the typical lifetime against dissociative recombination is only $\sim 10$ seconds.
The Xe ion's lifetime could be much longer, possibly as long as 10 days, its lifetime against radiative recombination. 
In all likelihood the reaction with O$_2$ is the actual sink in a CO$_2$-rich atmosphere.
For Xe$^+$ to last an hour, the O$_2$ density must be less than
$2\times 10^6$ cm$^{-3}$, $\sim 1$ ppm at the homopause. %with $K_{zz}=2\times 10^6$ cm$^2$s$^{-1}$.
% This potential sink highlights the need for rapid vertical transport through the molecular ionosphere if Xe is to escape.
 The O$_2$ sink would be smaller in an N$_2$-CO$_2$-H$_2$ atmosphere and negligible in an N$_2$-CO-H$_2$ atmosphere.

The usual method of describing vertical transport in a 1-D model is through
 an eddy diffusivity $K_{zz}$ that acts to reduce the gradient of the mixing ratio.
As a practical matter eddy diffusion is straightforward to implement in a 1-D model and it is well-behaved numerically.
 Equation \ref{SN2} in Appendix A  provides an effective definition. 
 
Here we will define a $K'_{zz}$ that acts on the ions at altitudes above the neutral homopause.
The observed vertical winds and timescales, $v=10$ m/s and $t=4000$ seconds,
imply $K'_{zz} \sim v^2 t \approx 4 \times 10^{9}$ cm$^2$s$^{-1}$. 
Another way to construct $K'_{zz}$ is as turbulence.  In this case $K'_{zz}$ scales
as the product of the sound speed and a length scale,
multiplied by a scaling factor $\alpha$.
This scaling is used in the ubiquitous $\alpha$-disk model of astrophysical accretion disks, with the scale length
equated to the scale height.  The models work best with $\alpha$ of the order of 0.1 to 0.4 (King et al, 2007). 
Using this prescription, a turbulent $K'_{zz}$ might be of the order of
% $0.1-0.4 \times c_sH \approx 
$1\times 10^9 - 4\times 10^{9}$ cm$^2$s$^{-1}$.
Turbulence with $\alpha > 1$ implies supersonic winds, which is regarded as unsustainable.
On the other hand, concerted motions could yield higher $K'_{zz}$ than turbulence. 

We will use $K'_{zz} = 4\times 10^{9}$ cm$^2$s$^{-1}$ {\em for the ions} as a nominal model,
but it should be noted that such a high value of $K_{zz}'$ 
vastly exceeds the neutral eddy diffusivity $K_{zz} \sim 10^6$ cm$^2$s$^{-1}$ used in contemporary thermospheric 
chemistry modeling \citep{Salinas2016}.  
    
In addition to drag and gravity,
 the electric field that tethers the ions to the free electrons has to be big enough to balance half the weight of the ion,
such that the sum of the masses of the electron and the ion is $(m_i+m_e)/2\approx m_i/2$.  The electric
field generated by HCO$^+$ (29 amu) would therefore produce an upward force of $14.5m_{\mathrm{H}}g$,
which is a relatively small correction for an ion as massive as Xe$^+$,
but quite big for light ions like H$^+$ and O$^+$.
We include the electric force on Xe$^+$
 as a modification of the gravitational force by computing the mean mass of the ions $\mu^+$.

The governing equation for Xe$^+$ escape is developed in Appendix B as Eq \ref{Xe velocity}.
Equation \ref{Xe velocity} is a differential equation for the Xe ion velocity $v_j(r)$ that is solved by the shooting method.
The Xe ion velocity $v_j(r)$ is integrated upward from the lower boundary for all nine Xe isotopes.
The lower boundary velocity $v_j(r_0)$ is bounded by 0 and by the hydrogen velocity $u(r_0)$
(i.e., Xe cannot escape more easily than hydrogen).   
The velocity $v_j(r_0)$ at the lower boundary is iterated until either $v_j(r) \rightarrow u(r)$, in which case Xe$^+$ escapes,
or $v_j(r) \rightarrow 0$, in which case Xe$^+$ is hydrostatic and does not escape.

\subsection{Some numerical results}

We describe our results in terms of an escape factor
$\alpha_j$ of isotope $j$ with respect to hydrogen. 
% This is the same ``escape factor'' defined by \citet{Zahnle1990}, although they used ``$x_j$'' to denote it. 
The escape factor is equal to the ratio of the velocity of isotope $j$ to the velocity of hydrogen at the lower boundary, 
\begin{equation}
\label{escape factor}
\alpha_j \equiv {\phi_j\over n_j(r_0)} \div {\phi_{\mathrm{H}_2} \over n_{\mathrm{H}_2}(r_0)} = {v_j(r_0)\over u(r_0)} .
\end{equation}
The escape factor can be thought of as the relative probability that $^j$Xe escapes compared to hydrogen.

We could define an analogous escape factor between two isotopes by taking the ratio of $\alpha_i$ to $\alpha_j$,
but it is more natural in hydrodynamic escape to define a ``fractionation factor''  
as the difference between $\alpha_i$ and $\alpha_j$,
% Here we will compute fractionation factors using the adjacent isotopes $^{130}$Xe and $^{131}$Xe,
\begin{equation}
\label{fractionation factor}
\alpha_{ij} \equiv \alpha_{j} - \alpha_{i} .
%\alpha_{ij} \equiv \alpha_{130} - \alpha_{131} \geq 0.
\end{equation}
We will evaluate $\alpha_{ij}$ for $j=130$ and $i=131$.

\begin{figure}[!htb] 
  \centering
  \includegraphics[width=0.9\textwidth]{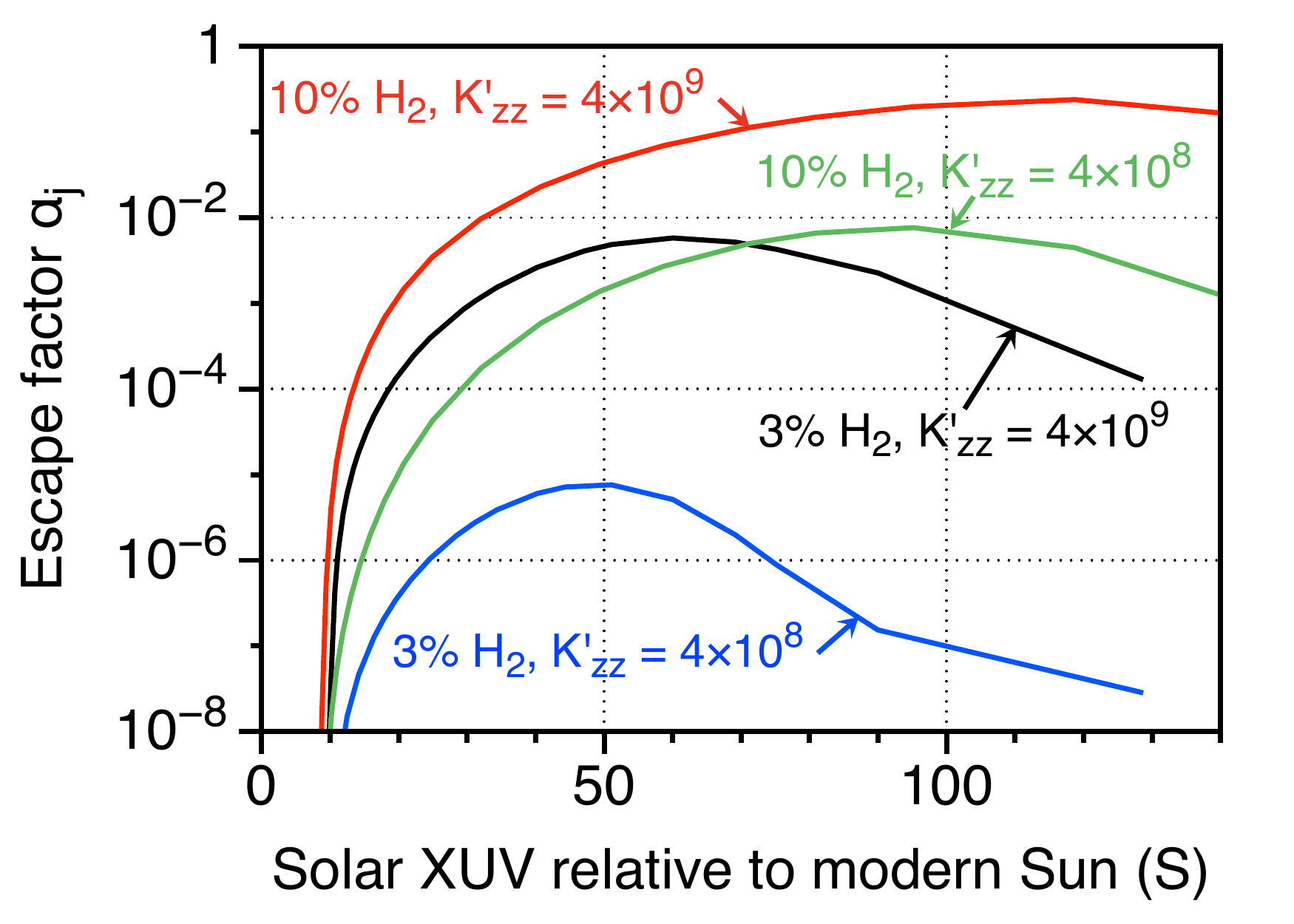} 
\caption{\small Xenon escape factors $\alpha_j$ for $j=130$ as a function of solar XUV irradiation $S$ for
selected values of $K'_{zz}$ and $f_{\mathrm{H}_2}$.
Xenon escape requires $S>10$, a level of solar activity expected of the active Sun before 2.5 Ga.}
\label{Figure13}
\end{figure}

\begin{figure}[!htb] 
  \centering
  \includegraphics[width=0.9\textwidth]{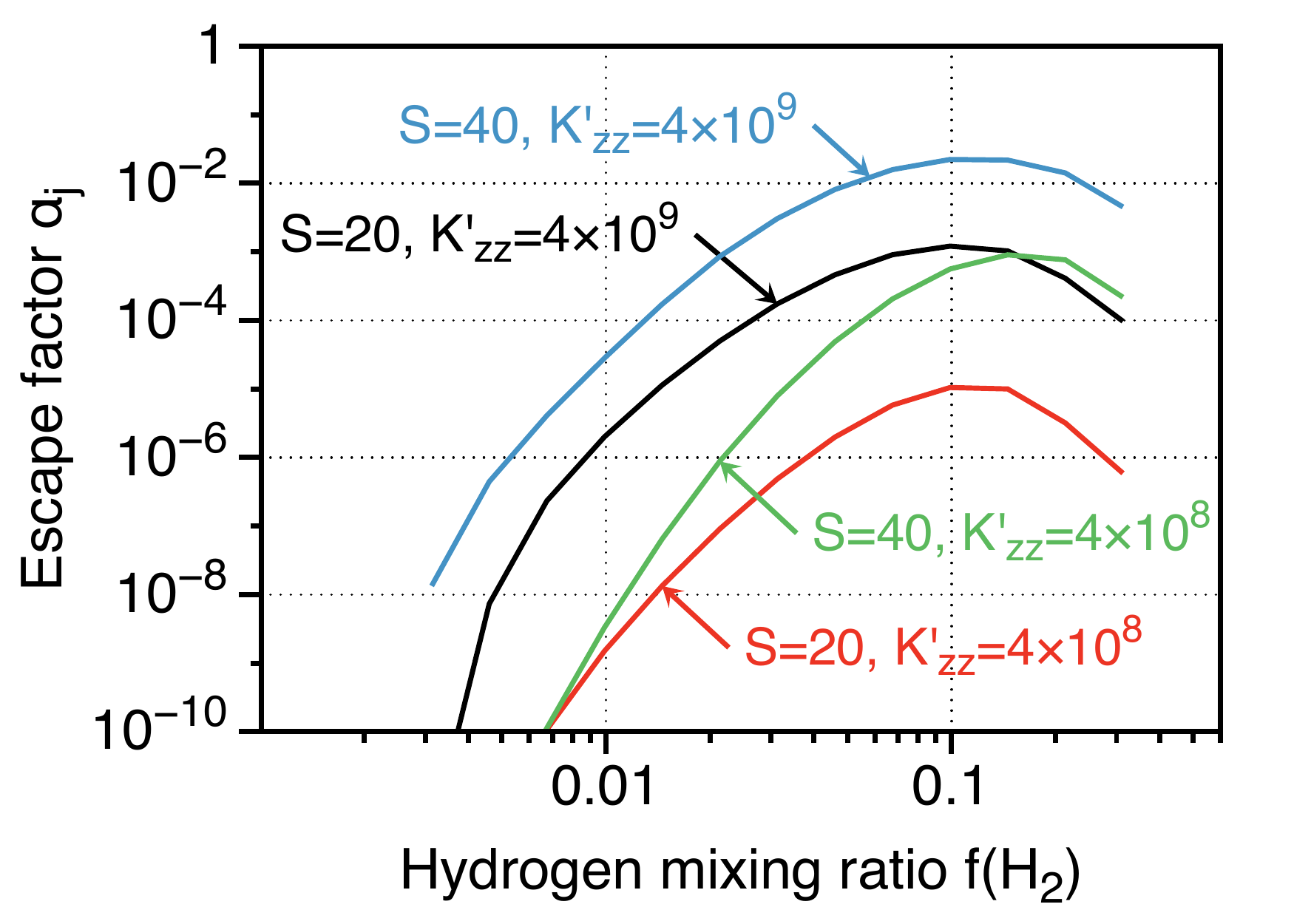} 
\caption{\small Xenon escape factors $\alpha_j$ for $j=130$ as a function of hydrogen mixing ratio for 
some selected values of $K'_{zz}$ and $S$.
The smallest hydrogen mixing ratio where Xe can escape is about 0.4\%,
and this only for the most favorable choices of $K'_{zz}$ and $S$.
Significant Xe escape in the Archean probably requires $f_{\mathrm{H}_2}>0.01$. 
 }
\label{Figure14}
\end{figure}

Figure \ref{Figure13} shows how the escape factor $\alpha_j$ varies as a function of $S$ for 
the nominal model ($f_{\mathrm{H}_2}=0.03$, $K'_{zz}=4\times 10^9$ cm$^2$s$^{-1}$) and some variants.
The minimum irradiation for significant Xe$^+$ escape  is $S> 10$, comparable to what is expected from the 
average Sun ca 3.5 Ga.
At very high levels of irradiation, Xe escape decreases because the Coulomb cross section
between ions decreases at high temperatures.

 \begin{figure}[!htb]
   \centering
    \includegraphics[width=0.9\textwidth]{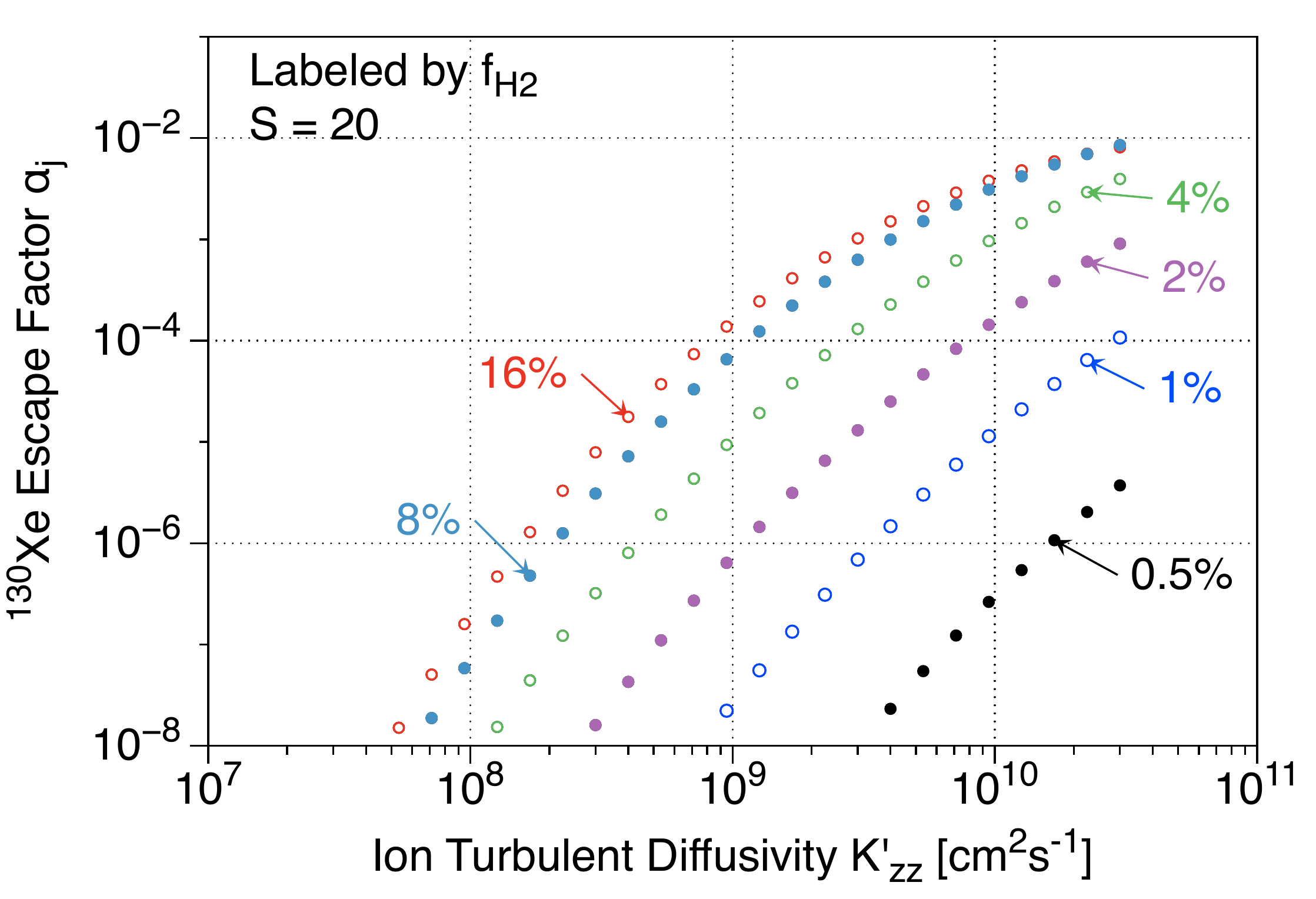} 
   \caption{Variation of Xe escape from a CO$_2$-H$_2$ atmosphere
    as a function of the modeling parameter $K'_{zz}$ for several
   different $f_{\mathrm{H}_2}$ for ``nominal'' value of $S=20$.    
   A plausible upper bound on  $K'_{zz} \sim 1\times 10^{10}$ cm$^2$s$^{-1}$ if $K'_{zz}$ derives from turbulence;
    it could be larger if transport is by large scale circulation, and it would be much smaller if transport
    were more akin to that amongst the neutrals.
   The figure illustrates the tradeoff between $K'_{zz}$ and $f_{\mathrm{H}_2}$.
} 
\label{Figure15}
\end{figure}

Figure \ref{Figure14} shows how the Xe escape factor $\alpha_j$ varies as a function of hydrogen mixing ratio $f_{\mathrm{H}_2}$ for 
the nominal model ($S=20$, $K'_{zz}=4\times 10^9$ cm$^2$s$^{-1}$) and some variants.
The minimum hydrogen mixing ratio at the homopause for significant Xe$^+$ escape is $\sim\!0.4\%$,
achievable when both $S$ and $K'_{zz}$ are very large; at the lower levels of irradiation expected during the
Archean, the lower bound on $f_{\mathrm{H}_2}$ is $\sim\!1\%$.

Figure \ref{Figure15} shows how the Xe escape factor $\alpha_j$ depends on the modeling parameter $K'_{zz}$
for the nominal $S=20$ and a variety of hydrogen mixing ratios.
The modeling parameter $K'_{zz}$ describes vertical transport by ions through the lower molecular ionosphere as a diffusivity,
a form well-suited to a 1-D model.
As discussed above, a plausible upper bound on $K'_{zz}$ is
 of the order of $c_s H_{\mathrm{CO}_2} \approx 1\times 10^{10}$ cm$^2$s$^{-1}$
  if $K'_{zz}$ derives from turbulence;
    $K'_{zz}$ could be larger if transport is by large scale circulation, and it could be much smaller if transport
    were more akin to that amongst the neutrals near the neutral homopause.
 
 \begin{figure}[!htb] %  figure placement: here, top, bottom
   \centering
  \includegraphics[width=1.0\textwidth]{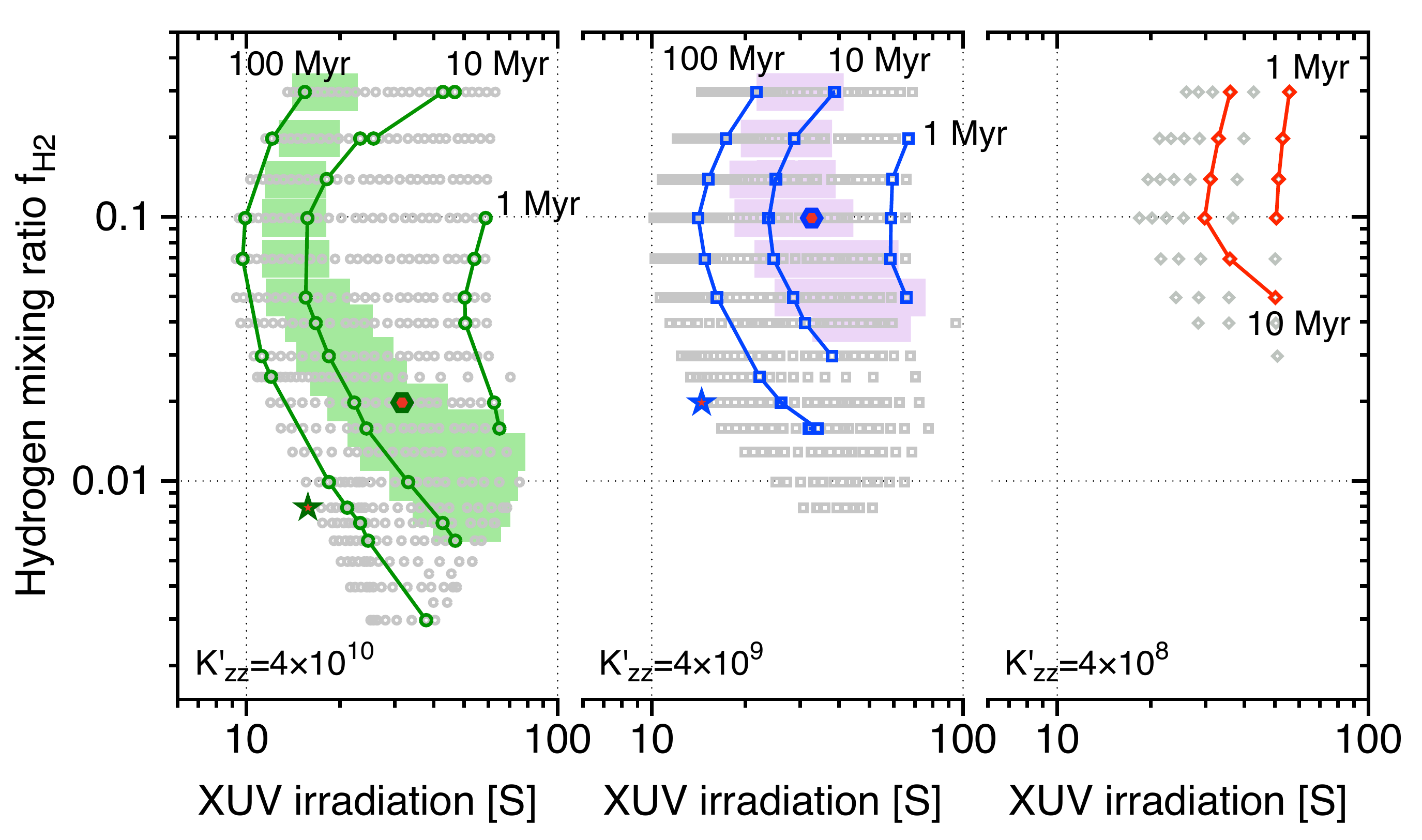} 
 \caption{\small Xenon escape as a function of irradiation $S$ and hydrogen mixing ratio $f_{\mathrm{H}_2}$,
for three values of eddy diffusivity $K'_{zz}$ that correspond to different rates of transport through the molecular ion barrier. 
The middle panel ($K'_{zz}=4\times 10^{9}$ cm$^2$s$^{-1}$) is the nominal case.
The high value $K'_{zz}=4\times 10^{10}$ cm$^2$s$^{-1}$ in the left-hand panel
exceeds what could plausibly be attributed to turbulence. 
The gray symbols mark out regions of the $S-f_{\mathrm{H}_2}$ plane where xenon escape is significant.
The shaded regions encompass models that at uniform conditions can simultaneously reproduce both Xe's
mass fractionation and depletion. 
The contours are labeled by how long it takes (elapsed time) 
for fixed conditions to generate the observed Xe isotopic fractionation if Xe escape is global.
It takes proportionately longer if Xe escape is channeled through polar apertures; i.e., the curve marked ``10 Myr'' 
corresponds to 100 Myrs if escape is restricted to 10\% of Earth's area.
 %values of $\alpha_{ij} > 1\times 10^{-5}$ correspond to Xe escape rates too great to have been 
%sustained globally over the entire Archean.
The stars and hexagons mark exemplary cases singled out for discussion in the text.
}
\label{swaths}
\end{figure}

Figure \ref{swaths} summarizes Xe$^+$ fractionation factors $\alpha_{ij}$ as contours on the $S-f_{\mathrm{H}_2}$ plane,
with the three panels corresponding to three values of $K'_{zz}$.
The figure graphically illustrates the tradeoffs between $S$, $f_{\mathrm{H}_2}$, and $K'_{zz}$.
Bounds on $f_{\mathrm{H}_2}$ depend on $K'_{zz}$.
The right hand panel shows that if $K'_{zz}$ is too small, Xe escape would be difficult in the Archean where $S<20$, 
while the left hand panel of Figure \ref{swaths} sets $K'_{zz}$ so high that the mixing length
would have to greatly exceed the scale height, which is probably not possible unless the flow were organized
by the geomagnetic field, for which characteristic length scales are long.   
The middle path is consistent with Xe escape in the Archean provided that the atmosphere contains $>1\%$ hydrogen (or $>0.5\%$ methane). 
 
\section{Fractionation}
\label{section four} 

Escape factors and fractionation factors are snapshots in time.
%Specifically, they describe conditions averaged over a time scale longer than the characteristic Xe escape time of
%$H\div v_j \sim 10^5\div \alpha_j$ seconds, here evaluated for H escape in the diffusion limit.
%We will find below in Section \ref{section four} that $\alpha_{j} > 5\times 10^{-6}$
%creates significant mass fractionation on a billion year time scale in a 1 bar atmosphere, which means that
%$\alpha_j$ and $\alpha_{ij}$ should be thought of as 1000-year averages. 
%
The observables --- the isotopic fractionation and total xenon loss --- 
are integrated quantities that depend on the cumulative histories of the solar XUV irradiation $\int{\!S(t)dt}$,
the mass and composition of the atmosphere, the presence or absence of a planetary magnetic field, and probably
several other things that we haven't addressed. 
% Fractionation will be discussed in Section \ref{section four}.

A convenient way to describe the fractionation between two isotopes $^i$Xe and $^j$Xe is to compare the ratio of their relative abundances at a later time $t_B$ to their relative abundances at an earlier time $t_A$,
\begin{equation}
 \xi_{ij}(t_A,t_B) \equiv {N_i(t_B)\over N_i(t_A)} \div {N_j(t_B) \over N_j(t_A)} ,
\label{four}
\end{equation}
where $N_j$ represents the total (column) reservoir of isotope $j$ (number per cm$^2$). 
When applied to hydrodynamic escape, the fractionation $\xi_{ij}(t_A,t_B)$ describes the relative depletion of isotope $i$ compared to isotope $j$.
We will take U-Xe \citep{Pepin1991,Pepin2006} as the initial composition at $t_A$; our results are insensitive to this choice.
In hydrodynamic escape, the rate of change of $\xi_{ij}(t)$ can be expressed in terms of hydrogen escape,
\begin{equation}
\label{rate_of_change}
{1\over \xi_{ij}}{\partial \xi_{ij} \over \partial t} 
= {1\over N_{i}}{\partial N_i \over \partial t} - {1\over N_j}{\partial N_j \over \partial t}
= {\phi_j\over N_j} - {\phi_i\over N_i} 
= -{\phi_{\mathrm{H}_2}\over N_{\mathrm{H}_2}}\alpha_{ij} .
\end{equation}
The fractionation $\xi_{ij}$ between isotopes $^j$Xe and $^i$Xe is the integral  
\begin{equation}
 \ln{\left\{ \xi_{ij}(t_A,t_B)\right\}} 
% = \int_{t_A}^{t_B}\! {\phi_{\mathrm{H}_2}\over N_{\mathrm{H}_2}} \left( \alpha_{j} - \alpha_{i} \right) dt
 = -\int_{t_A}^{t_B}\! {\phi_{\mathrm{H}_2}\over N_{\mathrm{H}_2}} \alpha_{ij}(t) dt .
\label{fractionation-1}
\end{equation}
Similarly, the total loss or depletion $\xi_{j}$ of the isotope $^j$Xe can be defined
\begin{equation}
\label{depletion-0}
 \xi_{j}(t_A,t_B) \equiv {N_j(t_B)\over N_j(t_A)} ,
\end{equation}
which can also be described by an integral over hydrogen escape
\begin{equation}
 \ln{\left\{ \xi_{j}(t_A,t_B)\right\}} 
 = -\int_{t_A}^{t_B}\! {\phi_{\mathrm{H}_2}\over N_{\mathrm{H}_2}} \alpha_{j}(t) dt .
\label{depletion-1}
\end{equation}
%The factors $\alpha_j(t)$ and $\alpha_{ij}(t)$ 
%are in general complicated functions of $f_{\mathrm{H}_2}$ and $S$.
Equation \ref{fractionation-1} for fractionation $\xi_{ij}$ and Eq \ref{depletion-1} for depletion $\xi_{j}$
presume that almost all of Earth's xenon is and was in the atmosphere;
i.e., it is presumed that the columns $N_j$ faithfully represent the global inventories.

\subsection{Constraints}

Xenon directly places two constraints on escape. 
The mass fractionation of the isotopes, $\xi_{ij}$ --- roughly 4\% per amu, about half of this
taking place after 3.5 Ga --- is the more secure and the more telling, so it takes primacy.
As we shall see, the most important aspect of this constraint is that escape was drawn out over one to two billion years,
which is why Xe escape implies the loss of a great deal of water. 

The second constraint is on the total xenon escape, $\xi_j$.
Total escape conflates the apparent 4-20-fold depletion with respect to Kr in meteorites (Figure 1)
with the missing daughter products of spontaneous fission of $^{244}$Pu;
it also lacks the historic record of evolution we have for the isotopes. 
%Earth is about 20-fold depleted in Xe with respect to Kr compared to carbonaceous chondrites, and about 4-fold depleted
% compared to the enstatite meteorites (Figure 1). 
Because Earth's original U-Xe is not found in meteorites, the depletions with respect to Kr in
meteorites are at best rough guides to what the true depletion might be.
Earth retains about 20\% of the Xe spawned by spontaneous fission of $^{244}$Pu 
 \citep[80 Myr half-life;][]{Pepin1991,Pepin2006,Tolstikhin2014}.
%an estimate obtained by assuming that the uranium content of the bulk silicate Earth is 21 ppb, that Earth's 
% Pu/U ratio is that of chondrites, and that 4.6\% of the $^{136}$Xe in air is fissiogenic  
%Given plutonium's considerable half-life, this factor 5 depletion might seem a lower limit
%(as Xe yet to form or degas to the atmosphere cannot escape), but 
There are enough uncertainties 
in Earth's U abundance and original Pu/U ratio, and in possible nonfractionating loss mechanisms like impact erosion)
 that we will simply treat $\xi_j \approx 0.25$ as a best guess.
 
A third constraint is the total oxidation of the Earth.
Hydrogen escape can oxidize the Earth from an initially reduced, Moon-like state \citep{Sharp2013}.
An upper bound on cumulative oxidation is obtained by adding up the ferric iron and the CO$_2$ contents of Earth.
To first approximation, the Fe$_2$O$_3$
inventory in the MORB-source mantle \citep[$>0.3$ wt\%,][]{Cottrell2011} can be balanced with the oxygen from
the escape of $\sim \! 1$ ocean of water,
\begin{equation}
\label{iron stochiometry}
  2\mathrm{FeO} + \mathrm{H}_2\mathrm{O} \rightarrow \mathrm{Fe}_3\mathrm{O}_2 + \mathrm{H}_2 .
\end{equation}  
% Carbon dioxide is another large reservoir of oxygen generated by hydrogen escape.  
\citet{Dauphas2014} estimate from CO$_2$/Nb systematics that there are $3.4\times 10^{22}$ moles
of carbon on Earth (the majority in the mantle). 
Carbon isotopes suggest that 80\% of Earth's C is in CO$_2$.
Although some carbon was accreted by Earth in the form of CO$_2$
(as ice in comets or as carbonate minerals in meteorites), most of it probably accreted in reduced
form, which we idealize as CH.  The stoichiometry of carbon oxidation by H$_2$O is 
\begin{equation}
\label{stochiometry}
  \mathrm{CH} + 2\mathrm{H}_2\mathrm{O} \rightarrow \mathrm{CO}_2 + {5\over 2}\mathrm{H}_2 .
\end{equation}
If we assume that 20\% of Earth's C was directly accreted as CO$_2$ and the other 80\% as CH,
the oxidation of carbon on Earth corresponds to the oxygen extracted from $\sim\!0.3$ oceans of water.
As there are alternative stories to explain how the mantle became oxidized \citep{Frost2008},
escape of one or two oceans of water can be regarded as an upper bound. 

A fourth constraint stems from the kinetics of Earth's oxidation.
If the escaping H$_2$ comes from H$_2$O, something must be oxidized and exported to the mantle,
which can be rate-limiting.
It is illustrative to consider a rough upper bound set by oxidation of ferrous to ferric iron.
This includes weathering of continents and seafloor.
For the former, presume that continents of modern mass and 7\% Fe 
were built, eroded, weathered, and subducted on a 1 Gyr time scale, with 100\% conversion of Fe$^{+2}$ to Fe$^{+3}$. This would correspond to an average H$_2$O sink of $7\times 10^{12}$ moles yr$^{-1}$. 
For the latter, presume that the upper 1\% of the seafloor was oxidized and that the mantle turned over in 1 Gyr.
This corresponds to an average H$_2$O sink of $1.4\times 10^{13}$ moles yr$^{-1}$.
The latter can be compared to an Fe$^{+2}$ source of $0.6-1.2\times 10^{13}$ moles yr$^{-1}$ estimated by scaling modern midocean ridge hydrothermal fluxes \citep{Ozaki2018}.
Note that \citet{Ozaki2018} infer much higher total Fe$^{+2}$ fluxes of order $7-21\times 10^{13}$ moles yr$^{-1}$, most of which is  biologically recycled.  The high recycled flux maps to high CH$_4$ production, which suggests that H$_2$ levels could change dramatically in response to biological forcing; e.g., blooms.

Summed, the {\em de novo} weathering source of H$_2$ probably did not exceed $2\times 10^{13}$ moles yr$^{-1}$, or 0.25 oceans per Gyr.
The equivalent H$_2$ escape flux --- $8\times 10^{10}$ molecules cm$^{-2}$s$^{-1}$ in photochemical units ---
corresponds to hydrogen mixing ratios $\leq 1\%$ in a 1 bar Archean atmosphere,
and is probably insufficient to support Xe escape unless the atmosphere were substantially thinner than 1 bar,
or the ferrous iron oxidation rate varied significantly in response to
climate fluctuations, variations in the biological production of CH$_4$, changes in tectonic style, or episodic continent building.

Two other sources of H$_2$ are worth mentioning.  A big mantle source of H$_2$ and other reduced gases may be plausible in the Hadean while metallic iron was still extant
but is less likely in the Archean when the mantle appears to have been only modestly more reduced than today.
A more exotic possibility is that H$_2$ and other reduced gases were injected episodically into the atmosphere from degassing of reduced impacting bodies
\citep{Kasting1990,Hashimoto2007,Schaefer2010}. 
A $10^{21}$ g chondritic body, comparable to the bigger Archean impact events documented by \citet{Lowe2018},
on reacting with Earth's oceans would inject of order $3\times 10^{18}$ moles of H$_2$ into the atmosphere (0.02 bars),
enough to support a burst of Xe escape.

The D/H ratio of Earth may provide some support for the hypothesis that Earth lost an ocean or more of water.  
  \citet{Zahnle1990} found that the escape factor for HD with respect to H$_2$ 
  in diffusion-limited hydrogen escape with $f_{\mathrm{H}_2}\ll f_{\mathrm{CO}_2}$ is 
  $\alpha_{\mathrm{HD}} \approx 0.8$.
  The resulting Rayleigh fractionation of the water that remains on Earth leads to a modest enhancement of the D/H ratio,
  \begin{equation}
  \label{D/H}
 {\mathrm{D}/\mathrm{H}(t_B) \over \mathrm{D}/\mathrm{H}(t_A) } 
   = \left( M_{\mathrm{H}_2\mathrm{O}}(t_A)\over M_{\mathrm{H}_2\mathrm{O}}(t_B)\right)^{1-0.8} .
  \end{equation}
  If for example Earth lost two oceans of water by diffusion-limited hydrogen escape, the resulting
  D/H enrichment would be of the order of $3^{0.2}=1.25$.  The predicted
   25\% enrichment in D/H is less than the $\sim\!40-60\%$ difference
  between Earth's D/H and the D/H ratios measured in most carbonaceous chondrites,
  and it is small compared to the scatter between the measured D/H ratios in the array of known possible sources 
  of Earth's water \citep{Alexander2011}.   
 \citet{Pope2012} reported that the D/H ratio of seawater at 3.8 Ga was 2.5\% lighter than today,
 which corresponds to the loss of about 13\% of an ocean according to Eq \ref{D/H}. 
     
\medskip

\subsection{Example 1}

To quantify concepts, consider first an order of magnitude estimate for Archean conditions,
using only the fractionation $\xi_{ij}$ as a constraint.
Assume that fractionation of 2\% per amu took place between 3.5 Ga and 2.5 Ga (Figure 3),
and assume a constant integrand in Eq \ref{fractionation-1}, which effectively means holding $\alpha_j$ constant, 
\begin{equation}
 \ln{\left\{ \xi_{ij}\right\}} = 0.02 \approx {\phi_{\mathrm{H}_2}\over N_{\mathrm{H}_2}} \alpha_{ij} \left(t_B-t_A\right).
\label{lalalala}
\end{equation}
Using Eq \ref{fit}, Eq \ref{lalalala} evaluates to
\begin{equation}
\label{tralala}
% \alpha_{ij} = 5\times 10^{-7} \left({N_a \over 1.4 \times 10^{25}}\right) \left( {1\,\mathrm{Gyr}\over t_A-t_B} \right)
 \alpha_{ij} = 5\times 10^{-7} p_{\mathrm{atm}} \left( {1\,\mathrm{Gyr}\over t_A-t_B} \right)
\end{equation}
where $p_{\mathrm{atm}}$ is the total atmospheric pressure in bars and
$N_a=1.4\times 10^{25}$ cm$^{-2}$ for 1 bar of CO$_2$.
For the particular case with $f_{\mathrm{H}_2}=0.02$ and $K'_{zz}=4\times 10^9$ cm$^2$s$^{-1}$,
the inferred $\alpha_{ij} = 5\times 10^{-7}$ corresponds to $S=15$ and $\alpha_j = 5\times 10^{-6}$.
This case is marked by the blue star in the middle panel of Figure \ref{swaths}.
Total Xe loss is only $\xi_j=0.67$, which means that Earth loses only one-third of its Xe between 3.5 and 2.5 Ga. 
In global escape, the required hydrogen loss (expressed as equivalent water) is
\begin{equation}
  M_{\mathrm{H}_2\mathrm{O}} =  1.3\times 10^{13}m_{\mathrm{H}_2\mathrm{O}}A_{\oplus}\left( {f_{\mathrm{H}_2}\over 0.02}\right)  \left(t_B-t_A\right)
   = 1.2\times 10^{24} \,\mathrm{grams} 
\end{equation}
which is 0.86 oceans of water (1 ocean is $1.4\times 10^{24}$ g). 
This is probably too much mantle oxidation after 3.5 Ga, given that we have not addressed the first billion years 
and the other half of xenon's fractionation.
With the same $S=15$, the extreme $K'_{zz}=4\times 10^{10}$ cm$^2$s$^{-1}$ works with $f_{\mathrm{H}_2}=0.008$,
which corresponds to 0.35 oceans lost during the Archean. 
This case is marked by the green star in the left panel of Figure \ref{swaths}.
In both cases, the trickle of Xe escape is very close to no Xe escape at all.
In a more realistic scenario, $\alpha_{ij} \gg 5\times 10^{-7}$ where and when Xe escapes,
averaged with a more typical state in which Xe does not escape ($\alpha_{j}=\alpha_{ij}=0$).

\subsection{Example 2}

Our second example is constrained to simultaneously match both the fractionation $\xi_{ij}$ and the depletion $\xi_{i}$.
In the special case of constant integrands in Eqs \ref{fractionation-1} and \ref{depletion-1}, the fractionation $\xi_{ij}$
and the depletion $\xi_j$ are related by 
 \begin{equation}
\label{Rayleigh}
{\alpha_{j} \over \alpha_{ij}} = - {\ln{\left( \xi_{j}\right)} \over \ln{\left( \xi_{ij}\right)}} \approx  - {\ln{\left( 0.25\right)} \over 0.04} \approx 35.
\end{equation}
Equation \ref{Rayleigh} is effectively Rayleigh fractionation. 
Models with $25<\alpha_{j}/\alpha_{ij}<45$ are marked out in shades of color on Figure \ref{swaths}.
In general, these solutions require high rates of escape, driven by high $f_{\mathrm{H}_2}$ or high $S$ or both.
 
Two models that fit these constraints are marked by hexagons on Figure \ref{swaths}, one 
with $K'_{zz}=4\times 10^{9}$ cm$^2$s$^{-1}$ and $f_{\mathrm{H}_2}=0.10$ (middle panel), 
the other with $K'_{zz}=4\times 10^{10}$ cm$^2$s$^{-1}$ and $f_{\mathrm{H}_2}=0.02$ (left panel).
Both use a relatively high solar EUV irradiation of $S=30$.
Both examples give $\alpha_j=1\times 10^{-2}$ and $\alpha_{ij} = 2.5\times 10^{-4}$.
For uniform conditions,
the time scale $(t_B-t_A)$ to evolve Xe fractionation is just 3 Myrs if escape is global and 
30 Myrs if escape is channelled through polar windows open over 10\% of Earth.
The hydrogen lost globally in 30 Myr is only 0.05 oceans with $K'_{zz}=4\times 10^{10}$, or 0.25 oceans with $K'_{zz}=4\times 10^9$.
This could be the whole story if hydrogen were present only in brief episodes that sum to 30 Myrs,
scattered over the two billion years during which Xe fractionation took place.

To quantify the previous statement,
consider a pulse of hydrogen into the atmosphere.  From Eq \ref{fit}, escape goes as 
\begin{equation}
\label{decay}
\frac{d N_{\mathrm{H}_2}}{dt} \approx \frac{2\times 10^{12} S}{\sqrt{1+0.006S^2}}\frac{N_{\mathrm{H}_2}}{N_A} .
\end{equation}
Equation \ref{decay} predicts that $f_{\mathrm{H}_2}$ exponentially decays on the time scale
\begin{equation}
\label{decay-2}
\tau_{\mathrm{H}_2} = \frac{N_A\sqrt{1+0.006S^2}}{2\times 10^{12} S}\frac{N_{\mathrm{H}_2}}{N_A} \approx 3\times 10^4 \mathrm{~years}
\end{equation}
for $S>10$ and 1 bar of CO$_2$.
It takes $\sim 1000$ episodes (summing to 30 Myrs total if channeled through polar windows that open over 10\% of Earth) 
that each raised $f_{\mathrm{H}_2}$ to 10\% (or 2\% for $K'_{zz}=4\times 10^{10}$) to account for Xe escape.  
Of course there would also be many more hydrogen excursions too small to drive Xe escape, 
or that took place at an inopportune time when $S$ was too small,
and it will be the number and magnitude of these other events that 
determines how much hydrogen escapes in total.

\subsection{Example 3}

Consider finally a schematic example that more explicitly invokes the geomagnetic field.
This is the scenario illustrated schematically in Figure \ref{cartoon2}. 
The magnetic field partitions escape into two regimes.
 Near Earth any plausible geomagnetic field is much stronger than the hydrodynamic wind, 
 and therefore controls it.
 This ordering of strengths is determined by comparing the magnetic pressure $B^2/4\pi$ 
 of the field to the ram pressure $\phi_{\mathrm{H}_2} m_{\mathrm{H}_2} u$ of the wind.
 Because the strength of a dipole field drops as $r^{-6}$, at large distances
 the polar wind eventually overcomes the field and stretches the field radially.  
 The polar field lines open out to space and both ions and neutrals can escape.
 This is the situation we have been modeling.

 The equatorial region, however, is quite different.
 Here the field lines remain closed and ions, which cannot cross the field lines, cannot escape.
The result is a relatively hot and dense quasi-static partially ionized plasma that exerts a pressure limited only by the native strength of the geomagnetic field.
Although neutrals can still escape by diffusing through the ions, escape is inhibited and is probably restricted to H and He. 
 A detailed examination of the inflated magnetosphere is beyond the scope of this study; what is important here
 is that (i) Xe cannot escape through the equatorial magnetosphere because ions are trapped, and (ii) there will be a flow of energy
 from the hot dense equatorial magnetosphere to the cooler, sparser, but free-flowing polar wind.

As it pertains to Xe escape, the net effect of a more realistic geomagnetic geometry 
is to subvert the constraint on $S$ in the Archean that would otherwise be inferred from solar analogues.
%Under these general conditions we would expect hydrogen escape to be enhanced at the poles
%but possibly suppressed, or at least less vigorous, over the other 90\% of the globe. 
If we take the polar wind to be described by the constant $f_{\mathrm{H}_2}=0.02$, high $K'_{zz}$ case from Example 3, 
the total hydrogen escape through the polar windows would be 0.33 oceans over 2 billion years.
If away from the poles we take $S=6$, typical of the mid-Archean,
the total equatorial hydrogen loss with constant $f_{\mathrm{H}_2}$ 
over 2 Gyr would be $\sim 1.5$ ocean.
In this scenario xenon escape occurs exclusively at the poles and only at rare times when $S$ was abnormally large,
while hydrogen escape is constant from an atmosphere that is always 2\% hydrogen.

%Finally, pathological models can be devised with substantially less hydrogen escape by invoking special circumstances.  E.g., the total hydrogen escape can be reduced to  as little as a few tenths of an ocean if $f_{\mathrm{H}_2}$ or $S$ or both were highly variable and only rarely large,  but when large, hydrogen escape took place at a very high rate.   

\section{Discussion}
\label{section five}

In this study we have argued that Xe escaped Earth as an ion and
we have quantified the hypothesis
in the context of a restricted range of highly idealized 
models of diffusion-limited hydrodynamic hydrogen escape from CO$_2$-H$_2$ atmospheres in the absence of
inhibition by a magnetic field.
These atmospheres develop cold lower ionospheres dominated by molecular ions enveloped by much-extended partially-ionized hydrogen coronae that flow into space,
a general structure shown schematically in Figure \ref{expt5}.

We identify three requirements for Xe escape:
(1) Xe atoms must become ionized and remain ionized. 
(2) The flow of H$^+$ to space must be great enough that collisions between the ions can push
the Xe$^+$ ions upwards faster than they can be pulled back by gravity.
(3) The Xe$^+$ ions must be transported by vertical winds through the cold lower ionosphere
where CO$_2$ is abundant to high enough altitudes that outflowing hydrogen can sweep them away.
Here we discuss each requirement in turn.

(1) Xenon is mostly ionized above the homopause.  This occurs in part
because Xe can be photo-ionized at wavelengths  
to which H$_2$ and H are transparent; in part because
charge exchange of neutral Xe with CO$_2^+$ is fast; and in part because
there is a dearth of recombination mechanisms faster than radiative recombination.
In particular, Xe$^+$ does not react with H, H$_2$ or CO$_2$.
This partiality to life as an ion would change if O$_2$ were abundant, as was certainly the case after
the GOE (Great Oxidation Event) ca 2.4-2.1 Ga \citep[][p.~257]{Catling2017}, 
or if small hydrocarbons (other than CH$_4$) were abundant, as could have been the case
if Archean Earth experienced a Titan-like pale orange dot phase \citep{Domagal-Goldman2008,Arney2016}.
This is because O$_2$ and hydrocarbons like C$_2$H$_2$ 
 charge exchange with Xe$^+$ \citep{Anicich1993} to render Xe neutral 
and subject to falling back to Earth.
If a Xe$^+$ ion reaches the hydrogen corona,
it is very likely to remain ionized because of charge exchange 
with H$^+$.

%Although the reaction was not known to occur, the strong evidence that Xe escapes from planetary atmospheres
%led us to suspect that the reaction might be reasonably fast.
%We therefore calculated the previously unknown rate of the 
% $\mathrm{H}^+ + \mathrm{Xe} \rightarrow \mathrm{H} + \mathrm{Xe}^+(^2\mathrm{P}_{1\over 2}) + 0.16 \mathrm{~eV}$
%charge exchange reaction (R16b) that produces the electronically excited $\mathrm{Xe}^+(^2\mathrm{P}_{1\over 2})$ ion.
% We determined that the rate of this exothermic but nearly resonant reaction is
% of the order of $k_{16b}=10^{-11}$ cm$^3$s$^{-1}$ at 300 K and $10^{-10}$ cm$^3$s$^{-1}$ at 800 K,
% rates that are fast enough to ensure that Xe remains ionized as it is dragged off by hydrogen.
%The competition between the charge exchange source and the radiative recombination sink 
% strongly favor the ion at warm temperatures.
%What this means is that a Xe ion is very likely to stay ionized if it reaches the warm hydrogen corona.

(2) The H$^+$ escape flux depends on the hydrogen mixing ratio $f_{\mathrm{H}_2}$ and the flux of ionizing radiation $S$.
In the Hadean, when $S$ sometimes exceeded $40$,
and if everything else were favorable, Xe may have been able to escape with hydrogen mixing ratios as small as 0.4\%.
Otherwise more hydrogen is needed.
The minimum $f_{\mathrm{H}_2}$ for Xe escape probably exceeded 1\% in the Archean when $S$ was smaller
(the lower atmosphere could be 1\% H$_2$ or 0.5\% CH$_4$ to the same effect, 
because CH$_4$ is photochemically converted to H$_2$ and other products).  
The minimum EUV flux for Xe escape is in the range $10<S<15$, 
high enough that Xe escape would have been rare after 3.0 Ga.
These specific properties of our model are likely to be shared by any model that uses hydrogen escape to drive Xe$^+$ into space. 
 A more realistic model of the polar wind might drive off Xe with a smaller value of $S$,
because other means of heating and ionizing would be available to a magnetically channeled polar wind. 

(3) Transport of Xe$^+$ through the cold molecular ionosphere (Figure \ref{expt5})
 is an unresolved issue.
If the molecular ions themselves flow upwards with the hydrogen, then all is well with Xe$^+$ escape,
but our model gives them no cause to do so. 
To describe vertical transport we introduced a third modeling parameter --- an ion diffusivity $K'_{zz}$ --- to supplement the physical parameters $S$ and $f_{\mathrm{H}_2}$.
By construction $K'_{zz}$ acts only on ions, which are strongly coupled to each other
by electromagnetic forces.
Although fundamentally $K'_{zz}$ is a modeling parameter, we can
estimate its magnitude from observations of vertical winds in the thermosphere
and by analogy to turbulent viscosity in astrophysical accretion disks.
Both suggest that $K'_{zz}$ could be of the order of $4\times 10^{9}$ cm$^2$s$^{-1}$.
This is quite high compared to eddy diffusion amongst the neutrals,
but we find that Xe cannot escape in our 1-D model if $K'_{zz}$ is much smaller than $4\times 10^{8}$ cm$^2$s$^{-1}$.
The dependence of Xe escape on $K'_{zz}$ is more or less separate from the dependence on $f_{\mathrm{H}_2}$;
the former affects transport through the lower ionosphere, whilst the latter affects transport in the escaping hydrogen corona.
Thus our conclusions regarding the minimum amount of hydrogen required in the atmosphere for Xe escape
are not greatly affected by $K'_{zz}$. 
% In reality we expect that Xe$^+$ transport at the base of the polar wind is diurnal, channeled and organized
% on large scales by the geomagnetic field.}

\subsection{Other omissions and future directions}

We have omitted much in order to carve out a tractable problem. 
We have already mentioned transport through the molecular ions, which likely will require a 3-D 
MHD model to properly address.
Three other important omissions are thermal conduction, O$_2$, and O$^+$ ions.

Thermal conduction will smooth the thermal transition between the hydrogen corona and the molecular ionosphere,
and it can slow hydrogen escape 
by transporting energy from the warm hydrogen corona to low altitudes where radiative cooling can be effective.
Negligible differences between our results and those obtained using a time-dependent hydrocode 
that includes thermal conduction (Figure \ref{Figure9a}) suggest that thermal conduction 
is not very important to hydrogen escape from Earth where $S>2.5$.
Our own numerical experiments with hydrostatic CO$_2$-H$_2$ atmospheres suggest that hydrogen-rich atmospheres
can be stabilized against escape by thermal conduction for $S<1$. 
This should prove an interesting topic for further research.   

Charge exchange with O$_2$ is a major threat to Xe$^+$.
The relatively oxidized CO$_2$-H$_2$ atmospheres we have considered here are prone to spawning
 photochemical O$_2$ at high altitudes, especially given vigorous hydrogen escape.
 The O$_2$ threat can be mitigated by a generally more reduced atmosphere.
% Stoichiometry implies that $n_{\mathrm{O}_2} \propto n^2_{\mathrm{CO}_2}$.
 Diluting CO$_2$ with CO or N$_2$ could go a long way toward reducing O$_2$. 
 O$_2$ could also be removed by a catalytic chemical cycle involving trace constituents, as apparently occurs on Venus.
  That Xe escape might require a more reduced atmosphere for early Earth than we have considered,
 perhaps leading to an upper bound on the amount of CO$_2$ in the Hadean or Archean,
 and the suggestion that Xe fractionation may have continued well into the Proterozoic in the face of O$_2$ \citep{Warr2018},
 makes this an interesting direction to take further research.
  
 Third, O$^+$, if present, will escape if Xe$^+$ can escape. 
  O$^+$ is a better carrier for Xe$^+$ escape than H$^+$ owing to its greater mass,
 it should be able to ionize Xe by charge exchange, and it is known to escape from Earth today \citep{Shelley1972},
 although it is not known how \citep{Shen2018}.
 Considerable O$^+$ escape would imply less oxidation of Earth, 
 but if the presence of O$^+$ were the signal of considerable amounts of O$_2$ \citep{Mendillo2018},
 there would be few Xe$^+$ ions present and escape of Xe would be negligible. 
 Because of this potential to undermine somewhat our conclusions on one hand, and
  the potential of helping Xe escape after the rise of oxygen on the other,
  we regard O$^+$ as calling for further research.
  
\subsection{Conclusions}

Our conclusion that the H$_2$ (or CH$_4$) atmospheric mixing ratio was at times at least 1\% or higher 
through the Archean is important and perhaps unexpected.
This conclusion is based on the minimum requirements for Xe to escape at any particular time and therefore appears to be 
relatively robust. 
In particular, it is not greatly affected by the uncertainties attending the transport of Xe through the molecular ionosphere. 
Either H$_2$ or CH$_4$ would be important to early Earth because either gas would help create favorable environments for 
the origin of life \citep{Urey1952,Tian2005b}. 
Both gases can provide greenhouse warming in the struggle against the faint young Sun \citep{Wordsworth2013a}.

 What is not clear is how often the H$_2$ (or CH$_4$) atmospheric mixing ratio was 1\% or higher.
 If much of the time, Earth could have lost more than an ocean's worth of hydrogen to space.
 This is comparable to the amount of oxygen stored in the mantle and crust as partner to Fe$^{+3}$ and carbon,
 but it pushes hard against realistic upper bounds on how rapidly iron can be oxidized and cycled back into the mantle \citep{Ozaki2018}.
The implication that Earth grew significantly more oxidized through the Archean may or may not be in conflict 
 with evidence that the oxidation state of magma sources has changed \citep{Aulbach2017} or not changed
 \citep{Nicklas2018}.
 However, the lost ocean is not a robust conclusion from our model, because the
 observed Xe fractionation can be generated by many different histories.
 In particular, histories that simultaneously account both for Xe's fractionation and its depletion require that Xe escape
occurred in short intense bursts rather than as a constant trickle.  Bursts of Xe escape can be caused by solar EUV variability,
or by varying amounts of atmospheric hydrogen, or perhaps by changes in Earth's geomagnetic field.
If hydrogen were less abundant ($<0.1\%$) at most times, a lost ocean in the Archean becomes an 
order of magnitude overestimate.  
 
\medskip

In this paper we have presented Xe escape as an ion in the context of recent discoveries of 
evolving Xe preserved in ancient rocks spanning the Archean \citep{Avice2018}.
Xenon escape in the Archean is a requirement that only escape as an ion can meet.  
Escape as an ion also explains how Xe escapes when Kr and Ar do not,   
 which neatly resolves the paradox that has been the bane of traditional models of Xe fractionation in hydrodynamic escape.
 The mass fractionation mechanism itself is the same as in traditional hydrodynamic escape models: it is the competition between collisions and gravity,
 with the key difference being that the collisions that drive Xe$^+$ outward are between ions and governed by the Coulomb force.

Although we have not dwelt on the Hadean, the ion-escape model works very well on the $\sim\!200$ Myr time scale
 of the early Hadean (or early Mars), when the Sun was a bigger source of EUV radiation. 
To make the ion-escape model work over the stretched out two billion year timescale discovered by \citet{Avice2018}
requires adding at least one other feature
(limiting escape to narrow polar windows, or limiting Xe escape to brief episodes), or dropping the
expectation that a single set of parameters should explain both Xe's fractionation and its depletion. 
 If in fact Earth's Xe fractionation took place in the deep Hadean, 
 such that the drawn out evolution of Xe isotopes documented by \citet{Avice2018} has other causes
 than the evolution of the atmosphere,
 the ion-escape model would still provide the best explanation of how Xe acquired its fractionation.



\appendix

\section{Hydrogen escape from a CO$_2$-rich atmosphere}
\label{appendix one}  % does not work as expected

The goal here is to develop a description of hydrogen escape in the presence of a static background of CO$_2$
suitable for investigating Xe escape.
% The general problem of irradiation-driven thermal escape from planetary atmosphere can get very complicated.
 Our equations are simplified from the self-consistent 5 moment approximation to multi-component hydrodynamic
flow presented by \citep{Schunk1980}. 
We merge this description with the description of two component diffusion given by \citet{Hunten1973}
to express collision terms in terms of binary diffusion coefficients
and to include parameterized Eddy diffusivity in the lower atmosphere.

%We make several other major simplifications. 
%These will be described in more detail as they come up.
% (i) We will assume either spherical symmetry
% or an idealized dipolar symmetry for the case of a strong dipolar geomagnetic field.
% (ii) We ignore diurnal cycles and latitudinal differences.
% (iii) We presume that H and H$_2$ flow outward at the same velocity $u$.  This greatly simplifies the mathematics with no major consequences.
% (iv) We ignore photodissociation of CO$_2$ into CO and O,
% which we justify by showing that CO$_2$ rapidly recombines when H$_2$ is present, and thus is likely to persist while H$_2$ persists.
% (v) We presume that CO$_2$ does not escape, and thus that hydrogen must diffuse through the CO$_2$.
% (vi) We neglect thermal conduction, which becomes a relatively small term in the energy budget 
% at the high levels of solar irradiation needed if Xe is to escape.
% (vii) We neglect the solar wind, collisional ionization by exogenous particles, 
% and energy flows between different regions of the magnetosphere.

\subsection{Chemistry}
\label{Chemistry}

We begin with the photochemistry of the H$_2$-CO$_2$ atmosphere.
We denote the three neutral species H, H$_2$, and CO$_2$ with the indices 1, 2, 3, respectively.
H$_2$ is photolyzed by EUV radiation.  It can be dissociated
\begin{equation} \tag{J2a}
\label{J2a}
\mathrm{H}_2 + h\nu\, (\lambda < 85\,\mathrm{nm}) \rightarrow \mathrm{H} + \mathrm{H}, 
\end{equation}
ionized
\begin{equation}
%  \tag{$J_{\mathrm{H}_2}b$}
\tag{J2b}\label{J2b} 
\mathrm{H}_2 + h\nu\, (\lambda < 80\,\mathrm{nm}) \rightarrow \mathrm{H}_2^+ + \mathrm{e}^- ,
\end{equation}
or dissociatively ionized
\begin{equation}
% \tag{$J_{\mathrm{H}_2}c$} 
\tag{J2c}\label{J2c} 
\mathrm{H}_2 + h\nu\, (\lambda < 69\,\mathrm{nm}) \rightarrow \mathrm{H} + \mathrm{H}^+ + \mathrm{e}^- 
\end{equation}
The relative yields of the three channels are, respectively, $0.56\!:\!0.35\!:\!0.09$ for the solar EUV spectrum \citep{Huebner1992}. 
All channels eventually lead to dissociation of H$_2$, so that the total photodissociation rate is
$J_2 =  J_{2a} + J_{2b} + J_{2c}$. 

The H$_2^+$ ion reacts quickly with any of the three major species.
The reaction with H$_2$ forms the H$_3^+$ ion, an important radiative coolant
 \begin{equation}
\tag{R1}\label{R1}
\mathrm{H}^+_2 + \mathrm{H}_2 \rightarrow \mathrm{H}_3^+ + \mathrm{H} \qquad \qquad k_1=2\times 10^{-9}\, \mathrm{cm}^3\mathrm{s}^{-1}.
\end{equation}
The sink on H$_3^+$ is dissociative recombination,
 \begin{equation}
\tag{R2}\label{R2}
\mathrm{H}^+_3 + \mathrm{e}^- \rightarrow \mathrm{H}_2 + \mathrm{H} \qquad \qquad k_2=1.1\times 10^{-5}T^{-0.7} \, \mathrm{cm}^3\mathrm{s}^{-1},
\end{equation}
which, like dissociative recombination generally, is very fast.
The reaction of H$_2^+$ with H, a charge exchange, is a source of H$^+$
 \begin{equation}
\tag{R3}\label{R3}
\mathrm{H}^+_2 + \mathrm{H} \rightarrow \mathrm{H}^+ + \mathrm{H}_2 \qquad \qquad k_3=8\times 10^{-10}\, \mathrm{cm}^3\mathrm{s}^{-1}.
\end{equation}
The reaction of H$_2^+$ with CO$_2$ is also fast,   
 \begin{equation}
\tag{R4}\label{R4}
\mathrm{H}^+_2 + \mathrm{CO}_2 \rightarrow \mathrm{HCO}_2^+ + \mathrm{H} \qquad \qquad k_4=2.4\times 10^{-9}\, \mathrm{cm}^3\mathrm{s}^{-1}.
\end{equation}
but the listed \citep{McElroy2012} product, HCO$_2^+$, 
appears likely to dissociatively recombine to form CO$_2$ and H, so that on net Reaction R4 does nothing
to CO$_2$ while dissociating H$_2$.  

Atomic H is photo-ionized by EUV radiation, 
\begin{equation}
% \tag{$J_{\mathrm{H}_2}c$} 
\tag{J1}\label{J1}
\mathrm{H} + h\nu\,(\lambda < 91.2\,\mathrm{nm}) \rightarrow \mathrm{H}^+ + \mathrm{e}^-
\end{equation}
Radiative recombination of H$^+$,
 \begin{equation}
\tag{R5}\label{R5}
\mathrm{H}^+ + \mathrm{e}^-  \rightarrow  \mathrm{H} + h\nu  \qquad \qquad k_5=1.6\times 10^{-10}T^{-0.7} \, \mathrm{cm}^3\mathrm{s}^{-1}
\end{equation}
is about 5 orders of magnitude slower than dissociative recombination of H$_3^+$,
hence H$^+$ is typically much more abundant than H$_3^+$.
% Nonetheless we will find that H$_3^+$ is often an important radiative coolant at those same high altitudes.
At low altitudes the sink of H$^+$ is a chemical reaction with CO$_2$
 \begin{equation}
\tag{R6}\label{R6}
\mathrm{H}^+ + \mathrm{CO}_2  \rightarrow  \mathrm{HCO}^+ + \mathrm{O} \qquad \qquad  k_6=3.5\times 10^{-9} \, \mathrm{cm}^3\mathrm{s}^{-1} .
\end{equation}
The formyl radical, HCO, is relatively easily ionized,
and thus HCO$^+$ is often abundant in astrophysical settings.
We find it abundant in CO$_2$-H$_2$ atmospheres.
The sink on HCO$^+$ is dissociative recombination,
\begin{equation}
\tag{R7}\label{R7}
\mathrm{HCO}^+ + e^-  \rightarrow \mathrm{CO} + \mathrm{H} \qquad \qquad k_7=1.1\times 10^{-5}T^{-0.7} \, \mathrm{cm}^3\mathrm{s}^{-1} .
\end{equation}
CO$_2$ can be photolyzed by both FUV and EUV.
The most important FUV channel is dissociation to the excited $\mathrm{O}(^1\mathrm{D})$ state
\begin{equation}
\tag{J3a} \label{J3a}
\mathrm{CO}_2 + h\nu\, (\lambda < 167\,\mathrm{nm}) \rightarrow \mathrm{CO} + \mathrm{O}(^1\mathrm{D}) .
\end{equation}
Although spin-forbidden, photo-dissociation to the ground state $\mathrm{O}(^3\mathrm{P})$ does occur at a low rate.
At 157 nm, the quantum yield of $\mathrm{O}(^3\mathrm{P})$ is of the order of $0.1$ \citep{Schmidt2013}.
At wavelengths greater than 167 nm, where CO$_2$'s absorption cross-section is small but nonzero,
we assume that the product is mostly $\mathrm{O}(^3\mathrm{P})$,
\begin{equation}
\tag{J3b}\label{J3b}
\mathrm{CO}_2 + h\nu\, (\lambda < 227\,\mathrm{nm}) \rightarrow \mathrm{CO} + \mathrm{O}(^3\mathrm{P}) .
% \qquad 0.2J_{\mathrm{CO}_2}.
\end{equation}
Higher energy photons produce a variety of additional outcomes \citep{Huebner1992}, the most important of which
is ionization,
\begin{equation}
\tag{J3c}\label{J3c}
\mathrm{CO}_2 + h\nu\, (\lambda < 89.9\,\mathrm{nm}) \rightarrow \mathrm{CO}_2^+ + e^- .
\end{equation}
The CO$_2^+$ ion can dissociatively recombine
\begin{equation}
\tag{R8}\label{R8}
\mathrm{CO}_2^+ + e^-  \rightarrow \mathrm{CO} + \mathrm{O} \qquad \qquad k_8=1.7\times 10^{-5}T^{-0.7} \, \mathrm{cm}^3\mathrm{s}^{-1}
\end{equation}
or it can react with H to form HCO$^+$
\begin{equation}
\tag{R9}\label{R9}
\mathrm{CO}_2^+ + \mathrm{H}  \rightarrow \mathrm{HCO}^+ + \mathrm{O} \qquad \qquad k_9=2.9\times 10^{-10} \, \mathrm{cm}^3\mathrm{s}^{-1} .
\end{equation}
% Dissociative recombination of HCO$^+$ completes the photodissociation of CO$_2$.
In this study we simplify the chemistry by omitting alternative paths that generate O$^+$ and O$_2^+$ from CO$_2^+$.
Oxygen ions in particular are interesting because they too will escape in any wind that can carry off Xe$^+$.
Because O$^+$ is more massive than H$^+$, it is more efficient at driving off Xe$^+$.
% can help Xe escape, but it is here where we chose to truncate the system.
%; it is well-known
% that the complexity of the H-C-O-N photochemical system has no known natural limits. 

The $\mathrm{O}(^1\mathrm{D})$ atoms produced by CO$_2$ photolysis react 
quickly with H$_2$ to form OH, which then reacts with CO to remake CO$_2$.
The reaction between $\mathrm{O}(^1\mathrm{D})$ and H$_2$
\begin{equation}
\tag{R10}\label{R10}
\mathrm{H}_2 + \mathrm{O}(^1\mathrm{D}) \rightarrow \mathrm{OH} + \mathrm{H} \qquad \qquad k_{10}=1.2\times 10^{-10} \, \mathrm{cm}^3\mathrm{s}^{-1}
\end{equation}
 is very fast. The $\mathrm{O}(^1\mathrm{D})$ can also be de-excited to
 the ground state by collisions with CO$_2$
\begin{equation}
\tag{R11}\label{R11}
\mathrm{CO}_2 + \mathrm{O}(^1\mathrm{D}) \rightarrow \mathrm{CO}_2 + \mathrm{O}(^3\mathrm{P}) \qquad \qquad k_{11}=1.1\times 10^{-10} \,\mathrm{cm}^3\mathrm{s}^{-1}
\end{equation} 
 or much less efficiently by collisions with atomic H \citep{Krems2006}
\begin{equation}
\tag{R12}\label{R12}
\mathrm{H} + \mathrm{O}(^1\mathrm{D}) \rightarrow \mathrm{H} + \mathrm{O}(^3\mathrm{P}) \qquad \qquad k_{12}=1\times 10^{-12} \,\mathrm{cm}^3\mathrm{s}^{-1} .
\end{equation} 
The reaction between H$_2$ and $\mathrm{O}(^3\mathrm{P})$ 
 \begin{equation}
\tag{R13}\label{R13}
\mathrm{H}_2 + \mathrm{O}(^3\mathrm{P}) \rightarrow \mathrm{OH} + \mathrm{H}   \qquad k_{13} = 3.5\times 10^{-13} \left(T/298\right)^{2.67} e^{-3160/T} \, \mathrm{cm}^3\mathrm{s}^{-1}
\end{equation}
has a significant temperature barrier and is slow at room temperature.
The reaction of CO and OH to reconstitute CO$_2$ is not particularly fast but lacks a significant activation barrier,
which makes it by far the most important sink of CO
 \begin{equation}
\tag{R14}\label{R14}
\mathrm{CO} + \mathrm{OH} \rightarrow \mathrm{CO}_2 + \mathrm{H} \qquad \qquad k_{14} = 1.5\times 10^{-13} \, \mathrm{cm}^3\mathrm{s}^{-1}.
\end{equation}
The reverse of R14 is endothermic and does not become fast until $T$ approaches 2000 K.
The net result of R10 (or R13) followed by R14 is reconstitution of CO$_2$ and dissociation of H$_2$. 
Hence CO$_2$ tends to persist while H$_2$ persists, although as noted CO$_2$ is dissociated by ion chemistry. 
Here we will overlook dissociation of CO$_2$ and we will assume that reaction R13 is negligible.
The net rate that CO$_2$ photolysis leads to dissociation of H$_2$ can then be written
\begin{equation}
\tag{R15}\label{R15}
\mathrm{CO}_2 + h\nu + \mathrm{H}_2 \rightarrow \mathrm{CO}_2 + 2\mathrm{H} 
\qquad\qquad J_3' \approx {J_{3a}k_{10}n_2\over k_{12}n_1 + k_{10}n_2 + k_{11}n_3} ,
\end{equation}
which takes into account the sources and sinks of $\mathrm{O}(^1\mathrm{D})$.

\medskip\noindent{\em Ion chemistry}

\begin{table}[htp]
\caption{Summary of H$_2$-CO$_2$ Chemistry}
\begin{center}
\begin{tabular}{llll}
label & reactants & products & rate [cm$^3$s$^{-1}$]\\
\hline
J1 & $\mathrm{H} + h\nu $ & $\rightarrow \mathrm{H}^+ + e^{-1} $ & $J_1$ \\
J2a & $\mathrm{H}_2 + h\nu $ & $\rightarrow \mathrm{H} + \mathrm{H} $ & $J_{2a}$   \\
J2b & $\mathrm{H}_2 + h\nu $ & $\rightarrow \mathrm{H}^+_2 + e^{-1} $ & $J_{2b}$   \\
J2c & $\mathrm{H}_2 + h\nu $ & $\rightarrow \mathrm{H} + \mathrm{H}^+ + e^{-1} $ & $J_{2c}$   \\
J3a & $\mathrm{CO}_2 + h\nu $ & $\rightarrow \mathrm{CO} + \mathrm{O}(^1\mathrm{D}) $ & $J_{3a}$   \\
 J3b & $\mathrm{CO}_2 + h\nu $ & $\rightarrow \mathrm{CO} + \mathrm{O}(^3\mathrm{P}) $ & $J_{3b}$   \\
J3c &  $\mathrm{CO}_2 + h\nu $ & $\rightarrow \mathrm{CO}^+_2 + e^{-1} $ & $J_{3c}$    \\
R1 & $\mathrm{H}^+_2 + \mathrm{H}_2 $ & $\rightarrow \mathrm{H}_3^+ + \mathrm{H} $ & $k_1=2\times 10^{-9}$     \\
R2 &  $\mathrm{H}^+_3 + e^{-1} $ & $\rightarrow \mathrm{H}_2 + \mathrm{H} $ & $k_2=1.1\times 10^{-5}T^{-0.7}$       \\
R3 &  $\mathrm{H}^+_2 + \mathrm{H} $ & $\rightarrow \mathrm{H}_2 + \mathrm{H}^+ $ & $k_3=8\times 10^{-10}$      \\
R4 &  $\mathrm{H}^+_2 + \mathrm{CO}_2 $ & $\rightarrow \mathrm{HCO}^+_2 + \mathrm{H} $ & $k_4=2.4\times 10^{-9}$      \\
R5 &  $\mathrm{H}^+ + e^{-1} $ & $\rightarrow \mathrm{H} + h\nu $ & $k_5=1.6\times 10^{-10}T^{-0.7}$      \\
R6 &  $\mathrm{H}^+ + \mathrm{CO}_2 $ & $\rightarrow \mathrm{HCO}^+ + \mathrm{O} $ & $k_6=3.5\times 10^{-9}$      \\
R7 &  $\mathrm{HCO}^+ + e^{-1} $ & $\rightarrow \mathrm{CO} + \mathrm{H} $ & $k_7=1.1\times 10^{-5}T^{-0.7}$       \\
R8 &  $\mathrm{CO}^+_2 + e^{-1} $ & $\rightarrow \mathrm{CO} + \mathrm{O} $ & $k_8=1.7\times 10^{-5}T^{-0.7}$       \\
R9 &  $\mathrm{CO}^+_2 + \mathrm{H}$ & $\rightarrow \mathrm{HCO}^+ + \mathrm{O} $ & $k_9=2.9\times 10^{-10}$      \\
R10 & $\mathrm{H}_2 + \mathrm{O}(^1\mathrm{D}) $ & $\rightarrow \mathrm{OH} + \mathrm{H}  $ & $k_{10}=1.2\times 10^{-10}$      \\
R11 &  $\mathrm{CO}_2 + \mathrm{O}(^1\mathrm{D}) $ & $\rightarrow \mathrm{CO}_2 + \mathrm{O}(^3\mathrm{P}) $ & $k_{11}=1.1\times 10^{-10}$      \\
R12 &  $\mathrm{H} + \mathrm{O}(^1\mathrm{D}) $ & $\rightarrow \mathrm{H} +  \mathrm{O}(^3\mathrm{P}) $ & $k_{12}=1\times 10^{-12}$      \\
R13 &  $\mathrm{H}_2 + \mathrm{O}(^3\mathrm{P}) $ & $\rightarrow \mathrm{OH} + \mathrm{H}  $ & $k_{13}= 3.5\times 10^{-13}\left(T/ 298\right)^{2.67}e^{-3160/T}$  \\
R14 & $\mathrm{CO}+ \mathrm{OH} $ & $\rightarrow \mathrm{CO}_2 + \mathrm{H}  $ & $k_{14}= 1.5\times 10^{-13}$      \\
\end{tabular}
\end{center}
\label{Chemistry Table}
\end{table}%

We solve for the five ions H$^+$, H$_2^+$, H$^+_3$, CO$_2^+$, and HCO$^+$,
which we denote as $x_i$, with $i$ going from 1 to 5, respectively.
% These are the ions that form consequent to photo-ionization of H, H$_2$, and CO$_2$.
Consistent with our neglecting O, CO, and O$_2$, we neglect O$^+$, O$_2^+$, and OH$^+$.
% which form in consequence to photo-ionization of O. 
We assume local photochemical equilibrium, which is a good approximation for the molecular ions,
less good for H$^+$ at high altitudes where it is effectively the only ion.
The equations are
$\partial x_i / \partial t = 0$ for all $i$; reaction rates are summarized in Table \ref{Chemistry Table}.
\begin{eqnarray}
{\partial x_1 / \partial t} &=& J_1n_1 + J_{2c}n_2 + k_3x_2n_1 - k_5x_1n_e - k_6x_1n_3 \\
{\partial x_2 / \partial t} &=& J_{2b}n_2 - k_1x_2n_2 - k_3x_2n_1 - k_4x_2n_3 \\
{\partial x_3 /\partial t} &=& k_1x_2n_2 - k_2x_3n_e \\
{\partial x_4 / \partial t} &=& J_{3c}n_3 - k_8x_4n_e - k_9x_4n_1 \\
{\partial x_5 / \partial t} &=& k_6x_1n_3 - k_7x_5n_e + k_9x_4n_1 \\
\sum x_i &=& n_e 
\end{eqnarray}
Dissociative recombination of H$_2^+$ is relatively slow; we omit it.
Despite being very short-lived, the molecular ions (chiefly HCO$^+$)
 are dominant at low altitudes (as illustrated in Figure \ref{Densities} in the text).
The important aspect of molecular ions in the present study is that they interact strongly
with Xe$^+$ by the Coulomb force, and hence Xe$^+$ escape depends in part on
transport by the molecular ions.

\subsection{Radiative processes}
\label{Radiation}
 
We divide the EUV and FUV spectrum $F_{\lambda}$ into sixteen wavelength bins that capture the essential colors for
a CO$_2$-H$_2$ atmosphere.  In particular, we bookkeep a channel between 80 nm and 91.2 nm for ionization of atomic hydrogen at wavelengths that do not ionize CO$_2$ or H$_2$,
and we account for CO$_2$'s anomalously low cross-section in the channel encompassing Lyman $\alpha$.  
 
 \subsubsection{Photolysis and optical depths}

The total optical depth as a function of wavelength $\tau_{\lambda}$ is formally given by integration over the 
sum of the opacities of the different species   
\begin{equation}
\label{optical depth}
\partial \tau_{\lambda}/\partial r = \sum_i n_i\sigma_{i,\lambda}
\end{equation}
where $\sigma_{i\lambda}$ is the cross section [cm$^2$] of species $i$ at wavelength $\lambda$.
Photolysis rates and photoionization rates $J_i$ are obtained as integrals of the form
\begin{equation}
\label{J}
      J_i = S \int{F_{\lambda} \sigma_{i \lambda} \exp{\left(-\tau_{\lambda}\right)}} d\lambda .
\end{equation}
where $F_{\lambda}$ refers to the modern mean solar XUV spectrum and 
$S$ refers to the enhanced ancient solar XUV flux relative to the modern Sun (Eq \ref{S} of the main text).
The units of $J_i$ are s$^{-1}$.
% where 
%where $F_{\lambda}$ represents the spectrum of the quiet Sun
%and the parameter $S$ is treated as a global amplification independent of wavelength.

 \subsubsection{Radiative heating}

The local radiative volume heating rate [ergs cm$^{-3}$s$^{-1}$] is given by
\begin{equation}
\label{heating}
            \Gamma_{\mathrm{h}} = S \eta_h \sum_i { n_i \sigma_{i\lambda}
            {F_{\lambda} hc\over \lambda}\exp{\left(-\tau_{\lambda}\right)} } %\quad \mathrm{ergs~}\mathrm{cm}^{-3}\mathrm{s}^{-1}
 \end{equation}
 where $h$ is Planck's constant and $c$ is the speed of light, and $\eta_h\approx 0.5$ is a heating efficiency \citep{Koskinen2014}.

The total radiative heating, the integral of $\Gamma_{\mathrm{h}}$, 
can be used to define an energy-limited flux.
The energy-limited flux is intended to be an upper bound that compares the energy incident in XUV radiation to the energy required to lift the atmosphere into space \citep{Watson1981}. 
Details are lumped together in a different efficiency factor $\eta \neq \eta_h$ that is often taken to lie between 0.1 and 0.6
\citep{Lammer2013,Bolmont2017}.
The energy-limited flux can be expressed either in terms of $F_{\mathrm{xuv}}$,
% that can be absorbed by hydrogen ($\lambda < 91.2$ nm), which is more usual,
or in terms of the total solar radiation absorbed above some arbitrary altitude, a definition that implicitly takes FUV into
account.  With FUV taken into account, $\eta$ can exceed unity.

We consider two forms of the energy-limited fluxes of H$_2$ molecules [cm$^{-2}$s$^{-1}$], 
\begin{equation}
\label{energy-limited-1}
   \phi_{\rm EL} = {S \eta hc  R_{\oplus}\over 4 GM_{\oplus} m_2 } \int^{91.2\,\mathrm{nm}}_0 \!\!  \int_{0}^{\infty} \lambda^{-1} F_{\lambda}\exp{\left(-\tau_{\lambda}\right)}\,  d\lambda \, d\tau_{\lambda}  ,
\end{equation}
where all the photons with $\lambda < 91.2$ nm are absorbed, and
\begin{equation}
\label{energy-limited-2}
   \phi'_{\rm EL} = {S \eta hc R_{\oplus}\over 4 GM_{\oplus} m_2 } \int^{200\,\mathrm{nm}}_0 \!\! \int_{0}^{\tau_{\lambda}(r_0)} \lambda^{-1} F_{\lambda}  \exp{\left(-\tau_{\lambda}\right) }\, d\lambda \, d\tau_{\lambda}  ,
\end{equation}
where $\tau_{\lambda}(r_0)$ represents the optical depths at the lower boundary.
In Equations \ref{energy-limited-1} and \ref{energy-limited-2}, $M_{\oplus}$ and $R_{\oplus}$ refer
to the mass and radius of Earth; $m_2$ is the mass of an H$_2$ molecule;
$hc/\lambda$ is the energy of a photon [ergs]; and $\tau_{\lambda}$
is the optical depth as a function of $\lambda$.
 Equations \ref{energy-limited-1} and \ref{energy-limited-2} are evaluated
 with $\eta=0.5$ to compare to the fluxes computed with the full model in Figures \ref{Figure9} and \ref{Figure9a} of the main text.  

 \subsubsection{CO$_2$ cooling}

The major coolant of the thermospheres of Earth, Venus and Mars is 
collisional excitation of the $\nu_2$ vibrational band of CO$_2$,
probably by atomic oxygen, followed by emission of a 15 micron photon. 
% Cooling by this mechanism has been studied for at least 50 years at a high level of expertise 
%\citep{Chamberlain1966,Dickinson1972,Kumer1974,Gordiets1982,Dickinson1986,Bougher1994,Lopez-Puertas1996,Shama2015}.
%  The theory loosely resembles the analogous problem of atomic line cooling familiar to astrophysical HII regions, but is more difficult because (i) a molecular band is more complicated than an atomic line; (ii) excitation is by collisions with any of many other molecules rather than by electrons; and (ii) the lines are optically thick and therefore radiative transfer needs to be taken into account explicitly. 
%
% Only the older publications present the theory in forms that are simple enough to use in general applications.
We use the relatively simple algorithm described by \citet{Gordiets1982}, based on work by 
\citet{James1973} and \citet{Kumer1974}, which prioritizes generality over fidelity.
 We have had to make some notational changes to avoid potential
confusion with other symbols used in this study; the correspondences to the original notation 
are given in the Table of Symbols (Appendix C).  The volume cooling rate [ergs cm$^{-3}$ s$^{-1}$] is
\begin{equation}
\label{CO2_cooling}
\Gamma_{\mathrm{CO}_2} = 2h\nu_2 n_{\mathrm{CO}_2} \times  \sum_j{k^{\ast}_jn_j}  \times \Theta\left(\Lambda,\tau\right) \exp{\left(-960/T\right)} % \quad\mathrm{ergs}\,\mathrm{cm}^{-3}\mathrm{s}^{-1}
\end{equation}
 where $h\nu_2 = 1.33\times 10^{-13}$ ergs. %multiplies the energy of the photon by a statistical weight.
The collisional excitation and de-excitation rate is summed over all the constituents present, $\sum_j{k^{\ast}_jn_j}$.
Atomic oxygen is usually regarded as dominant for Venus, Earth, and Mars, with a collisional excitation 
rate $k^{\ast}_{\rm O}$ on the order of $2\times 10^{-12}$ cm$^3$s$^{-1}$,
but factor two mismatches between what the models require and what is observed in the laboratory remain unresolved \citep{Sharma2015}.
For a hydrogen-rich atmosphere, excitation will be dominated by H$_2$ and H.
The rate for $k^{\ast}_{\mathrm{H}_2}$ is known to be rather high, $5\times 10^{-12}$ cm$^3$s$^{-1}$ at 300 K and $7.5\times 10^{-12}$ cm$^3$s$^{-1}$ at 200 K \citep{Sharma2015}; we use the lower of these.
We presume $k^{\ast}_{\rm H}=1\times 10^{-11}$ cm$^3$s$^{-1}$, a rate consistent with the thermodynamic reverse of R14
${\rm CO}+{\rm OH}$.
Excitation by CO$_2$ itself is negligible by comparison; we use $k^{\ast}_{\mathrm{CO}_2}=2\times 10^{-14}$ cm$^3$s$^{-1}$.

The parameter $\tau_{x}$ is a line center optical depth.
\begin{equation}\label{Kumer x}
% x =  \sigma_{\nu_2} n_{{\rm CO}_2} H_{{\rm CO}_2}, 
 \tau_{x} =  \sigma_{\nu_2} n_3 H_3, 
\end{equation}
where $\sigma_{\nu_2} = 6.4\times 10^{-15}$ cm$^{2}$ is an effective absorption cross section at the center of the $\nu_2$ band.
The parameter $\Lambda$ describes the competition between radiative and collisional de-excitation,
\begin{equation}\label{Kumer lambda}
 \Lambda = {A_{\nu_2}\over A_{\nu_2} + \sum_j{k^{\ast}_jn_j}} .
\end{equation}
The spontaneous emission rate is $A_{\nu_2} = 1.35$ s$^{-1}$. 
The function $\Theta\left(\Lambda,\tau_{x}\right) $ is the normalized chance that a photon escapes 
%\begin{equation}\label{Kumer F}
% \Theta\left(\Lambda,x\right) = {0.5\,\Lambda\left(E\left(x\right) + E\left(0.5x\right)\right) \over 1-\Lambda+0.5\,\Lambda \left(E\left(x\right) + E\left(0.5x\right)\right)} 
%\end{equation} 
\begin{equation}\label{Kumer F}
 \Theta\left(\Lambda,\tau_{x}\right) = {\left(E\left(\tau_{x}\right) + E\left(0.5\tau_{x}\right)\right)\Lambda \over 2-2\Lambda+\left(E\left(\tau_{x}\right) + E\left(0.5\tau_{x}\right)\right)\Lambda } 
\end{equation} 
%\[ L\left(x\right) = E\left(x\right) + E\left(0.5x\right) 
where we have used a curve fit to the tabulated values of an exponential function
$E(\tau_x)$ (Doppler broadening) in \citet{Kumer1974} that \citet{Gordiets1982} refer to,
\begin{equation}\label{Kumer E}
 E\left(\tau_{x}\right) \approx {8.54\over 17.15 + \tau_{x}^{1.08} } + {0.272\over 141.7 + \tau_{x}^{0.53} } 
\end{equation} 

The 4.3 micron $\nu_3$ band is stronger than the 15 micron $\nu_2$ band,
but usually less important for cooling because the temperature must be higher to excite it.
We use the same formalism as Eq \ref{CO2_cooling} with
parameters appropriate to the $\nu_3$ band: $h\nu_3 = 4.5\times 10^{-13}$ ergs, 
$\sigma_{\nu_3} = 2.47\times 10^{-14}$ cm$^2$, and $A_{\nu_3}=400$ s$^{-1}$. 
The excitation temperature is 3350 K.    
Because the $A$ value is higher, the 4.3 $\mu$m band is less easily collisionally quenched,
and hence it can effectively cool from deeper levels in the atmosphere than the 15 $\mu$m band. 

 \subsubsection{H$^+_3$ cooling}

The molecular ion H$_3^+$ is known to be an effective coolant.
It is important in the energy budgets of the solar system's giant planets, and 
\citet{Yelle2004} showed it to be a major part of the budget of highly irradiated exo-Jupiters.
It is reasonable to expect it be important for mildly irradiated H$_2$-rich ionospheres
we are considering here.
We use a curve fit to cooling rates computed by \citet{Neale1996} to obtain the volume cooling rate
\begin{equation}
\label{Neale}
\Gamma_{\mathrm{H}_3^+} = 4\pi x_3 \exp{\left( -a_1 + a_2\ln{\left(T\right)} - a_3\left(\ln{\left(T\right)}\right)^2 \right)}
\end{equation}
 where $a_1 = 118.85$, $a_2=21.488$, and $a_3=1.2308$.
 The fit is for $500 < T < 3500$, which covers the range of interest for the present application.
% Although not immediately obvious, Eq \ref{Neale} fits the gray relationship  
% $\Gamma_{\mathrm{H}_3^+} = 1.3\times 10^{-25} T^4$ ergs/,s$^{-1}$sr$^{-1}$molecule$^{-1}$ to better than a factor of two.
 
 \subsubsection{Cooling by H and H$^+$}

Cooling by H and H$^+$ can be important in the high altitude atomic ionosphere.  
The volume cooling rate for free-free cooling is
% \[ \Gamma_{f\!f} \approx 1.7\times 10^{-27}[\hbox{H}^+]n_eT^{1/2} \]
\begin{equation}
\label{free-free}
 \Gamma_{f\!f} \approx 1.7\times 10^{-27} {x_1} n_e T^{1/2} \quad \mathrm{ergs}\,\mathrm{cm}^{-3}\mathrm{s}^{-1}
\end{equation}
and for bound-free cooling is
% \[ \Gamma_{bf} \approx 2.5\times 10^{-25}[\hbox{H}^+]n_e \]
\begin{equation}
\label{bound-free}
 \Gamma_{bf} \approx 2.5\times 10^{-25} {x_1} n_e  \quad \mathrm{ergs}\,\mathrm{cm}^{-3}\mathrm{s}^{-1} .
 \end{equation}
%where $n_e\approx x_1$ represents the electron density,
These turn out never to be important.
Collisional excitation of Ly$\alpha$ is very important 
at temperatures on the order of $\sim 1\times 10^4$ K \citep{Murray-Clay2009},
but is negligible for $T<8000$ K.
We will not encounter conditions where Lyman $\alpha$ cooling is important in this study.

 \subsubsection{Radiative effects of neglected species}

We have omitted N$_2$, NO, CO, O, OH, H$_2$O, and O$_2$.
Although N$_2$ itself does not radiate unless very hot, the lowest vibrationally excited states of N$_2$ are
resonant with the $\nu_3$ CO$_2$ asymmetric stretch at 4.3 $\mu$m,
and thus collisional excitation of N$_2$ leads directly to 4.3 $\mu$m CO$_2$ radiation (the mechanism behind CO$_2$ lasers).
This can be important where N$_2$ is more abundant than CO$_2$ and the gas is warm. 
Atomic oxygen and nitric oxide are important radiative coolants on Earth today \citep{Kulikov2007},
and would likely be important on early Earth as well.
 The 4.8 $\mu$m CO band %, like the stronger CO$_2$ 4.3 $\mu$m band, 
can be an effective coolant if the atmosphere is hot.
Rotational cooling by CO is important at low temperatures and low densities in astrophysical settings.
Water will be present at $\sim\!$100 ppm while H$_2$ persists, produced by 
 $\mathrm{H}_2 + \mathrm{OH} \rightarrow \mathrm{H}_2\mathrm{O}  + \mathrm{H} $
and destroyed by $\mathrm{H}_2\mathrm{O} + h\nu  \rightarrow \mathrm{OH} + \mathrm{H} $.
Water emission at 6.3 $\mu$m can be an effective coolant,
but water is also a much stronger FUV absorber than CO$_2$; exploratory numerical experiments suggest
that heating is probably more important than cooling. 
Molecular oxygen is a poor radiative coolant but a strong FUV absorber; its role is unambiguously to heat.

\subsection{Vertical structure equations}
\label{Equations}
  
We make several sweeping simplifications to the basic equations given by \citet{Schunk1980}.
As we are considering a relatively dense gas, we use a single temperature $T$ for all species.
We also assume steady state and spherical symmetry. 
 Densities, masses, and velocities of neutral species $i$ are denoted $n_i$, $m_i$, and $v_i$, respectively. 
% We will find that the escaping hydrogen is mostly neutral, so that the actual flow would neither be dipolar nor spherical,
% but something in between.     

\subsubsection{Continuity}
 Conservation of species $i$ can be written
\begin{equation}
\label{SN1}
{1\over r^{2}} {\partial \left( r^{2} n_i v_i \right) \over \partial r} = P_i - L_i
\end{equation}
where $P_i$ and $L_i$ are photochemical production and loss terms.
For hydrogen, we include only two terms: direct photolysis of H$_2$, and the splitting of
H$_2$ that takes place when O($^1$D) from CO$_2$ photolysis reacts with H$_2$. 
%We will assume that O($^1$D) reacts when it collides with H$_2$ but is de-excited to the much less reactive
%O($^3$P) when it collides with anything else.
  \begin{equation}
 \label{A5}
 {1\over r^{2}} {\partial \left(r^{2}v_1n_1\right) \over \partial r} = 2J_2 n_2 +2J'_{3}n_3
  \end{equation}
 \begin{equation}
 \label{A6}
 {1\over r^{2}} {\partial \left(r^{2}v_2n_2\right) \over \partial r} = -J_2 n_2 -J'_{3}n_3
 \end{equation}
All the equations are greatly simplified if we assume that both H and H$_2$ flow upwards with the same velocity $v_1=v_2=u$.
 With this useful simplification, total hydrogen conservation reduces to
 \begin{equation}
 \label{A7}
 {1\over r^{2}} {\partial \left(r^{2}u(0.5n_1+n_2)\right) \over \partial r} = 0 .
 \end{equation}
The integral of Equation \ref{A7} is the constant % $r^{2}_0\phi_{\mathrm{H}_2}$,
 \begin{equation}
 \label{A4}
 r_0^{2} \phi_{\mathrm{H}_2} = \left( 0.5n_1 + n_2 \right)u\,r^{2} ,
 \end{equation}
where we have defined $\phi_{\mathrm{H}_2}$ as the total hydrogen escape flux 
as measured at the lower boundary.

 When hydrogen is the only element escaping, it is useful to define a total hydrogen number
  density $n'=n_1+n_2$ and a mean molecular weight for hydrogen,
 \begin{equation}
 \label{A2}
\mu^{\prime} = {n_1m_1 + n_2m_2 \over n' }.
 \end{equation}
The prime denotes that $\mu^{\prime}$ refers only to the hydrogen and not to the whole atmosphere.
Equations \ref{A5} and \ref{A6} are summed for the change in $n'$
 \begin{equation}
 \label{A3}
 {1\over r^{2}} {\partial \left(r^{2}un'\right) \over \partial r} = J_2 n_2 +J'_{3}n_3 .
  \end{equation}
The mean molecular weight $\mu^{\prime}$ of the hydrogen varies in response to photolysis,
 \begin{equation}
 \label{B7}
  {1\over \mu^{\prime}}{\partial \mu^{\prime} \over \partial r} = - {J_2n_2 + J_3'n_3 \over \left( n_1 + n_2\right)u}.
\end{equation}
We will use Eq \ref{B7} to develop a planetary wind equation.

 \subsubsection{Conservation of momentum}
 
Starting from \citet{Schunk1980}, and after many simplifying assumptions,
 momentum conservation in spherical symmetry for a constituent $i$ in the presence of other constituents $j$ can be written
 \begin{equation}
 \label{B0}
n_im_iv_i {\partial v_i \over \partial r} + {\partial \left(n_ik_BT\right) \over \partial r} = -{GM_{\oplus} n_im_i\over r^2}  - \sum_j{\left( v_i-v_j\right) n_in_j{k_BT\over b_{ij} }} .
\end{equation}
 The collision terms are appropriate to the 5-moment approximation in the limit that all the species have the same temperature; 
 according to \citet{Schunk1980} these collision terms are valid for the Maxwell interaction
 potential (inverse fourth power) for arbitrary relative velocities, which provides some justification
 for using them in cases where $v_i-v_j$ might approach the sound speed.
 The mapping between the momentum transfer collision frequencies $\nu_{ij}$ used by \citet{Schunk1980}
 and the binary diffusion coefficients used by \citet{Hunten1973} is
 \begin{equation}
 \label{collision-rates}
  \nu_{ij} = {k_BT \over m_i}{n_j \over b_{ij} } .
 \end{equation}
 Unlike $\nu_{ij}$, the binary diffusion coefficient is symmetrical, $b_{ij}=b_{ji}$.
 
Eddy mixing in the lower atmosphere --- absent from Eq \ref{B0} --- is introduced following \citet{Hunten1973},
who implicitly defines the eddy diffusion coefficient $K_{zz}$ in terms of 
a total number density $n = \sum_i n_i$
and an appropriate average value ${\bar b}$ of $b_{ij}$.
\begin{equation}
\label{SN2}
{m_i v_i\over k_BT}{\partial v_i \over \partial r}  + {1\over n_i}{\partial n_i \over \partial r} + {1\over T}{\partial T \over \partial r} 
= - {GM_{\oplus}m_i\over r^2k_BT} - \sum_j {\left(v_i-v_j\right)n_j \over b_{ij} }
- {K_{zz}n \over {\bar b}}\left( {1\over n_i}{\partial n_i \over \partial r} - {1\over n}{\partial n \over \partial r} \right) .
\end{equation}
Equation \ref{SN2} converges on the correct limiting
expression for the bulk atmosphere's hydrostatic scale height when summed over the many constituents 
in the limit that the $u^2$ terms can be ignored,
 \begin{equation}
 \label{E4}
{1\over n} {\partial n \over \partial r} = -{1\over T} {\partial T \over \partial r}  
 -{GM_{\oplus}\mu \over r^2 k_BT} .
\end{equation}
% In Eq \ref{E4} the total atmospheric density $N = \sum_i n_i$, and the mean molecular mass $\mu = \sum_i m_in_i/N$.
The most important binary diffusion coefficient in the CO$_2$-H$_2$ system is (not surprisingly)
 $b_{23}$ between CO$_2$ and H$_2$:
 \begin{equation}
\label{b_23}
   b_{23} = 31.4 {T^{0.75} \exp{\left(-11.7/T\right)} \over k_B} \approx 1.2\times 10^{19}\left(T/200\mathrm{~K}\right)^{0.75} \quad \mathrm{cm}^{-1}\mathrm{s}^{-1}.
 \end{equation}
We will take ${\bar b} = b_{23}$. 
A homopause density $n_h$ can be defined by $n_h \equiv K_{zz}/{\bar b}$.
% In general the diffusivities of the different species are different and each species will have its own homopause.
% It should also be recalled that $K_{zz}$ is a modeling parameter cloaked in great uncertainties. 

% Using Eq \ref{E4}, Eq \ref{SN2} becomes
% \[
% {m_i v_i\over k_BT}{\partial v_i \over \partial r}  + 
% \left(1+{K_{zz}N\over {\bar b}}\right)
% \left( {1\over n_i}{\partial n_i \over \partial r} + {1\over T}{\partial T \over \partial r} \right)
% =  \phantom{- {GM_{\oplus}\over r^2k_BT}\left(m_i + {K_{zz}N\mu\over {\bar b}}\right)}
% \]
% \begin{equation}
% \label{SN2a}
% \phantom{%\left(1+{K_{zz}N\over {\bar b}}\right)
% \left( {1\over n_i}{\partial n_i \over \partial r} + {1\over T}{\partial T \over \partial r} \right)}
% - {GM\over r^2k_BT}\left(m_i + {K_{zz}N\mu\over {\bar b}}\right) - \sum_j {\left(v_i-v_j\right)n_j \over b_{ij} }
% \end{equation}

\medskip \noindent{\em The forces on hydrogen}

Here we wish to construct an expression that describes hydrogen escape.
First, we expand Equation \ref{SN2} into two equations, one for H ($i=1$) and the other for H$_2$ ($i=2$).
We next impose the approximation that H and H$_2$ flow as a single fluid with velocity $u$. 
With $v_1=v_2=u$ as an additional constraint,
neither Eq \ref{SN2} for H ($i=1$) nor Eq \ref{SN2} for H$_2$ ($i=2$) hold individually.
The mass-weighted sum of the two Eqs \ref{SN2} for H and H$_2$
 remains valid because the $b_{12}$ terms that couple H and H$_2$ cancel out. % for any $v_1-v_2$.
This sum is written
\[
{n_1m_1 + n_2 m_2 \over k_BT}{u\,\partial u \over \partial r} + {\partial n' \over \partial r}
+ {n'\over T} {\partial T \over \partial r} = 
-{GM\over r^2} { n_1m_1 + n_2 m_2 \over k_BT} 
\]
\begin{equation}\label{combined hydrogen femtoplus}
- {K_{zz}n\over {\bar b}} {\partial n' \over \partial r} +  {K_{zz}n\over {\bar b}}
{n'\over n}{\partial n\over \partial r}
 - u n_3\left( {n_1\over b_{13}} + {n_2 \over b_{23}} \right) .
\end{equation}
Subsequent derivations are clearer in terms of $\mu'$ and the hydrogen density $\rho'\equiv n' \mu'$ 
\[
{\rho' \over k_BT}{u\,\partial u \over \partial r} + {\partial \left(\rho'/\mu'\right) \over \partial r}
+ {\rho'\over \mu'}{1\over T} {\partial T \over \partial r} = 
-{GM\over r^2} { \rho' \over k_BT} 
- {K_{zz}n\over {\bar b}} {\partial \left(\rho'/\mu'\right) \over \partial r} 
\]
\begin{equation}\label{combined hydrogen femtoplus two}
+ {\rho'\over\mu'} {K_{zz}n\over {\bar b}}
{1\over n}{\partial n\over \partial r}
 - un_3\left( {n_1 \over b_{13}} + {n_2\over b_{23}} \right) .
\end{equation}
Using Eq \ref{E4} and gathering terms, 
Eq \ref{combined hydrogen femtoplus two} can be rewritten
\[
{\mu' \over k_BT} {u\partial u \over \partial r} 
+  \left(1 + {K_{zz}n\over {\bar b}}\right)\left( {1\over \rho'} {\partial \rho' \over \partial r} 
+  {1\over T} {\partial T\over \partial r}
 - {1\over \mu'}{\partial \mu' \over \partial r} \right)
  = -{GM \over r^2k_BT}\left(\mu' + {K_{zz}n\mu\over {\bar b}}\right)
\]
\begin{equation}
 -{un_3\over n_1+n_2}\left( {n_1\over b_{13} } +{n_2\over b_{23} }\right)  .
\end{equation}
Substituting for $\rho'$ from continuity Eq \ref{A3}
\begin{equation} \label{rho continuity}
 {1\over \rho'}{\partial \rho' \over \partial r} = {1\over u}{\partial u \over \partial r} - {2\over r} ,
\end{equation}
and using Eq \ref{B7} to replace $\partial \mu'/\partial r$, results in  
an equation for the velocity gradient $\partial u/\partial r$ that resembles the usual solar wind equation,
 \begin{equation}
 \label{E5}
 \left( u^2 - c'^2 \right) {1\over u}{\partial u \over \partial r}
 = -{GM_{\oplus} \over r^2}\left( 1 + {K_{zz} n \over b_{23}}{\mu \over \mu'} \right)  + {2 c'^2 \over r}
     -  {c'^2\over T} {\partial T \over \partial r} 
    - {n_3u\, k_BT \over m_2 b_{23}}
    - {c'^2\over u} {J_2n_2 + J'_{3}n_3 \over n_1 + n_2} 
\end{equation}
where the quantity
 \begin{equation}
 \label{E6}
  c'^2 \equiv {k_BT\over \mu'} \left( 1 + {K_{zz} n \over b_{23}} \right)
\end{equation}
takes on the role of the sound speed (squared).
There are some advantages to solving Equation \ref{E5} for $u(r)$ rather than solving for 
$n_1(r)$ and $n_2(r)$ directly.
First, there is only one variable. Second, the singularity (the critical point) in the
system at $u=c'$ is explicit.  Often one solves an equation like Eq \ref{E5} by
solving for the critical point's location, then evaluating the gradient $\partial u /\partial r$ using L'H{\^o}pital's rule,
then numerically integrating downward to the lower boundary. 
The critical point conditions are iterated until
the desired lower boundary conditions are met.  This works well for simpler systems but becomes cumbersome
for an equation like Eq \ref{E5} for which the L'H{\^o}pital's rule expression for $\partial u /\partial r$ at the critical
point is complicated.
Here we will integrate Eq \ref{E5} for $u(r)$ upward from the lower boundary
and iterate until the desired outer boundary conditions are closely approached.
% The process is discussed more fully in section \label{six} below.

Atomic and molecular hydrogen densities are obtained from Eq \ref{E5} and continuity
\begin{equation}
 \label{B9}
  {1\over n_2}{\partial n_2 \over \partial r} = - {1\over u}{\partial u \over \partial r}
      - {2\over r} 
    - {J_2n_2 + J'_3n_3 \over n_2 u} .
\end{equation}
and 
\begin{equation}
 \label{B10}
  {1\over n_1}{\partial n_1 \over \partial r}= - {1\over u}{\partial u \over \partial r}
      - {2\over r} 
    +  {2J_2n_2 +2 J'_3n_3 \over n_2 u} .
\end{equation}
% These and Eq \ref{E5} are integrated upwards from the lower boundary.

\medskip\noindent{\em The forces on carbon dioxide}

In this study, carbon dioxide is hydrostatic ($v_3=0$) and does not dissociate. 
Using Eq \ref{E4} to describe mixing with the lower atmosphere, Eq \ref{SN2} with $j=3$ reduces to
\begin{equation}\label{CCF two}
 \left(1 + {K_{zz}n\over {\bar b}}\right) \left({1\over n_3}{\partial n_3 \over \partial r}
+ {1\over T} {\partial T \over \partial r}\right) =
 -{GM_{\oplus}\over r^2k_BT}\left(m_3 + {K_{zz}n\mu\over {\bar b}}\right)
 + u\left( {n_1 \over b_{13}} +{n_2 \over b_{23}} \right) .
 \end{equation}
We then simplify Eq \ref{CCF two} by approximating $b_{13}=2b_{23}$ and expressing $u$ in terms of the total hydrogen flux
\begin{equation}\label{CCF three}
\left(1 + {K_{zz}n\over b_{23}}\right)\left( {1\over n_3}{\partial n_3 \over \partial r} 
+ {1\over T} {\partial T \over \partial r}\right) =
 -{GM_{\oplus}\over r^2k_BT}\left(m_3 + {K_{zz}n\mu\over b_{23}}\right)
  + {\phi_{\mathrm{H}_2}r_0^2\over b_{23}r^2}
 \end{equation}
 Equation \ref{CCF three} is used to integrate $n_3(r)$ from the lower boundary.

 \medskip

The diffusion-limited flux --- the upper bound on how fast hydrogen
can diffuse through a hydrostatic atmosphere of CO$_2$ --- can be recovered from 
Eq \ref{CCF three} in the limit of constant mixing ratios \citep{Zahnle1986,Hunten1987}.
 The $K_{zz}$ terms drop out. 
In the particular case of H$_2$ diffusing through CO$_2$, the diffusion limit is 
\begin{equation}
\label{diffusion-limited-1}
   \phi_{\rm DL} = { \left(m_3-m_2\right) GM_{\oplus} b_{23} f_{{\rm H}_2} \over r_0^2kT} , %\left(r\over r_0\right)^2 .
\end{equation}
 where $b_{23}$ is the binary diffusion coefficient between H$_2$ and CO$_2$.
For Earth, Eq \ref{diffusion-limited-1} evaluates to
\begin{equation}
\label{diffusion-limited-2}
   \phi_{\rm DL} = 3.0\times 10^{13} f_{{\rm H}_2}\left(200\over T \right)^{0.25} \quad {\rm cm}^{-2} {\rm s}^{-1}  .
\end{equation}
Equation \ref{diffusion-limited-2} is used in Figure \ref{Figure9}.
Important points about diffusion-limited escape are (i) it is proportional to the hydrogen mixing ratio
and (ii) it remains valid as an upper limit on how quickly two species can be separated, even
if both species escape.

\subsubsection{Conservation of Energy}
Energy equations can be written for each species in steady state and spherical symmetry \citep{Schunk1980}. 
%these equations can be written
\[
{3v_i\over 2}{\partial \left(n_ik_BT_i\right)\over \partial r} 
+ {5n_ik_BT_i\over 2} {1 \over r^{2}}{\partial \left(r^{2}v_i\right)\over \partial r}
 + {1 \over r^{2}}{\partial \left(r^{2}Q_i\right)\over \partial r}
\]
\begin{equation}
 \label{C0}
= \Gamma_i 
+ \sum_j {n_im_i\nu_{ij}\over m_i+m_j}\left( 3k_B\left(T_j-T_i\right) + m_j\left(v_i-v_j\right)^2\right)
\end{equation}
where $\Gamma_i$ represents the diabatic volume heating and cooling rates.
The collision terms between the species are written out.
Equation \ref{C0} as written applies to monatomic species.

The simplest approximation to Eq \ref{C0} assumes that all species share the same
temperature.  We then treat the fluid as a whole for the purposes of energy conservation,
\begin{equation}
\label{SN3}
{1\over r^{2}}{\partial  \over \partial r}\left\{ r^{2}\sum_i{n_iv_i m_i\left({v_i^2\over 2} + {\gamma_i\over \gamma_i-1}{k_BT\over m_i} - {GM_{\oplus}\over r} \right)}  + r^{2}\sum_iQ_i \right\} = \Gamma_{\mathrm{h}} - \Gamma_{\mathrm{c}}
\end{equation}
where $\gamma_i$ is the ratio of heat capacities of species $i$ ($\gamma=5/3$ for a monatomic gas);
  $Q=\sum_iQ_i$ represents the thermal conduction flux; and
$ \Gamma_{\mathrm{h}} $ and $ \Gamma_{\mathrm{c}} $ refer to total radiative volume heating and cooling rates, respectively.
Written in this form with $T_i=T$ for all $i$, 
the collisional terms in Eq \ref{C0} either zero out or cancel out.

In general, thermal conduction is a complicated transport phenomenon, a proper treatment of which
far exceeds the scope of this paper.
We will estimate the magnitude of the term using the familiar approximation  
\begin{equation}
\label{quirk}
 Q = - k_c{\partial T\over \partial r} ,
 \end{equation}
which is written in terms of a thermal conductivity $k_c$ ($k_c$ is a function of composition and temperature).
In passing we note that thermal conduction as approximated by Eq \ref{quirk}
implicitly assumes instantaneous transport of energy, when in fact energy transport by thermal conduction is
bounded by the speed of sound, which is an important consideration
for energy transport in a transonic planetary wind.

As above, we assume that H and H$_2$ flow outward with the same velocity $u$,
and that CO$_2$ is hydrostatic. 
The usual diatomic $\gamma=7/5$ for H$_2$ is assumed,
\[
{1\over r^{2}} {\partial \over \partial r}\left\{ 
   r^{2}n_1um_1\left({u^2\over 2} + {5\over 2}{k_BT \over m_1} - {GM_{\oplus} \over r}\right)
  + r^{2}n_2um_2\left({u^2\over 2} + {7\over 2}{k_BT \over m_2} - {GM_{\oplus} \over r}\right) \right\}
\]
\begin{equation} \label{C1}
 +{1\over r^{2}} {\partial (r^{2}Q) \over \partial r}
 = \Gamma_{\mathrm{h}} - \Gamma_{\mathrm{c}}  .
\end{equation}
The left hand side of Eq \ref{C1} describes the divergence of advected kinetic energy and heat, work done
lifting the gas out of the planet's potential well, and thermal conduction.
These are equated to the sum of radiative heating and cooling.

Expanded with substitutions, Eq \ref{C1} becomes
\[
 {\phi_{\mathrm{H}_2} r_0^{2}\over r^{2}}\left\{ m_2\left( {GM_{\oplus}\over r^2} + u {\partial u \over \partial r} \right)
+  {7k_B\over 2}{\partial T \over \partial r} \right\}
+ {3k_BT\over 2}\left( J_2n_2 + J_3'n_3\right)
+ {3k_Bn_1u\over 4}{\partial T\over \partial r} 
  \]
\begin{equation}
 \label{C2}
 + {1\over r^{2}} {\partial (r^{2}Q) \over \partial r}
 = \Gamma_{\mathrm{h}} - \Gamma_{\mathrm{c}}  .
\end{equation}
The terms involving $J_2$ and $n_1u$ arise because photolysis increases 
the heat capacity of the gas by increasing the total number.
Equation \ref{C2} can be recast as an expression for the temperature gradient,
 \[
 \left( {7k_B\over 2}\phi_{\mathrm{H}_2}  
 + n_1u{3k_B\over 4}n_1u \right) {\partial T \over \partial r}
 = {r^{2}\over r_0^{2}}\left(\Gamma_{\mathrm{h}} - \Gamma_{\mathrm{c}}\right)
 -m_2\phi_{\mathrm{H}_2} {GM_{\oplus}\over r^2} 
- m_2\phi_{\mathrm{H}_2} {u\partial u \over \partial r} 
\]
\begin{equation}
 \label{C3}
- {r^{2}\over r_0^{2}} {3k_BT\over 2}\left( J_2n_2 + J_3'n_3\right)
 - {1\over r_0^{2}} {\partial (r^{2}Q) \over \partial r} .
\end{equation}

At this point we make two additional simplifications to enhance numerical stability.
First, we drop the $u^2$ kinetic energy terms as small compared to the advected heat terms.
This approximation eliminates a sometimes numerically troublesome term
at little cost to the overall verisimilitude of the model. 
Second, we drop thermal conduction $Q$.
With these approximations, Eq \ref{C3} reduces to
 \begin{equation}
 \label{C4}
 {\partial T \over \partial r}
 = \frac{ \strut\displaystyle{r^{2}\over r_0^{2}}\left(\Gamma_{\mathrm{h}} - \Gamma_{\mathrm{c}}\right)
 -m_2\phi_{\mathrm{H}_2} {GM_{\oplus}\over r^2} 
 - {r^{2}\over r_0^{2}}{3k_BT\over 2} \left( J_2n_2 + J_3'n_3\right) }
{\strut\displaystyle {7k_B\over 2}\phi_{\mathrm{H}_2}  
 +{3k_B\over 4}n_1u } .
\end{equation}

Thermal conduction is important when the insolation is modest and the escape flux is small or negligible,
especially in cases where a hydrostatic atmosphere can be stabilized by thermal conduction \citep{Gross1972}.
For Earth and N$_2$-CO$_2$-H$_2$ atmospheres, hydrostatic solutions of this kind are possible for $S<1$.
But for a highly-irradiated planet like young Earth, ignoring $Q$ is an acceptable shortcut, because thermal conduction cannot
simultaneously be a big term in the energetics yet also span the dimensions of a much inflated thermosphere.
In particular, we do not expect thermal conduction to much exceed a 10\% effect for $S>10$, 
which we will find to be the minimum $S$ required for Xe escape.
Figure \ref{Figure9a}, which compares our results for hydrogen escape to 
those obtained using a hydrocode that
includes thermal conduction \citep{Kuramoto2013}, implies that thermal conduction
can be neglected for $S>2.5$ with no obvious consequences. 
% Therefore we omit $Q$ from the model.  
However, we do compute $Q$ from our solutions as an {\it a posteriori} check for self-consistency.

\subsection{Method of solution}

The system is solved with the shooting method, integrating upward starting from below the homopause.
The basic equations are Eq \ref{E5} for $u(r)$, Eqs \ref{B9} and \ref{B10} for $n_1(r)$ and $n_2(r)$,
 Eq \ref{CCF three} for $n_3(r)$, and Eq \ref{C4} for $T(r)$.
We seek a transonic solution that has just enough energy at the critical point to escape.
 The second outer boundary condition is in keeping with the
philosophy that nothing that happens beyond the critical point can influence the atmosphere at the lower boundary.

We assume a density $n(r_0)$, eddy diffusivity $K_{zz}$, and total hydrogen escape flux $\phi_{\mathrm{H}_2}$ at the lower boundary.
The temperature $T(r_0)$ is determined by solving the local energy balance between
absorbed XUV and FUV radiation and radiative cooling by CO$_2$, with the temperature gradient $\left(dT/dr\right)_0=0$. 
The nominal values of $n(r_0)$ and $K_{zz}$ are $1\times 10^{13}$ cm$^{-3}$ and $2\times 10^6$ cm$^2$s$^{-1}$,
respectively. 
The lower boundary density is larger than the nominal homopause
density $n_h \equiv b_{23}/K_{zz}=6\times 10^{12}$ cm$^{-3}$ in our nominal model.

The total H$_2$ mixing ratio at the lower boundary and the total hydrogen flux $\phi_{\mathrm{H}_2}$
are treated as independent free parameters.  
%These are the inputs to the model.
% In practice we express $\phi_{\mathrm{H}_2}$ as a fraction
% (less than one when CO$_2$ does not escape, which is the case here) 
% of the diffusion-limited flux.
We then solve for the solar irradiation $S$ required to support $\phi_{\mathrm{H}_2}$.
The irradiation $S$ is iterated until the hydrogen
velocity $u$ matches the speed of sound at the point where the specific energy of the wind $e_s$ 
\begin{equation}
\label{energy}
        e_s =  f_{\mathrm{H}_2} \left( {7k_B T\over 2} + m_{\mathrm{H}_2} \left({u^2\over 2} - {GM_{\oplus}\over r}\right)\right) 
         + f_{\mathrm{H}}\left( {5k_B T\over 2} + m_{\mathrm{H}}\left({u^2\over 2} - {GM_{\oplus}\over r}\right)\right) 
 \end{equation}
 is just sufficient to ensure escape with no further energy input.
 The second outer boundary condition is therefore
 \begin{equation}
\label{stopping condition}
  e_s(r_c) = {2GM_{\oplus}\over r_c} .
\end{equation}
Equation \ref{energy} omits CO$_2$ as negligible at $r_c$, and it neglects the chemical energy that would
 be released if the hydrogen were to recombine.

Optical depths and the quantities that depend on them --- radiative heating, radiative cooling,
and photolysis --- are problematic for the shooting method
because they should be computed by integrating inward from infinity.
This is inconvenient when shooting outward from deep within the atmosphere.
However, because optical depth appears only as an integrated quantity,
none of the computations in this study critically depend on an exact value of $\tau$.
All that is really required are plausible values that decrease monotonically as the shooting algorithm ascends.
Hence we make a local approximation to optical depth at each height using local densities and the local scale height,
\begin{equation} \tau_{i\lambda} \approx n_i \sigma_{i\lambda} H \end{equation} 
\begin{equation}  H = k_{\rm B}T/\mu g \end{equation} 
\begin{equation}  \tau_{i\lambda} \approx \sigma_{i\lambda}H n_i  \end{equation}  
where $\mu$ is the mean molecular mass averaged over the different constituents.
Comparison between our results for hydrogen escape and those obtained with a hydrocode that
self-consistently computes optical depths \citep[Figure \ref{Figure9a},][]{Kuramoto2013}
 shows no sign that our local approximation to $\tau$ is failing in any important way. 

% \medskip\noindent{\em Stopping conditions}

 \subsection{Results for hydrogen escape}
 \label{Wind Results}

It is helpful to illustrate
the properties of a particular model in some detail.
For this purpose we have chosen a model (hereafter referred to as the ``nominal'' model)
for illustration that lies well within the field of models for which Xe escape is predicted to take place.
The key parameters are a relatively high EUV flux ($S=20$) and a relatively high hydrogen
mixing ratio ($f_{\mathrm{H}_2}=0.03$).
Other nominal parameters are 
$n(r_0)=1\times 10^{13}$ cm$^{-3}$,
$K_{zz} = 2\times 10^{6}$ cm$^2$s$^{-1}$,
and spherical symmetry.
% and $\beta=2$ (spherical symmetry).
Our models of hydrogen escape are not very sensitive to the latter three parameters.
The computed hydrogen escape flux for this model is $\phi_{\mathrm{H}_2}=7.2\times 10^{11}$ cm$^{-2}$s$^{-1}$,
equivalent to 82\% of the diffusion-limited flux.

 \begin{figure}[!htb] 
  \centering
  \includegraphics[width=1.0\textwidth]{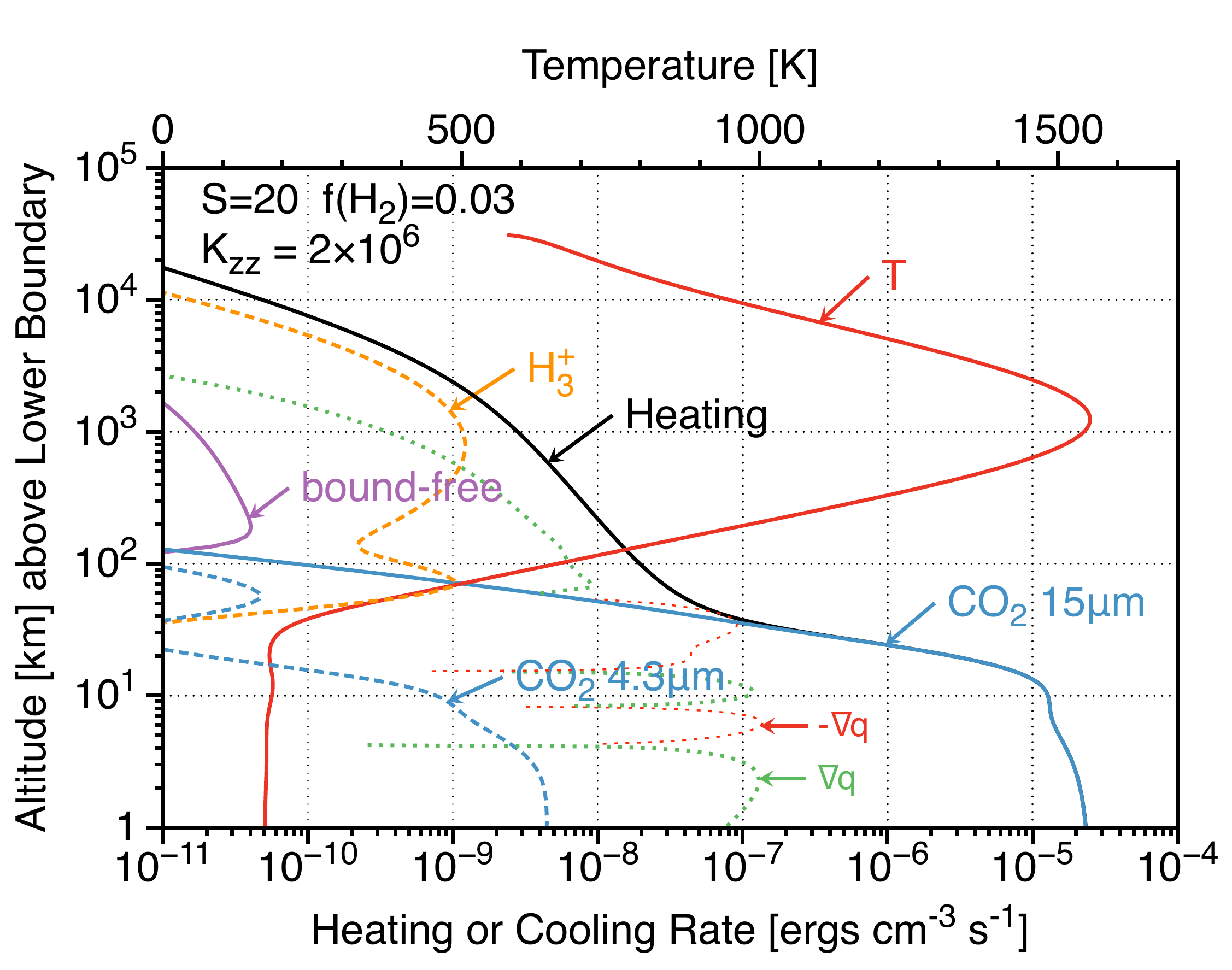} 
\caption{\small Volume heating and cooling rates in the nominal model as a function of altitude.
Altitude is measured from an arbitrary lower boundary where the total density $n(r_0)=1\times 10^{13}$ cm$^{-3}$.
The magnitude of the convergences and divergences of conductive heat flow --- the local volume heating or cooling due to
thermal conduction --- are also shown.}
\label{Figure7}
\end{figure}

 Figure \ref{Densities} in the main text shows
 the temperature and the densities of the ions and neutrals as a function of altitude.
Figure \ref{Figure7} shows temperature and local volume heating and cooling rates as a function of altitude in the nominal model.
At low altitudes where CO$_2$ is abundant, radiative heating and radiative cooling are
in balance, the gas is cold, and very little energy is channeled into hydrogen escape.
At high altitudes there is some radiative cooling by H$_3^+$ but not enough to shut off hydrogen
escape.
Also shown are the magnitudes of local volume heating or cooling due to thermal conduction that would
be consequent to the computed temperature structure.  These terms are not actually included in
the model (they would of course be smaller if thermal conduction were included, because the 
atmosphere would adjust to smooth out the temperature profile); we plot local conductive heating and cooling
 here to show that thermal conduction, although not small, can be neglected at $S=20$ for Earth
 without too much loss of accuracy.  
 
\begin{figure}[!htb] 
  \centering
  \includegraphics[width=1.0\textwidth]{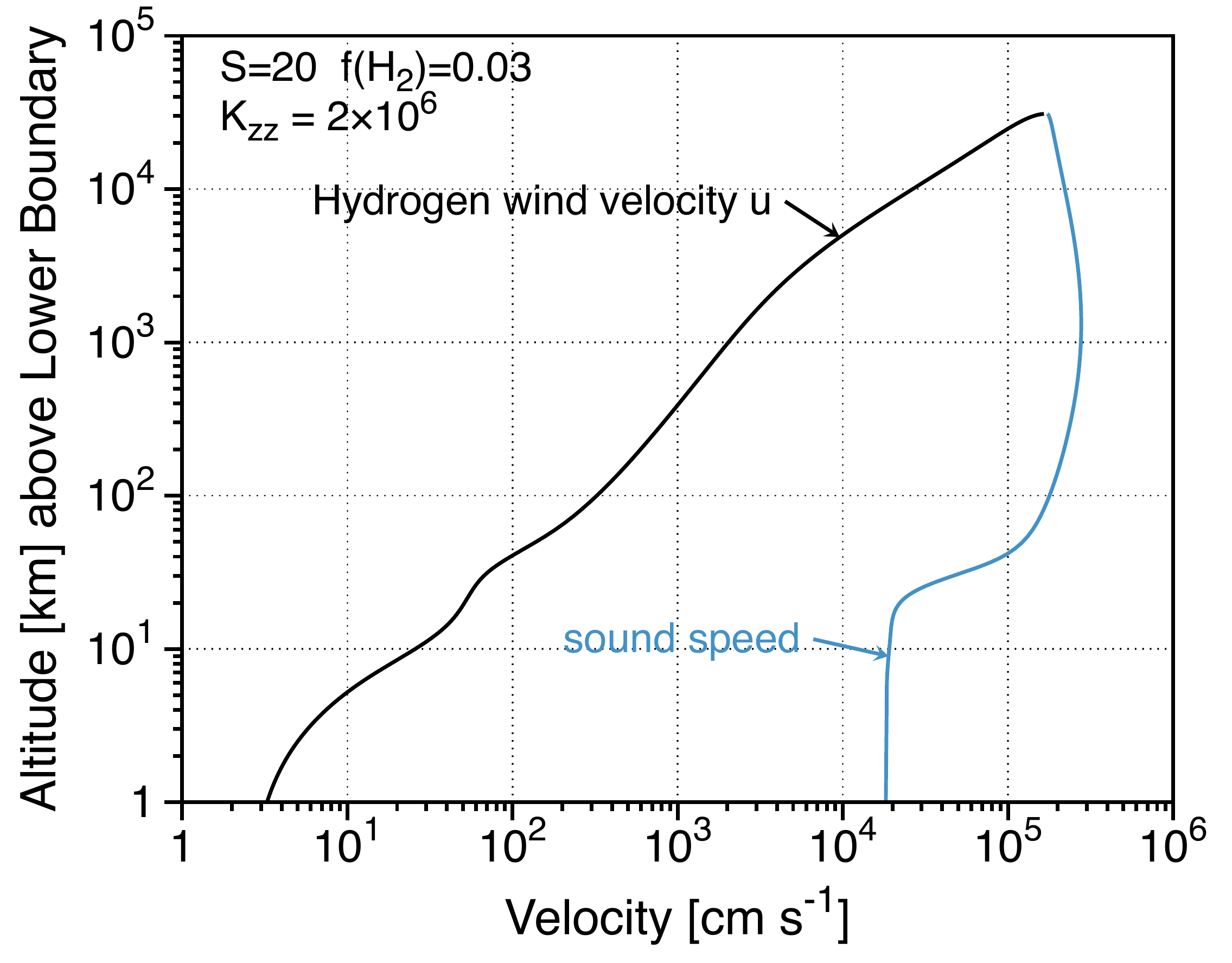} 
\caption{\small Flow velocity $u$ and sound speed $c_s$ as a function of altitude
in the nominal model.
Altitude is measured from an arbitrary lower boundary where the total density $n(r_0)=1\times 10^{13}$ cm$^{-3}$.
The model stops at the critical point where $u=c_s$.}
\label{Figure8}
\end{figure}

Figure \ref{Figure8} shows sound speeds and flow velocities in the nominal model as a function of altitude.
The flow velocity reaches the sound speed at the top of the model, which is the critical point.
In the approximation we are using in this study,
the gas at the critical point has exactly the minimum energy required to escape.
%The larger part of this energy is thermal, so that further expansion without any further heating would result in a 0 K gas at infinity.

Figure \ref{Figure9} in the main text shows the results of a parameter survey over $S-f_{\mathrm{H}_2}$ 
of hydrogen escape from the CO$_2$-H$_2$ atmosphere for Earth,
and Figure \ref{Figure9a} compares our results to model results previously obtained using different numerical 
approaches by \citet{Tian2005b} and \citet{Kuramoto2013}.

\section{Xenon Chemistry and Transport}
\label{appendix two}  

\subsection{Reaction rates}

To compute rates of the charge transfer reactions of H$^+$ with neutral Xe (reactions R16a and R16b), we used the Demkov approximation \citep{Demkov1964,Olson1972,Swartz1994}, a variation of the Landau-Zener model applicable to singly charged ions, to calculate the cross sections $\sigma_{16a}$ and $\sigma_{16b}$. The main assumption is that the charge-exchange reaction proceeds through non-local radial coupling between molecular states of the same symmetry. In contrast to the Landau-Zener model, where the coupling is assumed to be local at the internuclear distance $R_x$, the Demkov approximation assumes that the reaction takes place at the internuclear distance $R_x$ through a non-local interaction well-represented by an exponential form. 
As is conventional in atomic physics, 
throughout this section we will use Hartree atomic units (a.u.), in which the electron mass, the electric charge,
and the Coulomb constant are set to unity and the Planck constant is set to $2\pi$.
In a.u., the distance unit is the Bohr radius ($a_0=0.5291772 \AA$).
The energy unit, the Hartree, is $27.211385$ eV, and is exactly twice the ionization energy of H.
$R_x$ can be determined by equating $\Delta U(R_x)$, the separation of the adiabatic electronic potentials, with 
\begin{equation}
  \Delta H(R) = H_{11}(\mathrm{Xe} + \mathrm{H}^+) - H_{22}(\mathrm{Xe}^+ + \mathrm{H}) - \Delta E , 
\end{equation}
where $H_{11}(R)$ and $H_{22}(R)$ are incoming and outgoing adiabatic electronic interaction potentials, respectively, and $\Delta E=\mathrm{I_P}(\mathrm{Xe})-\mathrm{I_P}(\mathrm{H})-E_\mathrm{exc}$ is the energy defect given by the difference of ionization potentials $\mathrm{I_P}$ of Xe and H, with $\mathrm{I_P}$(Xe) = 12.1298436 eV and $\mathrm{I_P}$(H) = 13.6057 eV, and $E_\mathrm{exc}$ is the energy of the final excited state of Xe$^+$. 
The rate coefficients were calculated assuming the electronic potential model of \citet{Sterling2011}. 
The incoming channel is % given by %(in atomic units, a.u.):
\begin{equation}
  H_{11}(R) = - \frac{\eta_\mathrm{Xe}}{2 R^4} + A e^{-(0.8+\zeta)R}
\end{equation}
where $R$ is the internuclear distance, $\eta_\mathrm{Xe}=27.2903$ is the polarizability of the Xe atom in a.u.\ \citep{Lide2012}, $\zeta=1$ is the exponent of a single orbital wave function \citep{Butler1980}, and $A$ is the damping amplitude, typically set to 25 \citep{Sterling2011}.
Similarly, the outgoing channel is given by
\begin{equation}
 H_{22}(R) = -\frac{\eta_\mathrm{H}}{2 R^4} + A e^{-(0.8+\zeta)R} ,
\end{equation}
where the dipole polarizability of the H atom $\eta_\mathrm{H}$ is $9/2$ in a.u. 
For the potential $H_{11}$, a quadrupole polarizability term $\pm Q_q/(2R^3)$ must be included since Xe is not initially in the ground $S$ state \citep{Gentry1977,Sterling2011}, resulting in
\begin{equation}
  \frac{\eta_\mathrm{Xe}}{2 R^4} \rightarrow \frac{\eta_\mathrm{Xe}}{2 R^4} - \frac{Q_q}{2R^3} ,
\end{equation}
where the sign is selected to keep $H_{11}(R_x)-H_{22}(R_x)-\Delta E>0$. 
We use $Q_q=-13.2071$ a.u., which equals $-17.764$ Debye \AA \citep{Lide2012}. 

The analyzed charge exchange process is a ``type I'' electron capture, where the H $1s$ electron is transferred without perturbing the core electron arrangement of the ionic projectile. Consequently, $\Delta U(R_x)$ term can be estimated from the empirical fit \citep{Butler1980}
\begin{equation}
% \Delta U(R_x) = 27.21 R^2_x e^{-\beta R_x}, 
 \Delta U(R_x) = 27.21 R^2_x e^{-\beta R_x} \; \mathrm{eV} \;, 
\end{equation}
with $\beta^2 = 2 I_P(\mathrm{Xe})$.
The listed expressions are sufficient to calculate the distance $R_x$ where the charge exchange takes place.
Correct asymptotic energies for the two possible exit channels of the reaction Xe+H$^+$ are $\Delta E_{1/2}=0.16$ eV for Xe$^+$ ($^2P_{1/2}$) and $\Delta E_{3/2}=1.47$ eV for Xe$^+$ ($^2P_{3/2}$). 
% $U(R_x)=0.00498$ Hartree; $U(R_x)=0.06889$ Hartree
For the two cases, we computed $R_x=10.619$ a.u. and $U(R_x)=0.1356$ eV, and $R_x=6.935$ a.u. and $U(R_x)=1.8746$ eV, respectively. 
For the averaged asymptotic shift, $\Delta E_\mathrm{avg} = 1.03403$ eV, we obtained $R_x=7.933$ $a_0$, confirming the result reported in Table 1 in \citet{Sterling2011}. 

The transition probability for the charge exchange process is given by $p=e^{-w}$ \citep{Demkov1964,Swartz1994}, with 
\begin{equation}
 w = \left[1 + \exp \left(\frac{2\pi^2 \Delta U(R_x)}{h \beta v} \right) \right]^{-1} ,
\end{equation}
where $h$ is Planck's constant and $v$ is the relative velocity of the interacting particles.
The charge exchange cross section is given by \citep{Olson1972,Swartz1994}
\begin{equation}
 \sigma_x(E)=\pi R_x^2 \int_{1}^{\infty} \frac{dx}{x^3} \frac{4 e^{-\delta x}}{(1+e^{-\delta x})^2} , 
\end{equation}
where
\begin{equation}
 \delta = \frac{\pi^2 \Delta U(R_x)}{h \beta v} .
\end{equation} 

\begin{figure}[!htb] 
  \centering
  \includegraphics[width=1.0\textwidth]{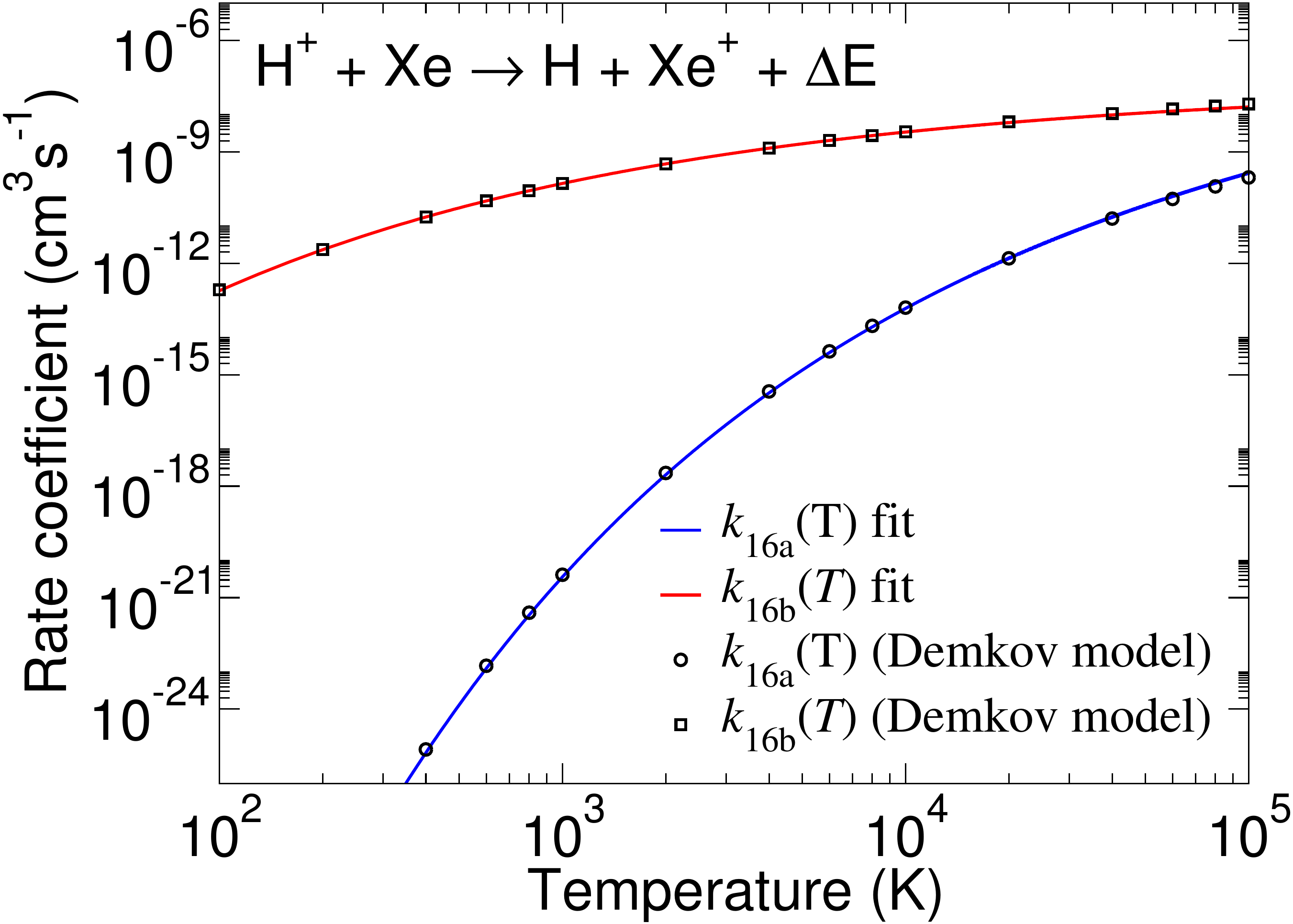}
 \caption{\small Calculated rate coefficients for the charge transfer reactions
 $\mathrm{H}^+ + \mathrm{Xe} \rightarrow \mathrm{H} + \mathrm{Xe}^+(^2\mathrm{P}_{3\over 2})$ (R16a)
and $\mathrm{H}^+ + \mathrm{Xe} \rightarrow \mathrm{H} + \mathrm{Xe}^+(^2\mathrm{P}_{1\over 2})$ (R16b).
Squares denote the calculated rate coefficients, the lines show the best fits to $k_{16a}$ and $k_{16b}$ for $50<T<10^5$ K.
The curve fits are listed in Table \ref{Xenon Chemistry Table}.}
\label{CXrate}
\end{figure}

Finally, the charge exchange rate coefficient at temperature $T$ can be expressed as $k(T)=\langle \sigma_x v \rangle$, where $\langle \ldots \rangle$ implies averaging over a Maxwell-Boltzmann distribution of initial collision velocities.
Numerically evaluated rates are shown in Figure \ref{CXrate} for temperatures between $100-10^6$ K. 
We constructed fits to $k(T)=A T^b \exp(-c/T^d)$, with two sets of coefficients depending on the exit channel: 
%$A=3.83017\times10^{-8}$ cm$^3$s$^{-1}$K$^{-b}$, $b=0.038557$, $c=55.7906$ K$^{d}$, $d=0.326$ (for $\Delta E_{1/2}$) and 
%$A=2.37109\times10^{-9}$ cm$^3$s$^{-1}$K$^{-b}$, $b=0.342427$, $c=294.089$ K$^{d}$, $d=0.3332$ (for $\Delta E_{3/2}$).
$A=3.83017\times10^{-8}$ cm$^3$s$^{-1}$, $b=0.038557$, $c=55.7906$ K$^{d}$, $d=0.326$ (for $\Delta E_{1/2}$) and $A=2.37109\times10^{-9}$ cm$^3$s$^{-1}$, $b=0.342427$, $c=294.089$ K$^{d}$, $d=0.3332$ (for $\Delta E_{3/2}$).
% We calculated the cross-sections using either two exit channels or a single exit channel, but found no significant improvement for the two exit channel model due to the large fine structure splitting. 
The uncertainty of rates calculated using the Demkov approximation has been shown to be less than a factor of three for rates larger than about $10^{-9}$ cm$^3$s$^{-1}$ \citep{Butler1980,Kingdon1996}, which is good enough for our purposes.

\medskip

Xenon reaction rates used here are listed in Table \ref{Xenon Chemistry Table}.
\begin{table}[htp]
\caption{Summary of Xe Chemistry}
% \begin{center}  % not sure what this does, but it has a biggish effect
\begin{tabular}{llll}
label & reactants & products & rate [cm$^3$s$^{-1}$]\\
\hline
JXe & $\mathrm{Xe} + h\nu $ & $\rightarrow \mathrm{Xe}^+ + e^{-1} $ & $J_{\mathrm{Xe}}$  \\
R15 & $\mathrm{CO}^+_2 + \mathrm{Xe} $ & $\rightarrow \mathrm{CO}_2 + \mathrm{Xe}^+ $ & $k_{15}=6\times 10^{-10}$   \\
R16a & $\mathrm{H}^+ + \mathrm{Xe} $ & $\rightarrow \mathrm{H} + \mathrm{Xe}^+\!\left(^2\mathrm{P}_{3\over 2} \right) $ & $k_{16a}=2.37\times 10^{-9} T^{0.342} e^{-294/T^{0.333}}$   \\
R16b & $\mathrm{H}^+ + \mathrm{Xe} $ & $\rightarrow \mathrm{H} + \mathrm{Xe}^+\!\left(^2\mathrm{P}_{1\over 2} \right) $ & $k_{16b}=3.83\times 10^{-8} T^{0.386} e^{-55.8/T^{0.326}}$   \\
R17 & $\mathrm{Xe}^+ + \mathrm{O}_2 $ & $\rightarrow\mathrm{Xe} + \mathrm{O}^+_2  $ & $k_{17}=1.2\times 10^{-10}$  \\
R17r & $\mathrm{O}_2^+ + \mathrm{Xe} $ & $\rightarrow \mathrm{O}_2 + \mathrm{Xe}^+  $ & $k_{17r}=3\times 10^{-10}e^{-500/T}$    \\
R18 & $\mathrm{Xe}^+ + e^{-1} $ & $\rightarrow \mathrm{H} + h\nu $ & $k_{18}=k_5$   \\
\end{tabular}
% \end{center}
\label{Xenon Chemistry Table}
\end{table}%

\subsection{Collision frequencies}

The ion-ion collision frequency for momentum transfer is \citep{Schunk1980}
 \begin{equation}\label{Coulomb one}
 \nu_{ij} \approx 1.27 \frac{z_i^2z_j^2A_{ij}^{0.5}}{A_i}  \frac{n_j}{T_{ij}^{1.5}}
 \end{equation}
 where $z_i$ and $z_j$ are the charges of the ions (integers),
 and $A_{i}$ and $A_{ij}$ are the mass and the reduced mass in amu, respectively.
 The mapping between momentum transfer collision frequencies $\nu_{ij}$ 
 and binary diffusion coefficients is given by Eq \ref{collision-rates}.
 For Xe$^+$ and H$^+$, % Eqs \ref{Coulomb one} and \ref{collision-rates} evaluate to 
 \begin{equation}\label{Coulomb two}
 b^{\scriptscriptstyle ++}_{1j} \approx \frac{k_BT^{2.5}}{1.27m_1} = 2.1\times 10^{15}\left(T/1000\right)^{2.5}
  \qquad \mathrm{cm}^{-1}\mathrm{s}^{-1}.
 \end{equation}
 The $b^{\scriptscriptstyle ++}_{ij}$ are much smaller (Table \ref{Xenon Table}) for the heavy molecular ions HCO$^+$ and CO$_2^+$
 because the reduced masses are much larger.  
 
  \begin{figure}[!htb] %  figure placement: here, top, bottom
   \centering
   \label{Figure16}
 \begin{minipage}[c]{0.54\textwidth}
   \centering
  \includegraphics[width=1.0\textwidth]{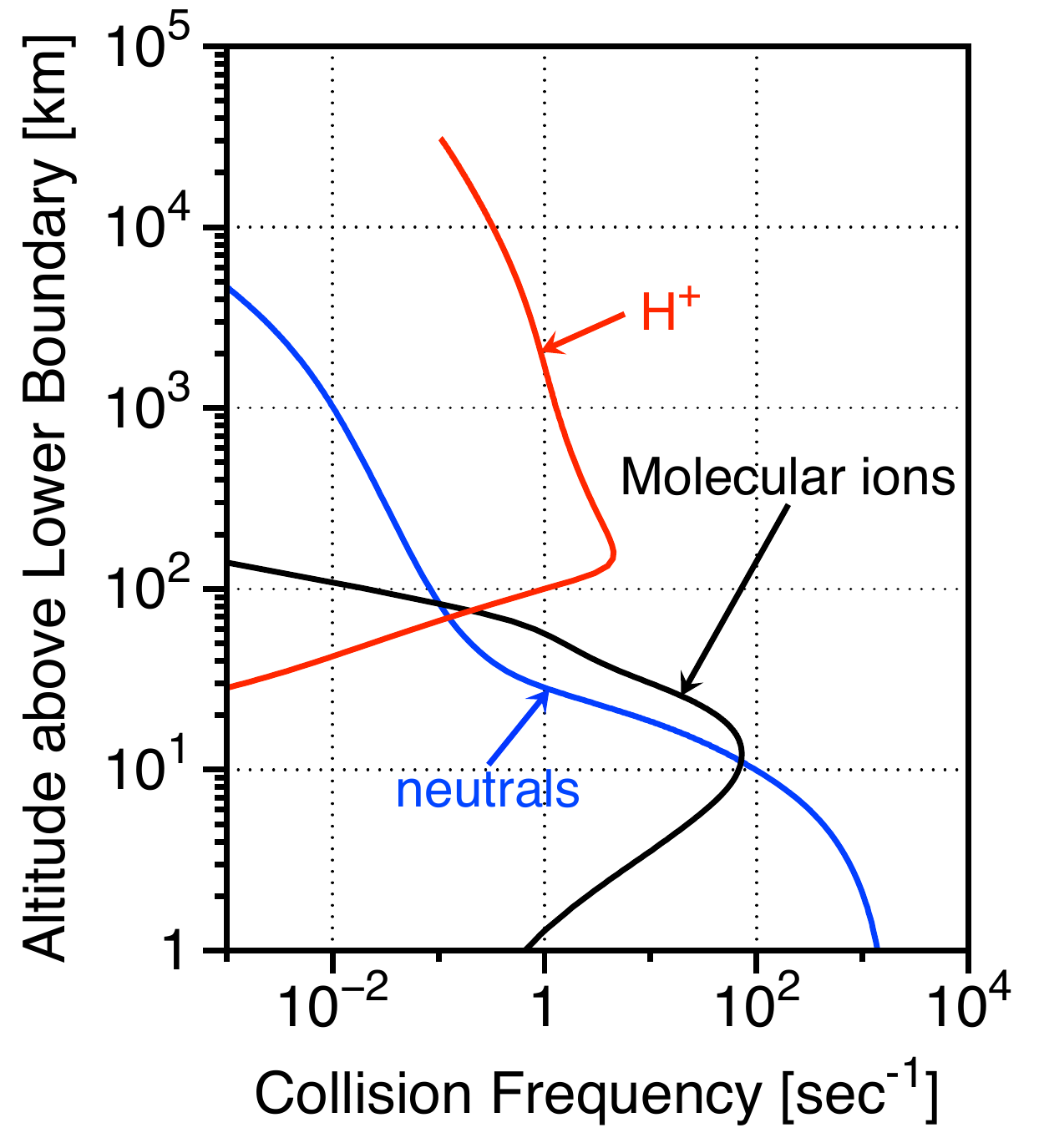} 
 \end{minipage}
\begin{minipage}[c]{0.45\textwidth}
   \centering
\caption{\small Momentum transfer collision frequencies 
$\nu_{ij}$ of Xe$^+$ with neutrals, molecular ions, and H$^+$ in the nominal model ($f_{\mathrm{H}_2}=0.03$, $S=20$).
Altitude is measured from an arbitrary lower boundary where the total density $n(r_0)=1\times 10^{13}$ cm$^{-3}$.
Xe$^+$ transport takes place in three regimes.  The homopause is at $\sim 4$ km.  Below the homopause Xe$^+$ transport is collisionally 
dominated by neutral molecules.  Between 10 km and 80 km (the baropause), the Xe$^+$ ions are collisionally dominated by molecular ions stemming from CO$_2$.  Above 80 km, Xe$^+$ is most responsive to the outbound flow of H$^+$.
 }
  \label{Collision}
 \end{minipage}
\end{figure}

Less important are interactions between ions and neutrals,
 but we include these for completeness.
% This interaction is affected by the dipole induced in the neutral by the charge of the ion.
The ion-neutral collision frequency for momentum transfer is \citep{Schunk1980}
\begin{equation}\label{induced one}
\nu_{in} = 2.21\pi \frac{n_nm_n}{m_i+m_n}\left(\frac{\eta_n e^2}{\mu_{in}}\right)^{0.5}.
 \end{equation}
Written as a binary diffusion coefficient,
\begin{equation}\label{induced two}
b^{\scriptscriptstyle +}_{in} = \frac{k_BT}{2.21\pi e\sqrt{\mu_{in} \eta_n}}
 \end{equation}
where $\mu_{in}$ is the reduced mass, $e$ the electric charge, and $\eta_n$ the polarizability of the neutral.
% Polarizabilities are usually given in cubic Angstroms.
% For atomic H the polarizability is exactly 0.6668 \AA$^3$. 
% Otherwise polarizabilities are not in general isotropic.
Unless the gas is cold, the ion-neutral cross sections are not much larger than the corresponding neutral-neutral
cross sections.  

\begin{table}[htp]
\caption{Binary diffusion coefficients for Xe}
\begin{center}
\begin{tabular}{llll}
symbol &~~$i,j$ & ~~$b(i,j)$ cm$^{-1}$s$^{-1}$ & reference \\
 $b_{23}$ & H$_2$, CO$_2$   & $4.0\times 10^{19}\left(T/1000\right)^{0.71}$  &  \citet{Mason1972} \\
 $b_{2j}$ & H$_2$, Xe   & $1.8\times 10^{19}\left(T/1000\right)^{0.71}$  &  \citet{Mason1972} \\
 $b_{1j}$ & H, Xe  &  $2.6\times 10^{19}\left(T/1000\right)^{0.71}$  & scaled from $b_{2j}$\\
 $b_{3j}$ & CO$_2$, Xe & $4.0\times 10^{18}\left(T/1000\right)^{0.75}$  & scaled from $b({\rm CO}_2,{\rm SF}_6)$\\
 $b^{\scriptscriptstyle +}_{1j}$ & H, Xe$^+$ & $3.9\times 10^{19}\left(T/1000\right)$ & Eq \ref{induced two}, $\eta_n$(H) $ = 0.6668$ \AA$\!^3$\\
 $b^{\scriptscriptstyle +}_{2j}$ & H$_2$, Xe$^+$ & $1.6\times 10^{19}\left(T/1000\right)$ & Eq \ref{induced two}, $\eta_n$(H$_2$) $= 0.9$ \AA$\!^3$ \\
 $b^{\scriptscriptstyle +}_{3j}$ & CO$_2$, Xe$^+$ & $3.3\times 10^{18}\left(T/1000\right)$ & Eq \ref{induced two}, $\eta_n$(CO$_2$) $= 2.9$ \AA$\!^3$ \\
 $b^{\scriptscriptstyle ++}_{1j}$ &   H$^+$, Xe$^+$   & $2.1\times 10^{15}\left(T/1000\right)^{2.5}$  & Equation \ref{Coulomb two}.\\
 $b^{\scriptscriptstyle ++}_{4j}$ &   CO$_2^+$, Xe$^+$   & $6.4\times 10^{13}\left(T/1000\right)^{2.5}$  & Equation \ref{Coulomb two}.\\
 $b^{\scriptscriptstyle ++}_{5j}$ &   HCO$^+$, Xe$^+$   & $8.9\times 10^{13}\left(T/1000\right)^{2.5}$  & Equation \ref{Coulomb two}.\\
  not used & H$^+$, Xe & $1.6\times 10^{19}\left(T/1000\right)$ & Eq \ref{induced two}, $\eta_n$(Xe) $ = 4.05$ \AA$\!^3$ \\
 \end{tabular}
\end{center}
\label{Xenon Table}
\end{table}%

Beginning from Eq \ref{SN2}, 
with appropriate substitutions and with the $v_j^2$ terms ignored,
an equation describing the several forces acting on Xe$^+$ is
\[
\left(1 +  \sum_i{K_{zz}n_i\over b^{\scriptscriptstyle +}_{ij}} + \sum_i{K'_{zz}x_i\over b^{\scriptscriptstyle ++}_{ij}} \right) \left( {1\over x_j}{\partial x_j \over \partial r} 
+ {1\over T} {\partial T \over \partial r}\right) =
 -\left({v_j} - u\right) \left( {n_1 \over b^{\scriptscriptstyle +}_{1j}} + {n_2 \over b^{\scriptscriptstyle +}_{2j}}\right)
\]
 \begin{equation}\label{Xe force}
 -{GM_{\oplus}\over r^2k_BT}\left(m_j - {\mu^{+}\over 2} +  \sum_i{K_{zz}n_i\mu\over b^{\scriptscriptstyle +}_{ij}} + \sum_i{K_{zz}'x_im_i^+\over b_{ij}} \right)   - v_j \sum_i {x_i \over b^{\scriptscriptstyle ++}_{ij}}
%  -v_j \left( {x_1 \over b^{\scriptscriptstyle ++}_{1j}} + {x_3 \over b^{\scriptscriptstyle ++}_{3j}} + {x_5 \over b^{\scriptscriptstyle ++}_{5j}} \right)
  -{v_j}{n_3 \over b^{\scriptscriptstyle +}_{3j}}
 \end{equation}
Save near the lower boundary, the upward force on $^j$Xe$^+$ is dominated by the ion-ion interactions. 
Equation \ref{Xe force} is recast as an equation for $\partial v_j / \partial r$
that is readily integrated numerically beginning at the lower boundary
\[
{1\over v_j}{\partial v_j \over \partial r} = {1 \over T}{\partial T\over \partial r} - {2 \over r} 
  + {GM_{\oplus}\over r^2k_BT} {m_j - \mu^+ + A_a\mu + A_b m^+_4 + A_c m^+_5 \over A_d}
\]
\begin{equation}\label{Xe velocity}
  + {v_j-u\over A_d} \left( {n_1\over b^{\scriptscriptstyle +}_{1j}} + {n_2\over b^{\scriptscriptstyle +}_{2j}}  \right)
   + {v_j\over A_d} {n_3\over b^{\scriptscriptstyle +}_{3j}}
   + {v_j-u\over A_d}{x_1\over b^{\scriptscriptstyle ++}_{1j}} .
 \end{equation}
where  
\begin{displaymath}
A_a = \sum_i {K_{zz}n_i\over b^{\scriptscriptstyle +}_{ij}} \qquad A_b = {K'_{zz}x_4\over b^{\scriptscriptstyle +}_{4j}} \qquad A_c = {K'_{zz}x_5\over b^{\scriptscriptstyle +}_{5j}} \qquad A_d = A_a + A_b + A_c + 1
\end{displaymath}
% The parameter $\beta=2$ in spherical geometry and $\beta=3$ in dipolar geometry.
%
Equation \ref{Xe velocity} is solved by the shooting method.
The velocity $v_j(r)$ is integrated upward from the lower boundary for all the Xe isotopes.
The lower boundary velocity $v_j(r_0)$ is bounded by 0 and by the hydrogen velocity $u(r_0)$
(i.e., Xe cannot escape more easily than hydrogen).   
The velocity $v_j(r_0)$ at the lower boundary is iterated until either $v_j(r) \rightarrow u(r)$, in which case Xe$^+$ escapes,
or $v_j(r) \rightarrow 0$, in which case Xe$^+$ is hydrostatic and does not escape.

 \begin{figure}[!htb] 
  \centering
  \includegraphics[width=1.0\textwidth]{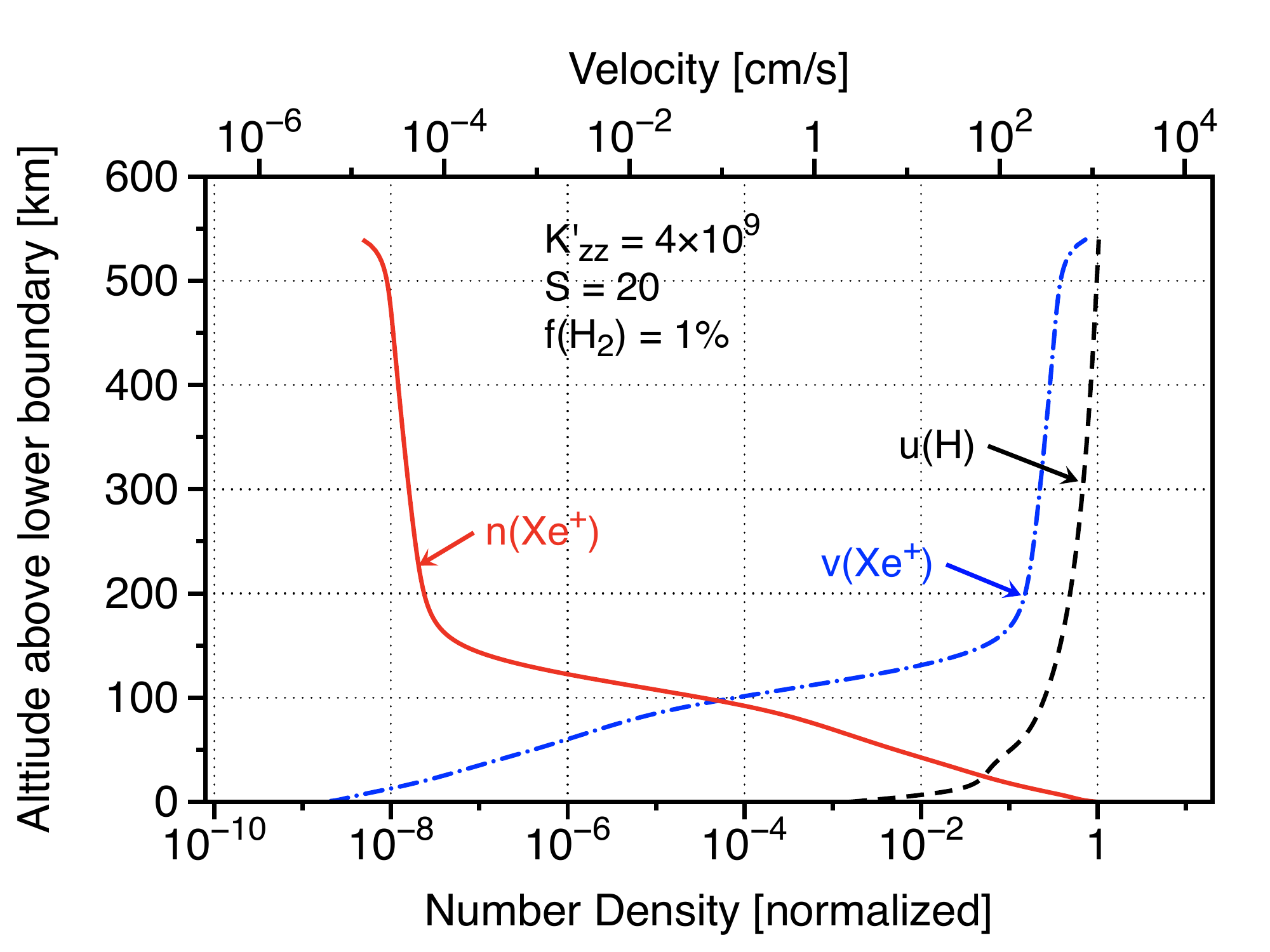} 
 \caption{\small An example of Xe$^+$ ions dragged to space by ionized hydrogen.
 Ion number density (red, labeled $n(\mathrm{Xe}^+)$) is normalized to unity at the lower boundary.
 Xe$^+$ velocities (blue, labeled $v(\mathrm{Xe}^+)$) are compared to the hydrogen velocity (black, labeled $u(\mathrm{H})$.
% Solid curves denote the nominal model
% ($S=20$, $f_{\mathrm{H}_2}=0.03$, $K_{zz}=2\times 10^6$, and $K'_{zz}=4\times 10^9$ cm$^2$s$^{-1}$).
% Three alternative models (dots and dashes) are each labeled according to how they differ from the nominal model.
% The blue curves stop at the altitude where the Xe$^+$ velocity merges with that of hydrogen.
% $K'_{zz}$ refers to eddy diffusivity among the molecular ions.
 Velocities and densities are mirror images because the flux $x_jv_jr^2$ is conserved by construction.
 Altitude is measured from an arbitrary lower boundary where the total density $n(r_0)=1\times 10^{13}$ cm$^{-3}$.
}
 \label{Figure12}
 \end{figure}

 Figure \ref{Figure12} documents the vertical structure of Xe$^+$ ions in the nominal model
 ($S=20$, $f_{\mathrm{H}_2}=0.03$, $K_{zz}=2\times 10^6$, and $K'_{zz}=4\times 10^9$ cm$^2$s$^{-1}$)
 and three variants
 ($S=40$, $f_{\mathrm{H}_2}=0.01$, and $K'_{zz}=4\times 10^8$ cm$^2$s$^{-1}$).
 In all four models Xe$^+$ is dragged to space by ionized hydrogen.
 The Xe$^+$ number densities $x_j$ are normalized to $x_j(r_0)=1$ at the lower boundary.
 The figure illustrates the sensitivity of Xe$^+$ to vertical transport amongst the molecular ions (parameterized by $K'_{zz}$),
 to the solar XUV flux $S$, and to the hydrogen mixing ratio at the lower boundary $f_{\mathrm{H}_2}$.

\clearpage 

\section{Table of Symbols}
\label{appendix three}  % does not work as expected

\setlongtables % keeps the width uniform across both pages
% \footnotesize{
\begin{longtable}{ll} 
%\caption{All Reactions}\\
% & Alphabetized Table of Symbols  \\
\hline
 Symbol & {\strut Definition} \\
\hline %\hline 
\endfirsthead
\hline
 Symbol & {\strut Definition} \\
\hline %\hline 
\endhead 
% \begin{table}[htp]
%\caption{Appendix: Symbols}
%\begin{center}
% \begin{tabular}{ll}  % left left
\hline
%$A$ & damping length in atomic units \\
$A_a$, $A_b$, $A_c$, $A_d$ & mixing parameters of neutral atmosphere and molecular ionosphere\\
$A_s$, $A_{st}$ & mass and reduced mass (amu)\\
$A_{\nu_2}$ & spontaneous emission rate for 15 $\mu$m ($\nu_2$) CO$_2$ band [s$^{-1}$]\\
$A_{\nu_3}$ & spontaneous emission rate for 4.3 $\mu$m ($\nu_3$) CO$_2$ band [s$^{-1}$]\\
$a$ & the power of XUV decay $S(t) \propto t^{-1}$\\
$a_0$ & Bohr radius, $a_0=0.529177 \AA$. $a_0\equiv 1$ in atomic units. \\
$a_1$, $a_2$, $a_3$ & curve fit parameters for $\Gamma_{\mathrm{H}^+_3}$ \\
$A_{\oplus}$ & area of Earth (cm$^2$)\\
$B, B_{\oplus}$ & geomagnetic field; $B_{\oplus}=0.31$ Gauss \\
$b_{ij}$ & binary diffusivity between species $i$ and $j$ [cm$^{-1}$s$^{-1}$] \\
$b^{\scriptscriptstyle +}_{ij}$ & binary diffusivity between an ion $i$ and a neutral $j$ [cm$^{-1}$s$^{-1}$] \\
$b^{\scriptscriptstyle ++}_{ij}$ & binary diffusivity between two ions $i$ and $j$ [cm$^{-1}$s$^{-1}$] \\
${\bar b} = b_{23}$ & the binary diffusivity pertinent to the homopause [cm$^{-1}$s$^{-1}$] \\
$c$ & speed of light [cm\,s$^{-1}$ \\
$c_s=\sqrt{\gamma k_B T /\mu}$ & speed of sound [cm\,s$^{-1}$ \\
$c'^2={\strut\displaystyle k_BT\over \strut\displaystyle \mu'}\left(1+{\strut\displaystyle K_{zz}n\over \strut\displaystyle {\bar b}} \right)$ & a kind of sound speed in hydrogen, squared\\
$E(\tau_x)$ & dimensionless function, Kumer \& James $E(x)$\\
$e$ & electron charge [$4.8\times 10^{-10}$ statcoulombs]\\
$e_s$ & specific energy of the gas [ergs\,g$^{-1}$]\\
$F_{\mathrm{fuv}}$ & total FUV irradiation ($91.2 < \lambda < 200$ nm) [ergs\,cm$^{-2}$s$^{-1}$]\\
$F_{\mathrm{xuv}}$ & total XUV irradiation ($\lambda < 91.2$ nm)[ergs\,cm$^{-2}$s$^{-1}$]\\
$F_{\lambda}$ & incident photon flux [cm$^{-2}$s$^{-1}$] in wavelength bin centered on $\lambda$ \\
$f_i=n_i/N$ & mixing ratio of species $i$ \\
$f_{\mathrm{H}_2}$ & total mixing ratio of hydrogen at the lower boundary\\
$G$ & universal gravitational constant [cm$^{3}$g$^{-1}$s$^{-2}$] \\
$H$ & scale height of atmosphere [cm]\\
$H_{11}$ ($H_{22}$) & incoming (outgoing) adiabatic electronic interaction potential [eV] \\
$h$ & Planck's constant [ergs\,s] \\
$\mathrm{I_P}$ & ionization potential [eV] \\
$J_{in}$ & photolysis rate of species $i$, pathway $n$ [s$^{-1}$] \\
$J'_{3}$ & weighted photolysis rate of CO$_2$ for splitting H$_2$ [s$^{-1}$] \\
$J_{\mathrm{Xe}}$ & photo-ionization rate of Xenon $i$ [s$^{-1}$] \\
$K_{zz}$ & Eddy diffusivity of lower atmosphere near homopause  [cm$^{2}$s$^{-1}$] \\
$K'_{zz}$ & Eddy diffusivity of ions in the molecular ionosphere [cm$^{2}$s$^{-1}$] \\
$k_B$ & Boltzmann constant [ergs\,K$^{-1}$] \\
$k_c$ & thermal conductivity [ergs\,cm$^{-1}$s$^{-1}$K$^{-1}$]  \\
$k^{\ast}_i$ & collisional de-excitation rate of CO$_2$ by species $i$ [cm$^{3}$s$^{-1}$] \\ 
$k_{n}$ & rate of reaction $n$ [cm$^{3}$s$^{-1}$] \\
$M_{\oplus}$ & mass [g] of Earth\\
$M_{\mathrm{H}_2\mathrm{O}}$ & equivalent mass of water corresponding to lost H$_2$ [g]\\
${\dot M_{sw}}$ & solar wind flux [g cm$^{-2}$ s$^{-1}$]\\
% \end{tabular}
%\end{center}
% \label{Symbols}
% \end{table}%
%$m_c$ & crossover mass for neutrals\\
%$m^{\scriptscriptstyle +}_c$ & crossover mass for ions\\
$m_i$ & mass of species $i$ [g] \\
$m^+_i$ & mass of ion $i$ [g] \\
$m_{\mathrm{H}_2} = m_2$ & mass of molecule H$_2$ [g]\\
$N_a$ & column density of atmosphere [number cm$^{-2}$]\\
$N_{\mathrm{H}_2}$ & column density of H$_2$ [number cm$^{-2}$]\\
$N_j$ & column density of $^j$Xe  [number cm$^{-2}$]\\
$n =\sum_i n_i$ & total number density  [cm$^{-3}$] \\
$n_e = \sum_i x_i$ & electron density [cm$^{-3}$] \\
$n_i$ & number density of species $i$ [cm$^{-3}$] \\
$Q$ & thermal conduction flux [ergs\,cm$^{-2}$s$^{-1}$] \\
$Q_q$ & quadrupole polarizability [Debye \AA] \\
$R_{\oplus}$ & radius [cm] of Earth\\
$R$, $R_x$ & internuclear distance, [$a_0$] in atomic units \\
$r$ & radial distance [cm] \\
$r_0$ & radius to lower boundary of the model [cm] \\
$S \equiv F_{\mathrm{xuv}}/F_{\mathrm{xuv}\oplus}$ & XUV/FUV irradiation relative to modern Earth \\
$T$ & temperature [K] \\
$t$ & time [Gyr] \\
$t_A$, $t_B$ & starting and ending times [Gyr] of a period of Xe fractionation\\
$u=v_1=v_2$ & uniform outward mean velocity of hydrogen [cm\,s$^{-1}$] \\
$v_i$ & outward mean velocity of species $i$ [cm\,s$^{-1}$] \\
$v_{sw}$ & solar wind velocity [cm s$^{-1}$]\\
% $v_t$ & Xe$^+$ terminal velocity [cm s$^{-1}$]\\
$x_i$ & number density of ion $i$ [cm$^{-3}$] \\
$y = f_2/f_3$ & ratio of H$_2$ to CO$_2$ at the lower boundary\\
$z=r-r_0$ & altitude [cm] \\
$z_s$, $z_t$ & ionic charge\\
$\alpha$ & the $\alpha$ parameter in accretion disk theory\\
$\alpha_j$ & escape factor for Xe isotope $j$\\
$\alpha_{ij}\equiv \alpha_{130} - \alpha_{131} $ & fractionation factor between Xe isotopes separated by 1 amu\\
%$\beta=2$ & spherical geometry\\ 
%$\beta=3$ & dipolar geometry along the pole itself\\ 
$\beta=$ & \\ 
$\Gamma_{\mathrm{h}}$ & total volume radiative heating [ergs\,cm$^{-3}$s$^{-1}$] \\
$\Gamma_{\mathrm{c}}$ & total volume radiative cooling [ergs\,cm$^{-3}$s$^{-1}$] \\
$\Gamma_{\mathrm{H}^+_3}$ & broadband $\mathrm{H}^+_3$ cooling [ergs\,cm$^{-3}$s$^{-1}$] \\
$\Gamma_{bf}$ & bound-free cooling by $\mathrm{H}^+$ [ergs\,cm$^{-3}$s$^{-1}$] \\
$\Gamma_{f\!f}$ & free-free cooling by $\mathrm{H}^+$ [ergs\,cm$^{-3}$s$^{-1}$] \\
$\Gamma_{15} $ & CO$_2$ cooling at 15 $\mu$m [ergs\,cm$^{-3}$s$^{-1}$] \\
$\Gamma_{4.3} $ & CO$_2$ cooling at 4.3 $\mu$m [ergs\,cm$^{-3}$s$^{-1}$] \\
$\gamma_{i}$ & ratio of specific heats of species $i$ \\
$\eta$ & net heating efficiency for energy-limited escape\\
$\eta_h$ & radiative heating efficiency \\
$\eta_n$ & dipole polarizability [cm$^3$]\\
$\Theta\left(\Lambda,\tau_{x}\right)$ &  dimensionless function, Kumer \& James $F\left(\lambda,x\right)$\\
$\Lambda$ & dimensionless parameter, Kumer \& James $\lambda$\\
$\lambda$ & wavelength of light \\
$\mu = \sum_i n_i m_i/n$ & mean molecular mass [g] \\
$\mu_{ij}$ & reduced mass between two species $i$ and $j$ [g]\\
$\mu' = \rho'/(n_1+n_2)$ & hydrogen mean molecular mass [g] \\
$\mu^{\scriptscriptstyle +} = \sum_i x_i m_i/n_e$ & mean molecular mass of ions [g] \\
$\nu_{ij}$ & collision frequency for momentum transfer [s$^{-1}$] \\
$\nu_{in}$ & ion-neutral collision frequency for momentum transfer [s$^{-1}$] \\
$\nu_{st}$ & ion-ion collision frequency for momentum transfer [s$^{-1}$] \\
$\xi_j$ & depletion (total escape) of Xe isotope $j$\\
$\xi_{ij}$ & fractionation (per amu) between $^{130}$Xe and $^{131}$Xe.\\
$\rho'=n_1m_1+n_2m_2 $ & total hydrogen density [g\,cm$^{-3}$] \\
$\sigma_x$ & charge exchange cross-section [$a_0^2$ in a.u.] \\
$\sigma_{i\lambda}$ & absorption cross-section species $i$ at wavelength $\lambda$ [cm$^2$] \\
$\sigma_{\nu_2}$ & effective cross section for 15 $\mu$m ($\nu_2$) CO$_2$ band [cm$^2$]\\
$\sigma_{\nu_3}$ & effective cross section for 4.3 $\mu$m ($\nu_3$) CO$_2$ band [cm$^2$]\\
$\tau_{i\lambda}$ & optical depth of species $i$ at wavelength $\lambda$ \\
$\tau_{\lambda} = \sum_i \tau_{i\lambda}$ & total optical depth at wavelength $\lambda$ \\
$\tau_x$ & line center optical depth, Kumer \& James $x$\\
$\phi_i $ & flux of species or isotope $i$ [cm$^{-2}$s$^{-1}$] \\
$\phi_{\mathrm{H}_2} = 0.5\phi_1 + \phi_2$ & total hydrogen escape flux [cm$^{-2}$s$^{-1}$] \\
$\phi_{\mathrm{lim}}$ & diffusion limited flux $\phi_{\mathrm{H}_2}$ [cm$^{-2}$s$^{-1}$] \\
\end{longtable}  

\section*{Acknowledgments}
The authors thank JCG Walker, JF Kasting, T Donahue, R Kirshner, DM Hunten, JB Pollack, RO Pepin, G Wasserburg, M Ozima, B Marty, M Pujols, G Avice, S Mukhopadhay, A Hoffman, N Dauphas, JH Waite, and C Reinhard for specific advice and encouragement to quantify this hypothesis.
DCC acknowledges support from Simons Foundation SCOL Award 511570, and NSF Frontiers in Earth System Dynamics award No. 1338810.
KJZ and DCC acknowledge support from NASA Astrobiology Institute?s Virtual Planetary Laboratory grant NNA13AA93A.
MG was supported by a NASA NPP fellowship.
%This research did not receive any specific grant from funding agencies in the public, commercial, or not-for-profit sectors.



\end{document}